\documentclass[letterpaper,twocolumn,10pt]{article}
\usepackage{usenix-2020-09}

\usepackage{times}
\usepackage{epsfig}
\usepackage{bbm}
\usepackage{mathrsfs}
\usepackage{multirow}
\usepackage{paralist}
\usepackage{mathtools}
\usepackage{lipsum}

\usepackage{url}            % simple URL typesetting
\usepackage{booktabs}       % professional-quality tables
\usepackage{nicefrac}       % compact symbols for 1/2, etc.
\usepackage{threeparttable}
\usepackage{color}
\usepackage{subfigure}
\usepackage{multicol}

\usepackage{listings}
\usepackage{wrapfig}

\usepackage{pifont}
\usepackage[utf8]{inputenc}
\usepackage{subfigure}
\usepackage{comment}
\usepackage{amsmath,amssymb,amsfonts}
\usepackage{makecell}
\usepackage{indentfirst}
\usepackage{shorttoc}
\usepackage{caption}
\usepackage{bbm}
\usepackage{soul}
\usepackage{cite}
\usepackage{algorithmic}
\usepackage{graphicx}
\usepackage{textcomp}
\usepackage{tabularx} 
\usepackage{collcell}

\usepackage{yi_math}
\usepackage{array}
\usepackage{caption}
\usepackage{hhline}
\usepackage[linesnumbered, ruled, vlined]{algorithm2e}
\usepackage[table,xcdraw]{xcolor}
\usepackage{diagbox}
\usepackage{tikz}
\newcommand*\circled[1]{\tikz[baseline=(char.base)]{
            \node[shape=circle,fill,inner sep=1.5pt] (char) {\textcolor{white}{#1}};}}

\def\BibTeX{{\rm B\kern-.05em{\sc i\kern-.025em b}\kern-.08em
    T\kern-.1667em\lower.7ex\hbox{E}\kern-.125emX}}
\definecolor{deepred}{rgb}{0.631,0.102,0.102}
\definecolor{skyblue}{HTML}{126da2}
\definecolor{mildyellow}{HTML}{FFF2CC}
\newcommand{\AlgName}{\textsc{Meta-Sift~}}

\newenvironment{packeditemize}{
\begin{list}{$\bullet$}{
\setlength{\labelwidth}{8pt}
\setlength{\itemsep}{0pt}
\setlength{\leftmargin}{\labelwidth}
\addtolength{\leftmargin}{\labelsep}
\setlength{\parindent}{0pt}
\setlength{\listparindent}{\parindent}
\setlength{\parsep}{0pt}
\setlength{\topsep}{3pt}}}{\end{list}}

\usepackage{authblk}

\makeatletter
\newcommand{\printfnsymbol}[1]{%
  \textsuperscript{\@fnsymbol{#1}}%
}
\makeatother

\begin{document}

\title{
% Is it Still Feasible to Sift Out a Clean Data Subset \\in the Presence of Data Poisoning?
\AlgName: How to Sift Out a Clean Subset in the Presence of Data Poisoning?
\\


% \and
% \and
% \and
% \and
% \and
}
% \author{Anonymous Authors}

\author[1,2]{Yi Zeng\thanks{M. Pan and Y. Zeng contributed equally. Corresponding \href{mailto:yizeng@vt.edu}{\textbf{Y. Zeng}}, \href{mailto:lingjuan.Lv@sony.com}{\textbf{L. Lyu}} or \href{mailto:ruoxijia@vt.edu}{\textbf{R. Jia}}. Work partially done during Y. Zeng's internship at Sony AI.
% \textbf{Yi Zeng} and \textbf{Minzhou Pan} contribute equally.
}\printfnsymbol{1}}
\author[1]{Minzhou Pan\printfnsymbol{1}}
\author[1]{Himanshu Jahagirdar}
\author[1]{Ming Jin}
\author[2]{Lingjuan Lyu}
\author[1]{Ruoxi Jia}

\affil[1]{Virginia Tech, Blacksburg, VA 24061, USA}
\affil[2]{Sony AI, Tokyo, 108-0075, Japan}

\renewcommand\Authands{ and }
\vspace{-2em}
\maketitle

% \vspace{-30em}
\begin{abstract}
% \vspace{-1.5em}
% \ruoxi{Abs can be shortened. target at most 0.5 page for abs}
% Given the volume of data needed to train modern machine learning models, external suppliers are increasingly used. However, incorporating external data into training poses data poisoning risks, wherein malicious providers 
% manipulate their data to degrade model
% utility or integrity.
% Most data poisoning defenses presume access to a set of clean data 
% (referred to as the base set). 
% While this assumption has been taken for granted, given the fast-growing research on stealthy poisoning attacks,
% a question arises: can defenders really identify a clean subset within a contaminated dataset to support defenses? \ruoxi{tone down the question}
% \rev{
% As the demand for data to train modern machine learning models continues to escalate, the reliance on external suppliers has become increasingly prevalent. 
External data sources are increasingly being used to train machine learning (ML) models as the data demand increases. However, the integration of external data into training poses data poisoning risks, where malicious providers manipulate their data to compromise the utility or integrity of the model. Most data poisoning defenses assume access to a set of clean data (referred to as the base set), which could be obtained through trusted sources. But it also becomes common that entire data sources for an ML task are untrusted (e.g., Internet data). In this case, one needs to identify a subset within a contaminated dataset as the base set to support these defenses.
% }
% Given the amount of data required to train modern machine learning models, it is increasingly common to harvest data from external providers. However, incorporating external data sources into training poses data poisoning risks, wherein malicious providers manipulate their data to degrade model performance or control model prediction. Most existing defenses against data poisoning assume access to a set of clean data (referred to as the base set hereinafter). While this assumption has been taken for granted, given the fast-growing research on stealthy data poisoning techniques, an important question arises: can the defender really identify a clean subset within a contaminated dataset to support the defenses? 

This paper starts by examining the performance of defenses when poisoned samples are mistakenly mixed into the base set. We analyze five representative defenses that use base sets and find that their performance deteriorates dramatically
% (e.g., attack success rate over 80\%) 
with less than 1\% poisoned points in the base set. 
% Surprisingly, even one poisoned sample can nullify the effect of a state-of-the-art poisoned data detector. 
These findings suggest that sifting out a base set with \emph{high precision} is key to these defenses' performance. 
Motivated by these observations, we study how precise existing automated tools and human inspection are at identifying clean data in the presence of data poisoning. Unfortunately, neither effort achieves the precision needed that enables effective defenses. Worse yet, many of the outcomes of these methods are worse than random selection.

In addition to uncovering the challenge, we take a step further and propose a practical countermeasure, \AlgName. 
% We not only identify the challenge but also propose a practical solution termed \AlgName. 
Our method is based on the insight that existing poisoning attacks 
shift data distributions, resulting in high prediction loss when training on the clean portion of a poisoned dataset and testing on the corrupted portion. 
% use data manipulation techniques that cause shifts from clean data distributions. Hence, training on the clean portion of a poisoned dataset and testing on the corrupted portion will result in high prediction loss. 
Leveraging the insight, we formulate a bilevel optimization to identify clean data and further introduce a suite of techniques to improve the efficiency and precision of the identification.
Our evaluation shows that \AlgName can sift a clean base set with 100\% precision under a wide range of poisoning threats. The selected base set is large enough to give rise to successful defense when plugged into the existing defense techniques.

% In addition to uncovering the challenge of identifying a clean base set with high precision, we take a step further and propose \AlgName to resolve the challenge. Our approach is based on an insight that data manipulation techniques exploited by existing poisoning attacks inevitably result in a distributional shift from the clean data. Hence, training on the clean portion of the contaminated dataset and testing the trained model on the other corrupted portion will lead to a high prediction loss. We formulate a novel bilevel optimization problem to split the contaminated dataset in a way that training on one split and testing on the other leads to the highest prediction loss. We introduce a suite of techniques to improve the efficiency and precision of clean data selection. Our evaluation shows that \AlgName can robustly sift out a clean base set with 100\% precision under a wide range of poisoning attacks. The selected base set is large enough to give rise to successful defense when plugged into the existing defense techniques.

\end{abstract}

% \begin{IEEEkeywords}
% Data Poisoning, Backdoor Attacks, AI Security
% \end{IEEEkeywords}

% \vspace{-1.3em}
% \vspace{-1.2em}
\section{Introduction}
\label{sec:intro}
\vspace{-1.1em}
% Security in cyber systems traditionally demands isolation from the outside world via access control.
% % measures such as firewalls, user authentication, and data encryption. 
% On the other hand, constructing high-performance machine learning~(\textit{ML}) systems necessitates a vast amount of open world data. Data gathering is generally outsourced to anonymous and unverified third parties. This lack of reliable supervision exposes ML systems to malicious behavior on account of data poisoning, which has been remarked as the top concern in the industry when using ML as a service \cite{kumar2020adversarial}. 

Constructing high-performance machine learning~(\textit{ML}) systems requires large and diverse data. The data-hungry nature will inevitably force individuals and organizations to leverage data from external sources, the beginning of which is already evident. For instance, CLIP \cite{radford2021learning}, the state-of-the-art image representation, is learned from 400 million image-text pairs collected from the Internet. Various data marketplaces and crowd-sourcing platforms also emerge to enable data exchange at scale. While incorporating external data sources into training has clear benefits, it exposes ML systems to security threats on account of data poisoning attacks, in which attackers modify training data to degrade model performance or control model prediction. In fact, data poisoning has been remarked as the top security concern regarding ML systems in the industry~\cite{kumar2020adversarial}.

% In contrast to traditional cyber systems' security, which typically isolates systems from the outside world via access control measures like firewalls, passwords, or data encryption, machine learning~(\textit{ML}) systems are designed to embrace the outside world. In particular, training high-performing ML models necessitates a vast amount of data. Thus, practitioners are forced to outsource anonymous and unverified third parties. The lack of reliable supervision over data collection opens the door for malicious behaviors in ML systems via data poisoning and has been remarked as the top concern in the industry when using ML as a service \cite{kumar2020adversarial}. 

%Security in cyber systems traditionally demands isolation from the outside world via access control measures like firewalls, user-authentication or data encryption. On the other hand, training high-performance/large-scale machine learning~(\textit{ML}) systems requires easy access to a vast amount of open public data. Data collection is generally outsourced to anonymous and unverified third parties. This lack of reliable supervision exposes ML systems to malicious behavior; data poisoning has been remarked as the top concern in the industry when using ML as a service. 

In this paper, the term ``data poisoning'' will be used in a broad sense, referring to attacks that involve training data manipulation. In particular, it includes both the attacks that interfere only with training data \emph{and} backdoor attacks that embed a backdoor trigger during the training time and further inject the trigger into test-time inputs to control their corresponding predictions.
Within the scope of this paper, we divide existing data poisoning attacks into three categories based on the attribute being manipulated:
\begin{compactitem}[$\bullet$]
% \begin{itemize}
% \vspace{-.5em}
    \item \textbf{Label-only attacks} that only alter labels, such as targeted~\cite{tolpegin2020data} and Random Label-Flipping attacks \cite{ren2018learning} aimed at degrading model utility; 
    \item \textbf{Feature-only attacks} that only manipulate features without changing the labels, such as feature collision attacks \cite{shafahi2018poison} and clean-label backdoor attacks \cite{turner2019label,zeng2022narcissus};
    \item \textbf{Label-Feature attacks} that change both feature and label, such as standard backdoor attacks~\cite{gu2017badnets,chen2017targeted,nguyen2021wanet}.
    % would fall into this category.
    % \vspace{-.5em}
\end{compactitem}
% \end{itemize}
% In particular, many of these poisoned data can infiltrate the training set by simply being posted on the web \cite{goldblum2022dataset} or being delivered directly to a dataset aggregator, e.g., a chatbot \cite{wakefield_2016}, or a large scale public benchmark dataset (ImageNet) \cite{anonymous}.

Intensive efforts have been invested in mitigating data poisoning. The types of defenses in the prior work range from identifying poisoned samples in a training set~\cite{zeng2021rethinking} (\textbf{Poison Detection}) to detecting whether a model has been trained on a poisoned dataset~\cite{xu2021detecting} (\textbf{Trojan-Net Detection}) to removing backdoors from a poisoned model~\cite{wang2019neural,zeng2021adversarial} (\textbf{Backdoor Removal}) to redesigning training algorithms to prevent poisoning from taking effect~\cite{ren2018learning} (\textbf{Robust Training})
% \llv{maybe put Poison Detection and Robust Training closer, as they are related. suggest number these types for clarity.}. 

Most existing defenses assume that \emph{the defender can access a set of clean data} (referred to as the \emph{base set} hereafter). Despite the prevalence of the assumption in existing literature, focused discussion about its validity is lacking. If the defender were capable of collecting a set of clean samples from trusted sources of data, then this assumption could be met easily. However, it has become increasingly common to learn solely from untrusted data sources, such as training with the data scraped from the Internet or purchased from specific vendors. In that case, the defender needs to identify a clean subset within the poisoned dataset to form the base set. \emph{Many important questions remain unclear}: 
\noindent\fcolorbox{deepred}{mildyellow}{\begin{minipage}{0.97\columnwidth}
How does the defense performance change if the identification is imperfect and some poisoned data are mixed into the base set? Are there any existing automated methods that can reliably identify a clean base set in the presence of various types of poisoning attacks? Can human inspection fulfill the need? If not, how can we reliably identify enough clean samples to support those defenses?
\end{minipage}}
% \emph{How does the defense performance change if the identification is imperfect and some poisoned data are mixed into the base set? Are there any existing automated methods that can reliably identify a clean base set in the presence of various types of poisoning attacks? Can human inspection fulfill the need?}

%\vspace{0.2em}
\noindent
\textbf{\ul{Takeaway \#1: Defense performance is sensitive to the purity of the base set.}} We start by examining the sensitivity of defense performance to the ratio of poisoned points in the base set. We study five representative defense techniques that rely on access to a base set. The techniques considered either achieve state-of-the-art performance or are popular baselines. We find that their performance degrades significantly (e.g., attack success rate exceeding 80\%) with less than 1\% of poisoned points in the base set. Surprisingly, even a single poisoned point is sufficient to nullify the effect of a state-of-the-art poisoned data detector. These findings suggest that the ability to sift out a base set with \emph{high precision} is critical to successfully applying these defenses.

\begin{figure}[t!]
  \centering
     \vspace{-2em}
\includegraphics[width=0.9\linewidth]{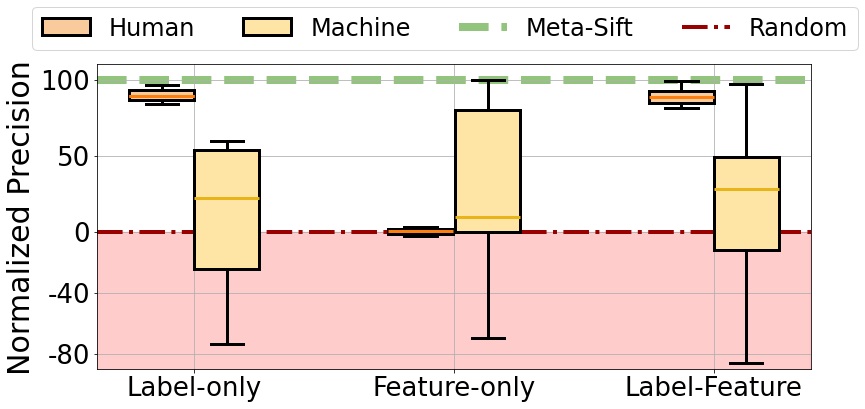}
% \includegraphics[width=\linewidth]{figs/compare.png}
%   %\vspace{-2em}
\vspace{-1em}
  \caption{A comparison of the normalized precision of existing automated methods (Machine), Human, and \AlgName in sifting out a clean subset from a poisoned CIFAR-10.
  Both human, machine-based, and \AlgName results are normalized with the poison ratio to ensure comparability. 
%   \ruoxi{do we also normalize the results for meta-sfiter? if so, we should also include it} 
  A larger value indicates a stronger filtering capability.
%   The larger the value is, the stronger the filtering capability of obtaining a clean subset of the method. 
  The \scalebox{0.9}{\colorbox[HTML]{FFCCCC}{\textbf{red}}} region depicts the filtering capability worse than random selection.
%   \ruoxi{fix the y-axis label to be "normalized precision"}
  }
  \label{fig:compare}
  \vspace{-1.5em}
\end{figure}

\noindent
\ul{\textbf{Takeaway \#2: Both existing automated methods and human inspection fail to identify a clean subset with high enough precision.}} We investigate how precise existing automated methods and human inspection can be in identifying clean data in the presence of data poisoning and the result is illustrated in Figure \ref{fig:compare}.
The precision of both humans and existing automated methods varies a lot across different attack categories. Humans are proficient at identifying poisoned samples that involve label changes, including Label-only attacks and Label-Feature attacks, and outperform existing automated methods by a large margin. However, humans still miss many poisons and cannot realize a 100\% success rate in sifting out a clean base set.   
Notably, for these two attack categories, several automated methods even underperform the random baseline.
% and hence receive a negative rating of filtering capability. 

On the other hand, for Feature-only attacks, human inspection results in a precision close to the random baseline.
% Feature-only attacks are the worst case for both automated methods and human.
As these attacks inject small perturbations only to the features while not changing the overall semantics,
human experts perform worse than most automated methods. 
% In particular, the performance is not better than random selection. 
\emph{This finding is in direct contrast to the traditional wisdom that treats human supervision as the final backstop of data poisoning.} Besides being time-consuming and cost-intensive, human inspection becomes less trustworthy in identifying poisoned data given the fast-growing research on stealthy attacks.
Overall, both existing automated methods and human inspection cannot reach the level of precision required to enable successful defense. 
% \ruoxi{takeaway two needs to be revised after the figure is fixed.}

% The fact that existing methods cannot reliably sift out clean subsets in an attack-agnostic manner is problematic.
% Security concerns arise when such poisoned data (e.g., labeling errors, poisoning, and backdoor attacks) infiltrate the validation set.
% The corrupted instances in the validation set might mislead the training algorithm by marking the poisoned training points as high-value points.
% This impedes the user from noticing potential security risks (e.g., observing a specific class's accuracy as too low, identifying possible targeted Label-Flipping attacks, etc.). 
% Even worse, as shown in our evaluation (TABLE \ref{table:fail}), if the poisoned samples sneak into the validation set, a small amount can largely degrade the defensive performance, or backdoor your backdoor defenses.
%\newline

% \noindent
% \underline{\textbf{Our Contributions.}}
% In this work, we identify this overlooked problem of sifting out a clean subset given a poisoned dataset. Our insight on why existing methods fail to reliably sift out clean subsets is that their filtering criteria is not general enough. For example, human inspection is based on visual and semantic mismatch and is not applicable to detecting Feature-only attacks (Figure \ref{fig:compare}). 

%\vspace{0.2em}
\noindent
\ul{\textbf{Takeaway \#3: $\AlgName$--- a scalable and effective automated method to sift out a clean base set.}} We propose \AlgName to sift a clean subset from the poisoned set. Our approach is based on a novel insight that
% Intuitively, an ML model trained over clean samples will react poorly to unseen poisoned inputs and vice versa, irrespective of the underlying poisoning attack.
% Take the dirty-label backdoor attack---an instance of Label-Feature attacks---as an example. When a clean model observes a (backdoored) sample with a manipulated label, it returns a larger loss than when it observes a benign sample. 
% The reason is that the observed (backdoored) sample with a manipulated label does not share a close distribution to the clean data originally trained the model. 
% Similarly, a model trained \textit{only} on poisoned samples would return large losses to predict clean samples. 
% This behavior can be expected for Feature-only and Label-only attacks, too. 
% The high-level idea of our proposed algorithm is to identify a clean subset by \textbf{splitting the contaminated dataset in a way that training on one split and testing on the other leads to the highest prediction loss}.
data manipulation techniques exploited by existing poisoning attacks inevitably result in a distributional shift from the clean data. Hence, training on the clean portion of the contaminated dataset and testing the trained model on the other corrupted portion will lead to a high prediction loss. We formulate a bilevel optimization problem to split the contaminated dataset in a way that training on one split and testing on the other leads to the highest prediction loss. However, this splitting problem is hard to solve exactly as it has a combinatorial search space and at the same time, contains two nested optimization problems. To address the computational challenge, we first relax it into a continuous splitting problem, where we associate each sample with a continuous weight indicating their likelihood of belonging to one of the splits and then optimize the weights via gradient-based methods. Secondly, we adapt the online algorithm that was originally designed for training sample reweighting~\cite{xu2021faster} to efficiently solve the continuous relaxation of the bilevel problem. Furthermore, we adopt the idea of ``ensembling'' to improve the precision of selection. In particular, we propose to apply random perturbations to each point, run the online algorithm on each perturbed version to obtain a weight, and aggregate the weights for final clean data selection.
% bilevel problems are known to be hard to solve and worse yet, the splitting problem is combinatorial in nature 
% bilevel optimization is known to be costly to solve. 
% To address the computational challenge, we first relax the binary splitting problem into a continuous one (which can take advantage of gradients to search for the solution). 
% Notably, the online algorithm we proposed only requires going through the whole dataset twice, i.e., only two epochs of training suffice.
% Inspired by state-of-the-art meta-learning references \cite{xu2021faster}, we adopt a novel solution with gradient samplers that adaptively select part of the model structures while constructing the hypergradient using the chain rule.
% Remarkably, the proposed method can conduct effective data splitting after \textit{only} two epochs of optimization, i.e., it only requires going through the whole dataset twice for parameter updating.
% With the efficient design, we can aggregate the selection results from multiple independent runs of the optimization and thus obtain reliable results. 
% \ruoxi{give more dtails about adaptive selection of hypergradient computing; what is it? what is the problem that it's solving} 
Our evaluation shows that \AlgName can robustly sift out a clean base set with 100\% precision under a wide range of poisoning attacks. The selected base set is large enough to give rise to successful defense when plugged into the existing defense techniques. It is worth noting that \AlgName significantly outperforms the existing automated methods (illustrated in Figure \ref{fig:compare}) while being orders of magnitude faster (Table \ref{tab:cifar}, \ref{tab:imagenet}, \ref{tab:gtsrb}, \ref{tab:pubfig}).

Our \ul{contributions} can be summarized as follows:
\begin{packeditemize}
    \item We \textbf{identify an overlooked problem} of the accessibility of a clean base set in the presence of data poisoning. 
    % Many existing defenses preassume such a base set; however, it might not be valid under nowadays attacks.
    % in existing data poison defenses, which may significantly impair the integrity of both existing data poisoning defenses and ML models;
    \item We \textbf{systematically evaluate the performance of existing automated methods and human inspection} in distinguishing between poisoned and clean samples;
    \item We \textbf{propose a novel splitting-based idea} to sift out a clean subset from a poisoned dataset and \textbf{formalize it into a bilevel optimization problem}.
    \item We \textbf{propose \textsc{Meta-Sift}}, comprising an \textbf{efficient algorithm} to solve the bilevel problem as well as \textbf{a series of techniques to enhance sifting precision}.
    % dataset splitting problem based on an intuitive, fundamental, and general observation of clean and poisoned subsets. The proposed formulation \textbf{only assumes that poisoned samples are among the minority} in the dataset.
    % \item We \AlgName to resolve the identified problem in an accurate and efficient way;
    \item We \textbf{extensively evaluate \AlgName} and compare with existing automated methods on four benchmark datasets under twelve different data poisoning attack settings. \textbf{Our method significantly outperforms existing methods} in both sifting \textbf{precision} and \textbf{efficiency}. At the same time, plugging our sifted samples into existing defenses achieves \textbf{comparable or even better performance} than plugging in randomly selected clean samples.
    % Thorough ablation is studied on the important choice point when designing the \AlgName.% Through?
    \item We \textbf{open-source the project} to promote research on this topic and facilitate the successful application of existing defenses in settings without a clean base set
    \footnote{\url{https://github.com/ruoxi-jia-group/Meta-Sift}}.
    % \footnote{\url{https://anonymous.4open.science/r/Meta_sift_sample_code}}.
    % 
\end{packeditemize}

\vspace{-1em}
\section{Sifting Out a Clean Enough Base Set is Hard}
\label{sec:human_study}
\vspace{-0.5em}
The ability to acquire a clean base set was taken for granted in many existing data poisoning defenses~\cite{wang2019neural,guo2019tabor,liu2018fine,zeng2021adversarial,xiang2020revealing, shu2019meta}. 
For instance, a popular Trojan-Net Detection strategy is to first synthesize potential trigger patterns from a target model and then inspect whether there exists any suspicious pattern \cite{wang2019neural,guo2019tabor}. Trigger synthesis is done by searching for a pattern that maximally activates a certain class output when it is patched onto the clean data. Hence, access to a clean set of data is indispensable to this defense strategy. Another example is defenses against Label-Flipping attacks (often referred to as mislabeled data detection in ML literature). State-of-the-art methods detect mislabeled data by finding a subset of instances such that when they are excluded from training, the prediction accuracy on a clean validation set is maximized. A clean set of instances are needed to enable these methods.

% \ruoxi{standardize the format of section/subsection titles, capitalize or not. remove period at the end of the titles}
\vspace{-1em}
\subsection{Defense Requires a Highly Pure Base Set}
\label{sec:success}
\vspace{-0.5em}

TABLE \ref{table:fail} summarizes some representative techniques that rely on access to a clean base set in each of the aforementioned defense categories, namely, Poison Detection, Trojan-Net Detection, Backdoor Removal, and Robust Training against label noise. These techniques either achieve the state-of-the-art performance (e.g., Frequency Detector~\cite{zeng2021rethinking}, I-BAU~\cite{zeng2021adversarial}, MW-Net~\cite{shu2019meta}) or are widely-adopted baselines (e.g., MNTD~\cite{xu2021detecting} and Neural Cleanse (NC)~\cite{wang2019neural}). In particular, MNTD is implemented as a base strategy in an ongoing competition for Trojan-Net Detection\footnote{\url{https://trojandetection.ai/}}. 
% The metrics for Poisoned Detection, 

% \ruoxi{Yi: please add details for the metrics. how are they defined? please also describe the baseline method in each category}
While conventionally, these defense techniques only report their performance based on a completely clean base set, given the fast-advancing research on stealthy attacks, it is possible that some poisoned samples may go unnoticed and get selected into the base set by mistake. Hence, it is critical to evaluate how the performance of these defenses depends on the ratio of the poisoned samples in the base set.

We adopt widely used metrics to measure defense performance for each defense category. Specifically, for Poison Detection, we use \textbf{Poison Filtering Rate (PFR)}, which measures the ratio of poisoned samples that are correctly detected. For Trojan-Net Detection, we follow the original work of MNTD and use the \textbf{Area Under the ROC Curve (AUC)} as a metric, which measures the entire two-dimensional area underneath the ROC curve\footnote{An ROC curve plots the true positive rate vs. the false positive rate at different classification thresholds}. 
\emph{The most naive baseline for poison detection and Trojan-Net detection is random deletion, which ends up with a PFR of 50\% and an AUC of 50\%.} 
The closer the performance of the defense in the Poison Detection and the Trojan-Net Detection category gets to 50\%, the weaker the defense is. For backdoor removal, we use the \textbf{Attack Success Rate (ASR)}, which calculates the frequency with which non-target-class samples patched with the backdoor trigger are misclassified into the attacker-desired target class. For Robust Training, we use the \textbf{Test Accuracy (ACC)}, which measures the accuracy of the trained model on a clean test set.
% Test Accuracy (ACC) is used for evaluating robust training methods. 
\textit{The baselines for Backdoor Removal and Robust Training are simply the deployment of no defenses at all.} We report ASR or ACC that is obtained directly from training on the poisoned dataset. The closer the performance of defense in these two categories gets to these baselines, the weaker the defense is.

We compare the resulting defense performance against standard attacks (e.g., BadNets \cite{gu2017badnets}, Random Label-Flipping) between clean and corrupted base sets (Table \ref{table:fail}). 
% \llv{any table or figure to support below observation?}
% As illustrated, with less than 3\% of adversarial points existing in the base set, the performance of those defenses can be largely weakened. 
% \ruoxi{Yi: in depth discussion of the result: what does largely weakened mean numerically? which defense category is more susceptible to noise in base set? Why is it the case? }
% We show the results of these popular or state-of-the-art defenses on mitigating standard attacks (e.g., BadNets \cite{gu2017badnets}, random Label-Flipping) with clean or corrupted base sets. As illustrated, 
For Poisoned Detection with Frequency Detector, even one poisoned example sneaking into the base set is sufficient to nullify the defensive effect, leading to a performance worse than the 
random baseline. For MNTD, with 1\% of poisoned examples mixed into the base set, the AUC drops by almost 40\%. Comparing the two techniques for Backdoor Removal, we can find that I-BAU is more sensitive to corruption of the base set than NC. Both techniques patch a trigger to partial samples in the base set to fine-tune the poisoned model, aimed at forcing the model to ``forget'' the wrong association between the trigger and the target label. Compared to NC, the design of I-BAU selects fewer samples in the base set to be patched with a trigger. Hence, the positive ``forgetting'' effect introduced by these samples is more likely to be overwhelmed by the negative effect caused by poisoned examples sneaking into the base set. This explains the larger sensitivity of I-BAU to corruption of the base set. For both techniques, less than 3\% of corruption in the base set is adequate to bring the ASR back above $60\%$. For Robust Training with MW-Net, 20 mislabeled samples in the base set can reduce the accuracy by about 10\%. Overall, we can see that base sets with high purity are crucial to enable the successful application of these popular defenses requiring access to base sets.

% Frequency Detector and MNTD, with 1\% of 
% For with less than 3\% of adversarial points existing in the base set, the performance of those defenses can be largely weakened. In particular, even one sample sneak into the validation set can ``backdoor'' the frequency detector to overlook almost all the poisoned samples. One MNTD, with 1\% of samples as the poisoned samples, the MNTD detector's AUC will drop by almost 40\%. In the backdoor removal case, the NC-protected model will remain an ASR above 80\% if 30 samples out of 1000 are poisoned. I-BAU is more sensitive to the number of poisoned samples in the base set, as only 0.8\% suffice to reduce I-BAU's efficacy by 69.24\%. I-BAU is more sensitive to the number of poisons in the base set because they used an unlearning rate (number of samples patched with the synthesized trigger but with ground-truth labels) of 1\%, and NC uses an unlearning rate of 20\%.
% With 20 label-noise samples sneaked into the base-set for MW-Net, the learning efficacy will drop by 10\% on the CIFAR-10. To conclude, many existing defenses to maintain their defense efficacy require a pure base set without corruption.

\begin{table}[t!]
\centering
  % \vspace{-1.5em}
\resizebox{0.9\columnwidth}{!}{
% \resizebox{0.76\columnwidth}{!}{
\begin{tabular}{l|c|c|cc|c}
\hline
 &
  \textbf{\begin{tabular}[c]{@{}c@{}}Poison\\ Detetcion\end{tabular}} &
  \textbf{\begin{tabular}[c]{@{}c@{}}Trojan-Net\\ Detetcion\end{tabular}} &
  \multicolumn{2}{c|}{\textbf{\begin{tabular}[c]{@{}c@{}}Backdoor\\ Removal\end{tabular}}} &
  \textbf{\begin{tabular}[c]{@{}c@{}}Robust\\ Training\end{tabular}} \\ \cline{2-6} 
\multirow{-2}{*}{} &
  \begin{tabular}[c]{@{}c@{}}Frequency\\ Detector\cite{zeng2021rethinking}\end{tabular} &
  \begin{tabular}[c]{@{}c@{}}MNTD\\ \cite{xu2021detecting}\end{tabular} &
  \begin{tabular}[c]{@{}c@{}}NC\\ \cite{wang2019neural}\end{tabular} &
  \begin{tabular}[c]{@{}c@{}}I-BAU\\ \cite{zeng2021adversarial}\end{tabular} &
  \begin{tabular}[c]{@{}c@{}}MW-Net\\ \cite{shu2019meta}\end{tabular} \\ \hline
\textbf{\begin{tabular}[c]{@{}l@{}}Task/\\ Settings\end{tabular}} &
  \begin{tabular}[c]{@{}c@{}}Detecting\\ BadNets;\end{tabular} &
  \begin{tabular}[c]{@{}c@{}}BadNets 5\%;\\ Target: 2;\end{tabular} &
  \multicolumn{2}{c|}{\begin{tabular}[c]{@{}c@{}}BadNets 5\%;\\ Target: 38;\end{tabular}} &
  \begin{tabular}[c]{@{}c@{}}20\% Random\\ Label-Flipping;\end{tabular} \\ \hline

\textbf{\begin{tabular}[c]{@{}l@{}}Base Set\end{tabular}} &
  \begin{tabular}[c]{@{}c@{}}100-CIFAR-10\end{tabular} &
  \begin{tabular}[c]{@{}c@{}}1000-MNIST\end{tabular} &
  
  \multicolumn{2}{c|}{\begin{tabular}[c]{@{}c@{}}1000-GTSRB\end{tabular}} &
  \begin{tabular}[c]{@{}c@{}}100-CIFAR-10\end{tabular} \\ \hline

\textbf{Metric }           & PFR ($\uparrow$ \%)    & AUC ($\uparrow$ \%)      & \multicolumn{2}{c|}{ASR ($\downarrow$ \%) } & ACC ($\uparrow$ \%)     \\ 

\hhline{-|-|-|-|-|-}
  
\rowcolor[HTML]{FFE599} 
{\begin{tabular}[c]{@{}c@{}}\textbf{Baseline}  \end{tabular}} & 
{\begin{tabular}[c]{@{}c@{}}Random: 50 \end{tabular}}& 
{\begin{tabular}[c]{@{}c@{}}Random: 50 \end{tabular}}      & 
\multicolumn{2}{c|}{{\begin{tabular}[c]{@{}c@{}}No Def: 97.43 \end{tabular}}}      & {\begin{tabular}[c]{@{}c@{}}No Def: 69.99 \end{tabular}}  \\ [-0.5pt]
\hhline{-|-|-|-|-|-}

\rowcolor[HTML]{D9EAD3} 
\textbf{\# poison}               & 0/100 & 0/1000 & 0/1000     & 0/1000     & 0/100 \\ \hhline{-|-|-|-|-|-}

\rowcolor[HTML]{D9EAD3} 
{\textbf{Original}} & 99.95 & 99.92   & 18.83        & 12.58      & 91.18  \\ \hhline{-|-|-|-|-|-}

\rowcolor[HTML]{FFCCCC} 
\textbf{\# poison }              & 1/100 & 10/1000 & 30/1000     & 8/1000     & 20/100 \\ \hhline{-|-|-|-|-|-}

\rowcolor[HTML]{FFCCCC} 
{\textbf{After}}    & 3.11  & 62.78   & 62.67       & 81.82      & 81.84  \\ \hhline{-|-|-|-|-|-}
\end{tabular}
}
\vspace{-.8em}
\caption{
Defenses in the case of using a corrupted base set. For each category of defense, we use different metrics according to these original works.
\scalebox{0.9}{\colorbox[HTML]{ffe599}{\textbf{Baseline}}} results are the settings with random guessing as defense or without defenses; \scalebox{0.9}{\colorbox[HTML]{D9EAD3}{\textbf{Original}}} results is the settings assuming an access to clean base set; \scalebox{0.9}{\colorbox[HTML]{FFCCCC}{\textbf{After}}} shows the results of using a contaminated base set. 
}
\label{table:fail}
\vspace{-1.5em}
\end{table}

\vspace{-1em}
\subsection{The Data Sifting Problem}
\vspace{-0.5em}
The sensitivity of defense performance to the purity of the underlying base set motivates us to study the \textbf{Data Sifting Problem}: \emph{How to sift out a clean subset from a given poisoned dataset?}
We highlight some unique challenges and opportunities towards answering this question.
\begin{packeditemize}
% \vspace{-.5em}
    \item (\ul{Challenge}) \textbf{High precision}: The empirical study in Section~\ref{sec:success} demonstrates that the defensive performance could drop significantly with a small portion of corruption in the base set. Hence, sifting out clean samples with high precision is crucial to ensure defense effectiveness.
    \item (\ul{Challenge}) \textbf{Attack-Agnostic}: In practice, the defender usually does not know the underlying attack mechanisms that the attacker used to generate the poisoned samples. Hence, it is important to ensure high precision across different types of poisoning attacks.
    \item (\ul{Opportunity}) \textbf{Mild requirement on the size of the sifted subset.} In contrast to the high requirement on the purity, the requirement of the size of the sifted subset is generally mild. The size of the base set required to enable an effective defense is usually much \textbf{smaller} compared to the size of the poisoned set. For instance, a clean base set of size less than 1\% of the whole poisoned dataset suffice to enable effective defenses~\cite{xiang2022post,zeng2021adversarial} .
    % the inspector or sifter does not know whether the given dataset is poisoned or not, and if poisoned, what kind of poisoning setting(s) was used;
    % \item \textbf{Validation-Set-Free}: solutions to the data sifting problem should be able to sift out the clean examples without additional knowledge on how the clean samples should be distributed;
    % \item \textbf{Robust to different attacks}: an ideal solution should be robust to different attack settings, thus requiring a more fundamental depiction of data poisoning attacks;
    % \item \textbf{Size is not a critical factor, but high precision is}: the base set size does not necessarily need to be large, as 10 to 1000 samples are already large enough to give rise to existing defenses \cite{xiang2022post,zeng2021adversarial}. However, the base set should be purely clean. Otherwise, many existing defenses can be unseated, as shown in TABLE \ref{table:fail}.
    % \vspace{-.5em}
\end{packeditemize}
Note that some attempts have been made in the prior work to lift the requirement on base sets in data poisoning defenses \cite{wang2020practical,gao2019strip,chen2019deepinspect,xu2021faster}. However, these works are focused on specific defense categories against specific attack settings. By contrast, when solving the data sifting problem one can obtain a highly pure base set that can be plugged into \emph{any} defense technique that requires the base-set-access. Hence, we argue that \textbf{solving the data sifting problem provides a more flexible pathway to address the base-set-reliance issue in current data-poisoning defense literature.}

\newpage
\vspace{-2.5em}
\subsection{How Effective are Existing Methods}
\vspace{-0.5em}
\label{sec:exist?}
We first consider automated methods that can potentially solve the data sifting problem. Note that the data sifting problem is similar to the traditional outlier detection problem, wherein the goal is to sift out the abnormal instances in a contaminated dataset. Ideally, if one could filter out all the abnormal instances perfectly, then the complement set can be taken to solve the data sifting problem. There are two \textbf{key differences} between data sifting and outlier detection. \emph{Firstly}, the data sifting problem is contextualized in data poisoning defense applications, where it is not necessary to sift out all the clean instances, but instead, a small subset of clean instances in the training set often suffices to support an effective defense. \emph{Secondly}, in outlier detection, it is often more important to achieve high recall at a given selection budget (i.e., a high proportion of true outliers is marked as outliers by the detection algorithm), whereas in the data sifting problem, high precision is the key to realize a successful defense (i.e., a high proportion of the ``marked-as-clean'' points is truly clean). 

At a technical level, the outlier detection algorithms are still applicable to approach the data sifting problem. In particular, existing outlier detection algorithms assign an ``outlier score'' to each data point, indicating their likelihood to be an outlier. To re-purpose these algorithms for data sifting, we select the points with the lowest ``outlier scores'' of each class to form the balanced base sets. We evaluate some representative outlier detection methods that do not rely on additional clean data to function and examine their potential to solve the identified problem. 
% Note that the data sifting 
% In this section, we will thoroughly evaluate and challenge whether existing statistical methods under extended settings can automatically sift out clean data from poisoned datasets in an attack-agnostic manner. 
% It is important to understand that no existing methods were specifically designed for sifting out a clean subset.
% Rather, most of the existing related schemes were designed for poison detection \cite{tran2018spectral}, i.e., looking at the one end of data-based scores to detect poisoned data rather than identifying which sample is clean. Can we identify clean data by looking at the other end of these scores?
% \noindent
% \underline{\textbf{automated Methods.}}
% In this section, we adapt and redesign existing state-of-the-art data selection methods 
% % \himanshu{yi? Synthesis?}
% that can sift out ``clean data'' in a validation-data-free manner: 
Specifically, we evaluate:
% the following methods:
\begin{packeditemize}
\vspace{-.5em}
    \item Distance to the Class-Means \underline{(\textit{DCM})}: We compute the mean of each class at the input-space (pixel-level) as the class center and assign the outlier score to a point based on the distance to its center. 
    % \ruoxi{average in pixel space or feature space, pls specify}
    % This is a natural extension of the clustering concept \cite{xu2015comprehensive}. The idea is that clean samples may be closer to the averaging point of each class. We compute the averaging point for each class as the class centers (10 in total) and obtain the ranking of each data point in the CIFAR-10 to its class centers. The top 100 samples from each class are then chosen for sifting results.
    
    \item Distance to Model-Inversion-based CM \underline{(\textit{MI-DCM})}: MI-DCM differs from DCM in the choice of the class center. Here, each class center is obtained by conducting the model inversion attack~\cite{fredrikson2015model} on the model trained on the entire dataset. Model inversion is a type of privacy attack aimed at reconstructing the representative points for each class from the trained model. 
    % \ruoxi{any reference for this outlier detection method? if so, pls add}
    
    % further considers a converged model's knowledge for synthesizing the class centers. The difference between MI-DCM and DCM is that MI-DCM's class centers are acquired by conducting model inversion \cite{fredrikson2015model} on each class. Model inversion is an extraction attack that recovers training sample information based on a pre-trained model. The optimization goal of a model inversion attack is to use the pre-trained model to find the input that most represents each class (which obtains the smallest loss based on an optimization process), which is aligned with the goal of finding ``class centers.''
    
    \item Spectral Filtering's Least Scores \underline{(\textit{SF-Least})}~\cite{tran2018spectral}: Spectral Filtering is an advanced outlier detection method in robust statistics. Its key idea is that the outliers will result in larger eigenvalues than expected eigenvalues of sample covariance. This idea has been applied to the detection of backdoored samples. To start with, we extract features for each sample using the model trained over the contaminated dataset
    and the outlier score of a sample is calculated based on the dot product between its feature and the top eigenvector of the sample covariance matrix.
    % a representative validation-data-free backdoor sample detection technique \cite{tran2018spectral}. \cite{tran2018spectral} adopts the observation that many backdoor attacks can leave detectable differences in the feature space of a pre-trained model. The original method conducts singular value decomposition of the covariance matrix of learned feature representations (features in the hidden layers) and uses this to compute an outlier score for each example. They then mark the top scoring samples as the outliers (poisons). We adapt this method to our case to select clean examples by choosing the points that scored the least values of each class (100 for each class).

    \item 
    Loss-based Poison Scanning
    \underline{(\textit{Loss-Scan})}~\cite{li2021anti}: Recent work has identified the difference between the losses of benign and backdoored samples \cite{li2021anti}. For most backdoor attacks, the losses of the poisoned sample would be lower than those of the benign sample during the early learning period (5 epochs). We adopt this notion and record the negative value of early training losses as the outlier score, i.e., the smaller the original loss, the greater the likelihood that the sample contains potentially hazardous information.
    
    \item Self-Influence-Function \underline{(\textit{Self-IF})}~\cite{koh2017understanding}: Influence functions are designed to measure the impact of each training point on the loss of the trained model and have been widely used to enhance training \cite{shao2021right} and repair mislabeled training data \cite{kong2021resolving}. In particular, Self-Influence-Function is a variant that measures the influence of a point on the training loss. The outlier score of a point here is taken to be the negative of the result returned by Self-Influence-Function.
    % Koh and Liang proposed interpreting an ML model's predictions by following them through their learning method and training data \cite{koh2017understanding}. They measure the impact of a training sample on a given loss value of a test sample, termed influence function (IF). This metric has also been used to enhance training \cite{shao2021right} and repair mislabeled training data \cite{kong2021resolving} based on calculated data values. We employ the Self-IF to suit our scenario in a validation-data-free manner, which calculates by replacing the test set with the training set and assessing data values based on how they contribute to the training process. The filtering result is obtained by selecting each class's top 100 score samples.
    % \vspace{-.5em}
\end{packeditemize}

\begin{table}[t!]
\centering
  % \vspace{-2em}
\resizebox{\columnwidth}{!}{
\begin{tabular}{c|c|ccccc} 
\hline
\diagbox[height=2.5em, width=11.5em]{\textbf{Poison attack}}{\textbf{Method}} & \multicolumn{1}{c|}{Random} & \multicolumn{1}{c}{DCM}                  & \multicolumn{1}{c}{MI-DCM$^*$} & \multicolumn{1}{c}{SF-Least$^*$} & \multicolumn{1}{c}{Loss-Scan$^*$} & \multicolumn{1}{c}{Self-IF$^*$}   \\ 
\hline
Targeted Label-Flipping\cite{tolpegin2020data} &    83.3                             & 92                                         & 93.3$\pm$0.5                           & \cellcolor[HTML]{FFCCCC}\textbf{82.7$\pm$0.5} & \cellcolor[HTML]{FFCCCC}\textbf{69.0$\pm$1.7}                           & \cellcolor[HTML]{FFCCCC}\textbf{72.0$\pm$0.8}                                       \\
Narcissus Backdoor\cite{zeng2022narcissus} &       90                                                         & \cellcolor[HTML]{FFCCCC}\textbf{90}                                           & \cellcolor[HTML]{FFCCCC}\textbf{0.0$\pm$0.0}                            & 94.00$\pm$0.0    & \cellcolor[HTML]{FFCCCC}\textbf{82.0$\pm$0.3}                       & \cellcolor[HTML]{FFCCCC}\textbf{88.3$\pm$0.9}                              \\
Poison Frogs\cite{shafahi2018poison} &         90                                                      & \cellcolor[HTML]{FFCCCC}\textbf{90} & \cellcolor[HTML]{FFCCCC}\textbf{72.7$\pm$2.9}                            & \cellcolor[HTML]{FFCCCC}\textbf{87.0$\pm$0.0}    & 92.4$\pm$0.7                        & 91.7$\pm$0.5                              \\
BadNets Backdoor\cite{gu2017badnets} & 66.7                                                              & \cellcolor[HTML]{FFCCCC}\textbf{63} & \cellcolor[HTML]{FFCCCC}\textbf{62.0$\pm$0.0}                       &76.0$\pm$0.8                      & 67.0$\pm$0.5      & 70.3$\pm$0.5                              \\
\hline
\hline
\textbf{Overhead (seconds)} &       NA                                                        & 10  & \cellcolor[HTML]{ffe599}\textbf{5411+300}                           & \cellcolor[HTML]{ffe599}\textbf{5411+15}                           & 210 & \cellcolor[HTML]{ffe599}\textbf{5411+19832}                              \\
\hline
\end{tabular}
}
\vspace{-.8em}
\caption{Automated methods' precision for sifting out a size-1000 CIFAR-10 subset (each class 100 samples). Precision is calculated using $\#$ correct clean samples in the selected subset divided by the total $\#$ samples for the respective target class (100). 
% We include random selection as the baseline to evaluate and compare. 
The \scalebox{0.9}{\colorbox[HTML]{FFCCCC}{\textbf{red}}} highlights the methods that perform even worse than random selection. The \scalebox{0.9}{\colorbox[HTML]{ffe599}{\textbf{yellow}}} highlights the methods that are not time-efficient. The methods with $^*$ denote that the setting includes randomness (e.g., resulting from deep learning model training); thus, we report the corresponding means and the standard deviations.
% \ruoxi{expalin the metrics}
}
\label{table:machine}
\vspace{-1.5em}
\end{table}

% \vspace{-.5em}
TABLE \ref{table:machine} shows the sifting results of existing methods on CIFAR-10. We include four popular attack strategies from the three categories of data poisoning in our evaluation, i.e., Targeted Label-Flipping, Narcissus backdoor, Poison Frogs, and BadNets backdoor. Settings are detailed in Appendix \ref{sec:detailed_atk_0}.

As shown in TABLE \ref{table:machine}, none of the existing automated methods can sift out a clean subset from a poisoned dataset with high enough precision to support effective defense (recall the results in Table \ref{table:fail} even the purity of 97\% can already significantly weaken the defenses under the BadNets attack). These methods employ simple criteria to calculate the outlier scores, which can be evaded by advanced attack techniques. Also, we observe that the sifting precision of each method varies largely over different attacks. Worse yet, many of these approaches produce results even inferior to random selection. These negative results motivate us to analyze the commonality of existing poisoning strategies in order to devise a sifting strategy effective under different attacks. Lastly, note that MI-DCM, SF-Least, and Self-IF require a well-trained model using the contaminated dataset and incur high computational costs. For instance, in our experiments, such training requires 5411 GPU seconds (on one RTX 2080 Ti).
% Lastly, note that MI-DCM, SF-Least, and Self-IF all require complete training a model on the contaminated dataset, and therefore incur large computational costs. For instance, in our experiments, such training requires 5411 GPU seconds.
% As a result, these methods %require a pre-trained model 
% are computationally expensive.

%\newline
\begin{figure}[t!]
  \centering
  % \vspace{-0.5em}
  \vspace{-2em}
  \includegraphics[width=0.6\linewidth]{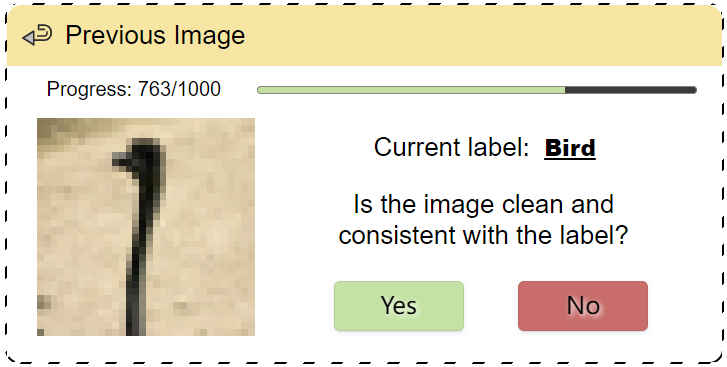}
  \vspace{-1em}
  \caption{The user interface designed for evaluating humans' ability to identify clean data. 
%   We use it to evaluate how humans would perform in the detection of poisoned samples and contribute to sifting out a clean validation set. Users are asked to determine whether the shown sample ($image+label$) is clean and consistent with the label.
  }
  \label{fig:ui-human}
  \vspace{-1.5em}
\end{figure}

\begin{figure*}[t!]
  \centering
  \vspace{-2em}
  \includegraphics[width=0.82\linewidth]{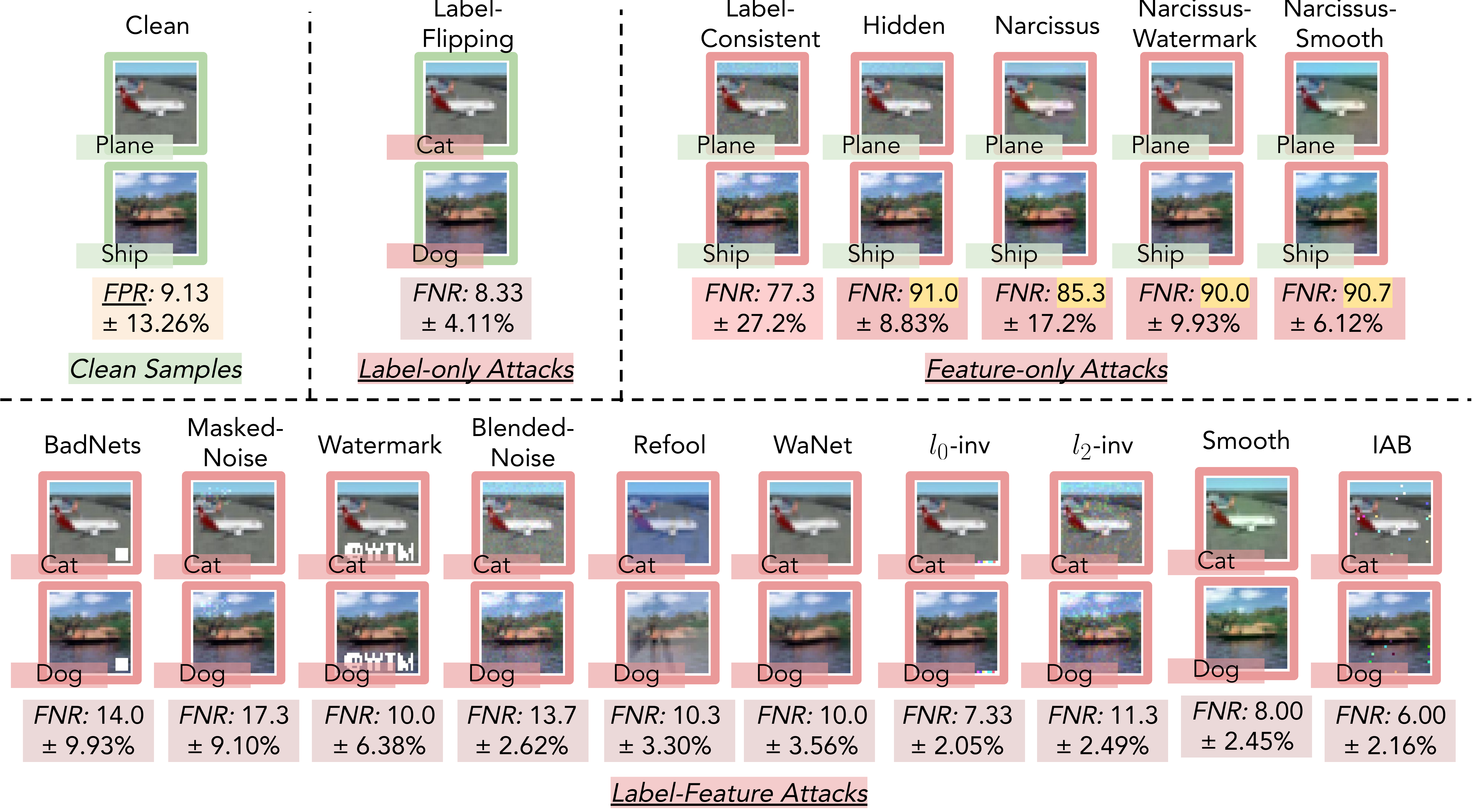}
  \vspace{-1em}
 \caption{Human inspection results regarding data poisoning attacks. 
The labels and images marked in \scalebox{0.9}{\colorbox[HTML]{FFCCCC}{\textbf{red}}} depict potential manipulations under that attack category, and the \scalebox{0.9}{\colorbox[HTML]{d9ead3}{\textbf{green}}} represents that the attribute remains intact. The evaluation uses 16 poisoned datasets, with 10\% of data being manipulated following the existing attack designs.
Among the three different categories of attacks, we report the error rate of misclassifying clean samples into poisoned ones (FPR) or poisoned ones into clean samples (FNR). Each poisoned dataset is inspected by at least three human experts from three different professional data labeling companies. We report the average results with standard deviations. The \scalebox{0.9}{\colorbox[HTML]{ffe599}{\textbf{highlighted}}} FNRs remark on the dangerous settings where human inspection failed in identifying 85\% or more poisoned data points.
}
  \label{fig:visual}
  \vspace{-1.5em}
\end{figure*}

\vspace{-1em}
\subsection{How Effective is Human Inspection}
\vspace{-.5em}
\label{sec:human?}
% \noindent

Given that the automated methods fall short of sifting out a clean base set, we ask the question: can humans accurately differentiate between clean and poisoned instances? 

% \underline{\textbf{Human Intelligence.}}
Traditionally, human inspection is considered a final backstop of data poisoning. For instance, in the highly-cited survey of data poisoning~\cite{goldblum2022dataset}, it is argued that it is the lack of human supervision on the dataset that opens the door for poisoning attacks and the subtext is that data poisoning attacks can be prevented with human inspection in place. However, as more and more stealthy poisoning attacks are developed, can human still serve a final backstop for data poisoning?

% required capability, human intervention is inevitable.  We evaluate how human experts would perform in filtering out the corrupted samples and ensuring access to a clean validation set.

For completeness, we evaluate all three types of data poisoning: Label-only, Feature-only, and Label-Feature attacks. We consider the most representative and advanced attacks for each type. We extensively study 16 data poisoning attacks (1 Label-only attack, 5 Feature-only assaults, and 10 Label-Feature attacks). The details of the attacks and their settings are deferred to Appendix \ref{sec:detailed_atk} due to the space limit.
% For completeness, we evaluate all three types of data poisoning: Label-only, Feature-only, and Label-Feature attacks. We consider the most representative and advanced attacks for each type. In total, we extensively study 16 data poisoning attacks (1 Label-only attack, 5 Feature-only assaults, and 10 Label-Feature attacks), the details of the attacks and their settings are deferred to the appendix due to the short of space.
% \ruoxi{maybe the details of the attack algorithms considered can be deferred to th appendix}
We create 16 poisoned datasets in total using CIFAR-10, each corresponding to a different attack. The size of all poisoned datasets are fixed to 1000 and the poison ratio is set to be 10\%.
% Each different data poisoning attack is used while poisoning a size 1000 CIFAR-10 subset with a fixed poison ratio of 10\% 
% (i.e., 100 samples are manipulated in each dataset). To ensure comparability, we the poisoned samples in each dataset share the same indices.
 To ensure comparability, we use the same random seed to select the poisoned samples (i.e., the poisoned samples in each dataset share the same indices). 
Finally, each poisoned dataset is sent to 3 human experts for inspection and each human expert only inspects one dataset to avoid possible bias introduced by repeated inspection. In total, we recruit 48 human experts from 3 different data labeling companies. Figure \ref{fig:ui-human} illustrates a screenshot of the web portal that we develop for human experts to identify poisoned data. In particular, we provide the image itself 
% (potentially corrupted) 
and the label of that image to the human experts. They are asked to discern whether a given image is clean and consistent with the label.

% To ensure the generalizability of the results, we include 48 human experts from 3 different data labeling companies (3 human experts for each poisoned dataset) throughout the experiment. 
We use false positive rate (FPR) or false negative rate (FNR) to measure the human's capability of distinguishing between clean and poisoned samples. FPR is defined as follows:
\vspace{-.5em}
\begin{equation}
    FPR = \frac{\text{\# of clean samples flagged as poisoned}}{\text{Total size of the clean samples}}.
    \vspace{-.5em}
\end{equation}
FPR indicates the ratio of clean samples that are mistakenly flagged as poisoned.
Meanwhile, FNR is the metric we are more interested in, which measures the ratio of poisoned samples that bypass human inspection and sneak into the sifted dataset. FNR is defined as:
\vspace{-.5em}
\begin{equation}
    FNR = \frac{\text{\# of poisoned samples marked as clean}}{\text{Total size of the poisoned samples}}.
    \vspace{-0.5em}
\end{equation}

We report the average and standard deviation of both metrics based on the results from multiple human experts. The results are summarized in Figure \ref{fig:visual}. Surprisingly, 9.13\% of the clean samples are mistakenly recognized as the poisoned. Humans are more capable of capturing manipulations on labels than on features. For Label-only attacks, human experts can filter out 91.67\% of the poisoned samples. Such a result implies that despite the costs incurred, manual inspection can help avoid most label-flipped samples in the base set.

On average, the results on the evaluated 10 Label-Feature attacks are similar to those on the Label-only attacks, achieving an averaged FNR of 10.8\%.
% \textcolor{red}{XXX}
% As shown in Figure \ref{fig:visual}, human intelligence can have a relatively large variance even on inspecting a clean sample. In the experiment, averaging 9.13\% of the clean samples are mistakenly marked as poisoned and thus thrown away. The interesting observation is that the result of FPR on clean samples is quite similar to FNRs on attacks that manipulate labels, i.e., Label-only attacks and Label-Feature attacks. 
% For Label-only attacks, human intelligence can filter out, averaging 91.67\% of the corrupted samples. Such a result indicates manual inspection can be regarded as a practical approach to avoiding most label-flipped samples being in the validation set. The averaging results on the evaluated 10 dirty-label backdoor attacks are similar to the Label-only attacks. Such an observation indicates that human intelligence is relatively sensitive to the semantic mismatches between the image and the label. 
However, Label-Feature attacks are much more powerful than Label-Flipping attacks. A small amount of the poisoned samples (e.g., a poison ratio of 0.01\% suffices in existing literature \cite{carlini2021poisoning,pan2023asset}) may successfully mislead the downstream model. At the same time, a small amount of such poisoned samples in the base set (less than 3\% based on Table \ref{table:fail}) may already significantly hinder defense performance. Hence, while humans achieve a relatively low FNR on Label-Feature attacks, the missed poisoned samples may still harm the downstream model or compromise the defense.

% When the poisoned samples enter the base set, they can further deceive the model or hinder downstream defenses. In conclusion, human intelligence is not a reliable method for filtering clean data from backdoor-poisoned datasets, which contradicts to the assumption of most existing backdoor literature \cite{wang2019neural,guo2019tabor,liu2018fine,zeng2021adversarial,xiang2020revealing}.
Even worse, when it comes to attacks that only manipulate data features, human expert's performance degrades sharply. The averaged FNR is above 86\% as shown in Figure \ref{fig:visual}. Such a result is almost close to random guessing, which achieves a FNR of around 90\% as the poison ratio is 10\%. 
To conclude, humans have limited capability of identifying samples corrupted by feature manipulation.
Finally, it is worth noting that clean-label backdoor attacks---an instance of Feature-only Attacks---are particularly dangerous. They can not only largely evade human inspection, but also backdoor a system with an extremely small poison ratio with state-of-the-art techniques (e.g., 0.05\% for attacking Tiny ImageNet~\cite{zeng2022narcissus}).

% \rev{
% It is worth highlighting that our human study is performed on only one image dataset (CIFAR-10). An in-depth investigation of how an image's resolution and the underlying concept affect humans' capability of poison detection is an interesting future work.
It's noteworthy that our human study is conducted using only one dataset (CIFAR-10). Further exploration into the impact of image resolution and concept on humans' capability of poison detection is an interesting future work.
% }

\vspace{-1em}
\section{\AlgName}
\vspace{-1em}
\subsection{Reducing to a Splitting Problem}
\vspace{-.5em}
\label{sec:probform}

% We consider a realistic and common case where a user aims to obtain a clean subset of data in an attack-agnostic manner, the same as the setting we evaluated in Section \ref{sec:human_study} for existing methods and human intelligence. The user will be granted access to all of the labeled samples gathered for the training task. A small portion of the data collected could be poisoned with unknown corruption. 

% The problem setting of sifting out a high-value subset is standard in the existing ML pipeline when a user wants to create a validation or test set. The acquired validation set can also be used for general downstream defenses in AI security, e.g., mitigating label noise \cite{shu2019meta,xu2021faster}, detecting poisons \cite{li2020rethinking} or backdoored models \cite{wang2019neural,xu2021detecting}, or conducting backdoor removal \cite{zeng2021adversarial}. 
% %\newline

% \noindent
% \underline{\textbf{The Splitting Problem.}}
% In this paper, we propose to use a more fundamental idea to develop a new method for distinguishing between clean and poisoned samples, which we term as the splitting problem. 
This paper presents a new but intuitive idea to solve the data sifting problem. Given a contaminated dataset, one can expect that an ML model trained on the clean portion of the dataset will perform poorly on the other corrupted portion and vice versa. Take standard backdoor attacks as an example. The poisoned instances are constructed by first applying a backdoor trigger to some clean inputs and then changing their labels to some target classes. A model trained on the clean data would return large prediction loss when it observes such poisoned instances because they are mislabeled, and their features only contain a trigger that does not appear in the clean data. This insight applies to general data poisoning attacks because they all involve feature and/or label manipulation, thereby resulting in a distributional shift from the clean instances.
Thus, we solve the data sifting problem by finding a split of the given contaminated dataset such that the model trained on one split produces a large loss on the other.

% \begin{equation}

We formulate the splitting problem as a bilevel optimization. 
Formally, given the contaminated dataset $D = D_1\cup\ldots\cup D_K$,
% \ruoxi{the parentheses need to be removed}
where $D_k$ is the subset of $D$ but only containing samples with class label 
% $k\in (1,K)$,
% \ruoxi{
$k\in\{1,\cdots,K\}$,
% }
our goal is to divide each class-wise subset, $D_k$, into two splits $B_k$ and $D_k \setminus B_k$, such that
\vspace{-1.9em}
\begin{align}
\gB^* &= \argmax_{\gB}\sum_{k=1}^{K}\sum_{z_i \in D_{k}\setminus B_{k}}L\left(\theta^*(\gB),z_i \right),\label{equ:split_1}
\\
\text {s.t. } &\theta^*(\gB) = \argmin _{\theta} \sum_{k=1}^{K} \sum_{z_j \in B_{k}}L\left(\theta,z_j\right ), \label{equ:split_2}\\
& |B_1| = \ldots = |B_K|, \label{equ:split_same_size}\\
& |B_1| + \ldots + |B_K| \geq (1-\epsilon) |D|.  \label{equ:split_larger}
\end{align}

\vspace{-.5em}
\noindent
where $\gB=B_1\cup\ldots\cup B_K$ is the union of the class-wise training splits, 
% \ruoxi{the parenthesis can be removed}
$\theta$ represents the parameters of an ML model, and $\theta^{*}(\gB)$ denotes the parameters obtained from training on $\gB$. $L$ is the loss function, e.g., cross-entropy loss for classification tasks. The inner function (\ref{equ:split_2}) acquires a model that is trained over one split. The outer optimization (\ref{equ:split_1}) evaluates the model performance on the other split -- $D \setminus \gB$. The formulation seeks for the subset $\gB$ that leads to the highest loss when evaluated on $D\setminus \gB$. Additionally, (\ref{equ:split_same_size}) constrains the size of the split of each class-wise subset so that the union $\gB$ is a perfectly label-balanced subset of $D$. (\ref{equ:split_larger}) enforces the acquired $\gB$ to have at least $(1-\epsilon)|D|$ samples, where $\epsilon$ is an upper bound of the poison ratio. If we assume the number of poisoned samples in a given dataset to be smaller than that of clean samples (a standard assumption in robust statistics~\cite{steinhardt2018robust}), i.e., $\epsilon$  will be 0.5. Thus, (\ref{equ:split_larger}) is also encourages $\gB$ to be the clean split.

\vspace{-1em}
\subsection{Relaxation of the Splitting Problem}
% \vspace{-.5em}
% \rev{
Exactly solving Eqn. (\ref{equ:split_1}-\ref{equ:split_larger}) requires training on every possible $\gB$ and testing on their complement. This is clearly intractable.
% For example, consider CIFAR-10 which is a standard computer vision benchmark dataset containing 10 classes with each class contains 5,000 samples; then, $\sum_{n=1}^{5000}\binom{5000}{n}^{10}$ different models need to be trained to solve the optimal splitting problem.
% }
% \ruoxi{I suggest removing "for example ... solve the optimal splitting problem'' }
% \rev{
The computational challenge in part arises from the combinatorial nature of the outer optimization problem.
To address this challenge, we propose to relax Eqn. (\ref{equ:split_1}-\ref{equ:split_larger})  to a \emph{continuous} splitting problem.
% }
% \rev{
We start by rewriting (\ref{equ:split_1}-\ref{equ:split_larger}) as 
\vspace{-1.em}
\begin{align}
    \gV^* &= \argmax_{\gV} \sum_{k=1}^{K} \sum_{z_i \in D_k}
    % ^{\left | D_k \right |} 
    (1-v_{(k,i)})  L(\theta^*(\gV),z_i),  \label{eq:bilevel-set2_1}\\
    \text{s.t. } & \theta^*(\gV) = \argmin_\theta \sum_{k=1}^{K} \sum_{z_i \in D_k} v_{(k,i)} \cdot L(\theta,z_i), \label{eq:bilevel-set2_2}\\
& \|V_1\|_1 = \ldots = \|V_K\|_1, \label{equ:bilevel-set2_split_s}\\
& \|V_1\|_1 + \ldots + \|V_K\|_1 \geq (1-\epsilon) |D|,
% \ruoxi{$l_1$-norm for (9) and (10} 
\label{equ:bilevel-set2_split_larger}
\end{align}
% }
% \end{equation}
% \rev{

\vspace{-0.5em}
\noindent
where instead of directly optimizing the split, we optimize the tuple of binary variables indicating each sample's presence, $\gV = [V_1,\ldots,V_K]$ ,where $k\in\{1,\dots K\}$. 
% \ruoxi{change parentheses to \{\}}
Each $V_k=[v_{(k,1)},\ldots,v_{(k,|D_k|)}]$ encodes the splitting result of $D_k$, where $v_{(k,i)}=1$ if $z_i\in \gB$ (a class-$k$ sample $z_i$ assigned to the split for training) and $v_{(k,i)}=0$ if $z_i\in D\setminus \gB$ ($z_i$ assigned to the split for testing). $\|V_k\|_1$ 
% \ruoxi{subscript 1 to indicate 1 norm} 
computes the $l_{1}$ 
% \ruoxi{$l_1$} 
norm of $V_k$, i.e., $\|V_k\|_1 = \sum_{i=1}^{|D_k|} v_{(k,i)} = |B_k|$.
To solve Eqn. (\ref{eq:bilevel-set2_1}-\ref{equ:bilevel-set2_split_larger}), we relax the binary variables $v_{(k,i)}$ as continuous ones, i.e., $v_{(k,i)} \in [0,1]$. 
Hereinafter, $v_{(k,i)}$ represents the likelihood of $z_i$ being assigned to the split for training. This continuous relaxation enables us to leverage gradient-based methods to search for an approximate solution, i.e., each class-wise clean split $B_k$ can be obtained by collecting the samples with larger $v_{(k,i)}$.

However, when the size of the dataset is large, the outer optimization needs to optimize over a continuous space of the same large dimension, which could be slow. 
% We can directly obtain a differentiable formulation from Equation (\ref{eq:bilevel-set2}) by allowing $v_{n}$ to be a floating number (instead of being an indicator). By resolving the differentiable splitting problem, we can obtain a similar result as the ideal solution to Equation (\ref{equ:split}). In particular, if we initialize $V = \mathbbm{1}^{\left | D \right |}$, then we can obtain a clean subset by looking at the top-valued samples (they are with a larger split as all the samples' assigned values are dropping from 1).
Inspired by advances in meta learning~\cite{shu2019meta,xu2021faster}, we propose to further learn a 
% meta weight network (MW-Net) 
weight-assigning network
to assign weight to each point instead of directly optimizing the weights. We found that the number of parameters of such a network required to produce useful weights is much smaller compared to the data size. For instance, in our experiments, we found a network with 100 nodes is sufficient for high-quality weight assignment, but on the other hand, the size of the datasets considered in this paper is orders of magnitude larger. Hence, through adapting a weight-assigning network, we effectively reduce the number of variables that are optimized in the outer optimization.
% The advantage of using a parameteric network for weight assignment is

% \yi{below has been updated}
Specifically, let $\mathcal{S}(\cdot)$ denote the 
% MW-Net 
weight-assigning network
and let $\psi$ denote its parameters.
% of MW-Net. Let $\mathcal{S}(\cdot)$ denotes the function of assigning weights according to the MW-NET and the classifier model, i.e.,
Furthermore, 
% \ruoxi{
$\forall k$ and $z_i\in D_k$, 
% }
we let $v_{(k,i)} = \mathcal{S}(L(\theta,z_i);\psi)$
% \footnote{
% For conciseness, we will denote $v_n = \mathcal{S}(L_{n}(\theta);\psi)$ hereinafter
, i.e., the weight-assigning network determines the weight for $z_i$ based on its associated learning loss from $\theta$. For simplicity, we use $L_{i}(\theta)$ as a shorthand for $L(\theta,z_i)$. Then, the continuous splitting problem that we will solve can be expressed as:
% \begin{equation}
\vspace{-1.em}
\begin{align}
& \psi^* = \argmax_{\psi}  \sum_{i=1}^{\left | D \right |} (1-\mathcal{S} (L_i(\theta^*(\psi));\psi) ))  L_i(\theta^*(\psi)), \label{eq:bilevel-final_1}\\
\text{s.t. } & \theta^*(\psi) = \argmin_\theta  \sum_{i=1}^{\left | D \right |}  \mathcal{S}  (L_i(\theta);\psi ) L_i(\theta). \label{eq:bilevel-final_2}
\end{align}

\vspace{-.5em}
\noindent
Since samples from each class are weighted by a shared weight-assigning network, in (\ref{eq:bilevel-final_1}, \ref{eq:bilevel-final_2}), we no longer specify which class $z_i$ is from. Moreover, note that the optimization problem above does not have explicit constraints to ensure that the training split is class-balanced and contains the majority of the samples. Instead, we will adopt two heuristics to enforce these constraints. \textbf{1)} We apply $\gS(\cdot)$ to each sample to obtain their weight and then select the same amount of highest-weight samples within each class; \textbf{2)} We initialize the 
% output of $\gS(\cdot)$ 
weights, $v_{(k,i)}$,
to be all ones, 
i.e., assigning all the samples to the training split and terminate the optimization with limited rounds. The limited round of the optimization will leave most of the samples with large $v_{(k,i)}$ values, thus resulting in an implicit satisfaction of the constraint of selecting more samples for the training split.
% \ruoxi{I am not sure how you can initialize the output of a weight network because during the optimization, it should be the weight network parameters that got initialized.}
% \rev{
% Noting that in (\ref{eq:bilevel-final_1}, \ref{eq:bilevel-final_2}) we dropped the notations related to class-wise processing. Nevertheless, (\ref{eq:bilevel-final_1}, \ref{eq:bilevel-final_2}) will still be able to obtain the approximated solution as to resolving (\ref{eq:bilevel-set2_1}-\ref{equ:bilevel-set2_split_larger}) when we: \textbf{1)} apply $\gS(\cdot)$ to select the same amount of samples for each $\gB_k$ (class-wise manner) for (\ref{eq:bilevel-final_1}), which is the same constraint detailed in (\ref{equ:bilevel-set2_split_s}); \textbf{2)} initialize the output of $\gS(\cdot)$ as all ones at the beginning of the optimization and ending the optimization with limited rounds. The limited round of the optimization will leave most of the sample end with larger $v_{(k,i)}$ values, thus resulting in an implicit satisfactory to the constraint to select more samples to the training split, i.e., (\ref{equ:bilevel-set2_split_larger}).
% }
%
Each hidden node in $\mathcal{S}(\cdot)$ is equipped with ReLU activation function \cite{agarap2018deep}, making it possible to approximate non-linear functions.
The output node produces a one-dimensional value indicating the weight assigned to the input to the weight-assigning network. The output node utilizes a Sigmoid activation function to ensure that the output is within the range $[0, 1]$.
% Regardless of the simplicity of $\psi$, this net is known as a universal approximator for practically any continuous function \cite{csaji2001approximation}.

\begin{figure*}[t!]
  \centering
   \vspace{-2em}
  \includegraphics[width=0.85\textwidth]{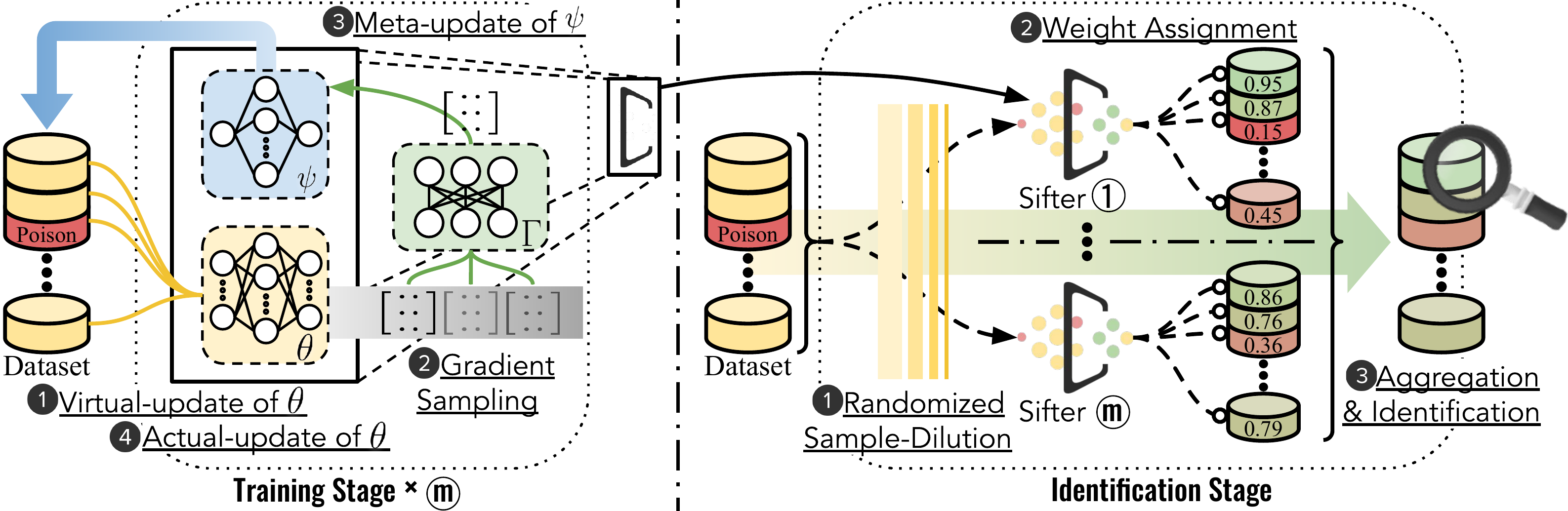}
  \vspace{-1em}
  \caption{The whole process of \AlgName consists of two stages: the \textbf{Training Stage} and the \textbf{Identification Stage}. 
Multiple ($m$) Sifters will be included during the \textbf{Identification Stage} to reduce the randomness resulting from SGD and randomized sample-dilution. 
As such, the \textbf{Training Stage} will be repeated $m$ times with different random seeds to obtain $m$ Sifters. 
In each Sifter, there are two different structures working as a pair: model $\theta$ and the weight-assigning network $\psi$.
In one iteration of the \textbf{Training Stage}, there are four steps: Virtual-update of $\theta$; Gradient Sampling using the meta-gradient-sampler $\Gamma$; Meta-update of $\psi$; then the Actual-update of $\theta$. 
After only one iteration, \textbf{Training Stage} will terminate. The trained Sifters will be adopted in the \textbf{Identification Stage} to assign weights to the diluted data from the dataset. Finally, \AlgName aggregates the results from multiple Sifters, and the clean samples will be sifted by inspecting the high-value end.
% Throughout the process, no additional validation data is required.
  }
  \label{fig:workflow}
  \vspace{-1.5em}
\end{figure*}

\vspace{-1.em}
\subsection{Overall Algorithm}
\vspace{-.5em}
\label{sec:overallalgo}
\noindent
\underline{\textbf{Overview.}} We now describe the full algorithm to sift out a clean subset, which we call $\AlgName$. 
% \rev{
Given a poisoned dataset and a selection budget, $\AlgName$ aims to select a subset that is most likely to be clean. The subset output by $\AlgName$ can then be used as a base set in different downstream defense algorithms to fulfill their corresponding defense goals.
% }
At a high level, \AlgName consists of two stages: \textbf{Training} and \textbf{Identification}. In the Training stage, we adopt an online algorithm to solve (\ref{eq:bilevel-final_1}, \ref{eq:bilevel-final_2}), and the algorithm will produce optimized parameters $\psi^*$ for $\gS(\cdot)$ as well as model parameters $\theta^*$ trained on the weighted samples. The definition of $\theta^*$ will be more clear after we explicate the algorithm. We will refer to the combination of $\gS(\cdot)$ parameterized by $\psi^*$ and the classification model parameterized by $\theta^*$ as a \textbf{Sifter}. To mitigate the randomness of batch selection in the online algorithm, we run the online algorithm independently for $m$ times, which results in $m$ Sifters. In the Identification stage, we feed the dataset into each Sifter and aggregate the weights produced by all Sifters to obtain the final weight for each sample in the dataset.
% By resolving Equation (\ref{eq:bilevel-final}), a pair of $\psi^*$ and $\theta^*$ will be obtained as one Sifter that can assign meta-splitting-values to each point in the dataset. 
% One Sifter's results can be biased as acquiring a pair of $\psi^*$ and $\theta^*$ takes much fewer rounds (2 epochs in our experiment\footnote{One epoch of normal SGD to warm-up $\theta$ over the whole dataset, and one epoch of updating the Sifter pairs ($\theta$ and $\psi$).}) of epochs than training a converged model (at least $200+$ epochs on CIFAR-10).
% To mitigate the randomness in the meta-splitting-value results (due to stochastic gradient descent (SGD) and randomized sample dilution), we will incorporate multiple Sifters and aggregate the average results to obtain the final result of \AlgName. 
% We term the training of Sifters in \AlgName as the \textbf{Training Stage} and the implementation and aggregation of the trained Sifters as the \textbf{Identification Stage}.
The full process of \AlgName is depicted in Figure \ref{fig:workflow}. 
% Next, we will provide the details for the algorithm design and implementation for the two stages.
% The following part of this section will break down the two stages in detail.
%\newline
% single bilevel is not enough
% lead to the multi-sifter structure (show preliminary results and the whole out-line)

% %\vspace{0.2em}
\noindent
\underline{\textbf{Training Stage.}}
We present how to train a single Sifter in \AlgName. 
The training process starts by ``warming up'' $\theta$---training $\theta$ on the entire dataset for one epoch. Essentially, we update $\theta$ while setting all weights to be one. This ``warm-up'' step helps promote $\mathcal{S}(\cdot)$ to assign large weight (close to $1$) to samples, which in turn, allows most of the samples to be placed into the split for training, as mentioned above.
% helps bias the weights towards one and therefore, 
% start with a warmed-up classifier model that is trained over the whole dataset for one epoch.
% Warming up $\theta$ can be regarded as training $\theta$ on a split where all the samples in $D$ are selected (weights with all 1s).
% The ``warm-up'' step can promote $\psi$ to assign larger values to data that belongs to the larger split, which normally is associated with clean samples.
Then, the Sifter (with $\theta$ and $\psi$) are updated via an online algorithm (illustrated in Figure \ref{fig:workflow}). At each round of the online algorithm, we perform the following four steps: \textit{Virtual-update of a warmed-up model $\theta$}, \textit{Gradient Sampling} using the meta-gradient-sampler $\Gamma$, \textit{Meta-update of $\psi$}, and finally the \textit{Actual-update of $\theta$}.
% and terminating the training process of one Sifter.

% \ruoxi{shorten the following algorithm description as it is not our main contribution. Just retain the main update steps and leave the detailed derivation to appendix.}
\scalebox{0.8}{\circled{1}}\textit{ Virtual-update of $\theta$.} 
% layer-wise gradient formulation of the one-epoch warmed-up downstream classifier, $\theta$. 
This step takes a virtual update of $\theta$, the same as the traditional stochastic gradient descent (SGD). Formally, a mini-batch of data $\{z_i\}_{i=1}^n$ is sampled, where $n$ is the batch size, and the following update is performed:
% $\left\{\left(z_{i} = (x_{i},y_{i})\right), 1 \leq i \leq n\right\}$ is sampled, where $n$ is the batch size. Updating $\theta$ regarding the inner optimization in Equation (\ref{eq:bilevel-final}) in a standard SGD would be formulated as: 
\vspace{-.5em}
\begin{equation}
    \theta^{\prime}=\theta-\alpha  \mathbf{g}_\theta,
    \vspace{-.5em}
\end{equation}
where $\mathbf{g}_\theta=\frac{1}{n}\sum_{i=1}^{n} \mathcal{S}\left(L_i\left(\theta\right);\psi \right) \frac{\partial L_i(\theta)}{ \partial \theta} 
% L_i\left(\theta\right)
$, representing the weighted gradient of the training loss with respect to $\theta$ and $\alpha$ is the learning rate.
% Without loss of generality, s
Suppose $\theta$ is a neural network consisting of $\mathcal{L}$ layers. Let the parameters in layer $l$ be denoted by $w_l$. 
% $\theta\left(\cdot ;\left\{w_{l}\right\}_{l=1}^{\mathcal{L}}\right)$, where $w_{l}$ represents the parameter for the $l$-th layer. 
Then, we can rewrite $\mathbf{g}_\theta$ as
\vspace{-1em}
\begin{equation}
\begin{aligned}
   \mathbf{g}_\theta = \bigg(\frac{1}{n}\sum_{i=1}^{n} \mathcal{S}\left(L_i\left(\theta\right);\psi \right)\bigg)\times [g_1,\ldots,g_{\mathcal{L}}],
\end{aligned}
\label{eq:layers}
\vspace{-.8em}
\end{equation}
where $g_l = \frac{\partial L_i\left(\theta\right)}{\partial w_l}$, representing a scaled gradient with respect to the parameters in layer $l$.

% We can decompose the computation of $\mathbf{g}$ as:
% \begin{equation}
%     \mathbf{g} = \frac{1}{n} \sum_{i=1}^{n} \psi\left(L\left(\theta,z_i\right)\right) \times g_l,
% \label{eq:layers}
% \end{equation}
% where $g_l = \frac{\partial L\left(\theta,z_i\right)}{\partial w_l}$ denotes the gradient of the $l$-th layer.

\scalebox{0.8}{\circled{2}}\textit{ Gradient Sampling.} 
% layer-wise gradient mask with a list of meta-gradient-samplers, $\Gamma$.
Inspired by \cite{xu2021faster}, we parallelly train a list of $u$ gradient-samplers to select partial gradients from the last $u$ layers, $\{g_{\mathcal{L}-u},\ldots,g_\mathcal{L}\}$ for subsequent update on $\psi$. The goal of gradient sampling is to accelerate and achieve more stable results on the subsequent update on $\psi$ \cite{xu2021faster}.

In particular, we associate a different gradient-sampler $\gamma_l(\cdot)$ for each layer $l\in \{\mathcal{L}-u,\ldots,\mathcal{L}\}$. $\gamma_l$ takes as input the gradient $g_l$ and returns a value indicating whether $g_l$ is selected. The reason why each layer has a different gradient-sampler is that their corresponding gradients are of different dimensions and  $\gamma_l(\cdot)$'s input dimension needs to be tailored to the size of $g_l$.
% for gradient aggregation and in preparing for efficient computation of the meta-update of $\psi$.
% The process is done by passing each $g_l$ through a gradient sampler, $\gamma_l(\cdot) \in \Gamma$. The output to a gradient sampler is the discrete activation status $\gamma_l(g_l) \in\{0,1\}$.
% The reason why $\mathcal{L}$ gradient samplers are required is that each $g_l$ is of a different shape, and thus each $\gamma_l(\cdot)$'s input structure would need to change accordingly.
Following~\cite{xu2021faster}, we implement each gradient-sampler as a two-layer fully-connected neural network. The activation units in the first and second layer are 
PReLU \cite{he2015delving} and ``Gumbel-softmax'' \cite{jang2016categorical}, respectively.

The $l$-th gradient sampler, $\gamma_l$, can be updated using the gradient of the outer optimization objective w.r.t. the gradient sampler's parameters. 
The dependency of the outer optimization objective on the sampler's parameters is via $\psi$. The gradient can be calculated automatically via auto-differentitaion in standard deep learning frameworks~\cite{paszke2019pytorch}.
% Notably, this process is automated by passing the computation graph through the gradient sampler to the update of $\psi$. 
% The gradient can be automatically calculated via backpropagating the outer optimization's loss function via automatic differentiation in the backward mode \cite{paszke2019pytorch}. 
We then update each sampler according to the respected gradients before selecting layers.
% The first fully connected layer is followed by a PReLU \cite{he2015delving} and the second layer is followed by the ``Gumbel-softmax'' \cite{jang2016categorical}. 
The gradient of a layer is selected if the output of the respect sampler exceeds $0.5$. We denote the collection of the weights for layers that are selected by a gradient sampler as
\vspace{-.5em}
\begin{align}
    w_\mathcal{C}=\{w_l: \gamma_l(\frac{\partial L_i\left(\theta\right)}{\partial w_l}) >0.5\}.
    \label{eq:sample}
\end{align}
% The final result of whether to turn on the gradient of a specific layer is determined by a threshold equal to $0.5$. 

% As $\mathbf{g}$ passes through the samplers, when conducting the meta-update of $\psi$, the gradient sampler's gradients will be automatically obtained via the reverse mode of automatic differentiation \cite{griewank2008evaluating}, which is supported by deep learning frameworks like PyTorch \cite{paszke2019pytorch}, TensorFlow \cite{abadi2016tensorflow}, or JAX \cite{frostig2018compiling}.
% Finally, by applying the gradient samplers, $\Gamma$, to all layers, we can aggregate the actual gradient for meta-update as:
% \begin{equation}
%     \widehat{\mathbf{g}} = \frac{1}{n} \sum_{l=1}^{\mathcal{L}} 
%     \mathbbm{1}\left[\gamma_{l}(g_l)=1\right]
%     \sum_{i=1}^{n} \psi\left(L\left(\theta,z_i\right)\right) \times g_l,
% \label{eq:sample}
% \end{equation}
% where $\mathbbm{1}\left[r_{l}=1\right]$ is the indicator function. In other words, the meta-update of $\psi$ will only incorporate the gradient from the layers that had been turned on by $\Gamma$. As shown in \cite{xu2021faster}, by using the gradient sampler, one can not just compute the hypergradient faster but with better performance as the gradient incorporated will be more stable.

\vspace{-.5em}
\scalebox{0.8}{\circled{3}}\textit{ Meta-update of $\psi$.} 
% Updating the MW-Net, $\psi$, with the hypergradient considering only the $\Gamma$ selected layers. 
In this step, we update the parameter $\psi$ of the weight-assigning network by calculating the gradient of the outer optimization objective function with respect to $\psi$ and performing the gradient descent: 
% \begin{align}
$
    \hat{\psi} = \psi -\beta \mathbf{g}_\psi,
$
% \end{align}
%%%new summarizing
% \rev{
where $\mathbf{g}_\psi$
\vspace{-.8em}
\begin{equation}
\resizebox{\linewidth}{!}{
\ensuremath{
\begin{aligned}
\mathbf{g}_\psi &= \frac{\partial}{\partial \psi} \bigg (\frac{1}{n}\sum_{i=1}^n
(1-\mathcal{S} (L_i(\theta');\psi )) L_i(\theta^{\prime}) \bigg) \\
&=\frac{1}{n} 
\sum_{i=1}^{n}\bigg( -\frac{\partial \mathcal{S}(L_i(\theta');\psi)}{\partial \psi}  L_i (\theta^{\prime} ) %\\
+( 1-\mathcal{S}(L_i( \theta');\psi))
\frac{\partial L_i(\theta^{\prime})}{\partial\psi}
\bigg).
\end{aligned}
}}
\label{eq:metaupdate}
\vspace{-.5em}
\end{equation}
$\frac{\partial L_i(\theta^{\prime})}{\partial\psi}$ is often referred to as \emph{indirect gradient}, and expands as
\vspace{-1em}
\begin{align}
% \resizebox{0.4\linewidth}{!}{
% \ensuremath{
% \begin{aligned}
\frac{\partial L_i(\theta^{\prime})}{\partial\psi}
% & = \frac{\partial L_i(\theta^{\prime})}{\partial \theta} \frac{\partial \theta'}{\partial \psi}\\
&= \frac{\partial L_i(\theta^{\prime})}{\partial \theta} \bigg(-\frac{\alpha}{n} \sum_{j=1}^n \frac{\partial \mathcal{S}\left(L_j\left(\theta\right);\psi \right)}{\partial \psi} \frac{\partial L_j(\theta)}{\partial \theta}\bigg).
% \end{aligned}
% }}
\end{align}
% }

\vspace{-.8em}
\noindent
% \rev{
The process of calculating the indirect gradient of $\psi$ has been shown to be computationally intensive \cite{liu2021investigating,grazzi2020iteration}. However, in our case, by using the gradient samplers, we only consider the selected layers to calculate the indirect gradient, i.e., 
\vspace{-.8em}
\begin{align}
    \frac{\partial L_i(\theta^{\prime})}{\partial\psi} = \frac{\partial L_i(\theta^{\prime})}{\partial w_\mathcal{C}} \bigg(-\frac{\alpha}{n} \sum_{j=1}^n \frac{\partial \mathcal{S}\left(L_j\left(\theta\right);\psi \right)}{\partial \psi} \frac{\partial L_j(\theta)}{\partial w_\mathcal{C}}\bigg).
\label{eqn:sample_grad}
\end{align}
% }
% Due to the page limit, we defer the detailed derivation of the sampled gradient to Appendix \ref{sec:meta-gradient}.

\vspace{-.8em}
\noindent
Note that $\mathbf{g}_\psi$ does not need to be calculated manually; instead, it can also be computed via automated differentiation.

\scalebox{0.8}{\circled{4}}\textit{ Actual-update of $\theta$.}
% Updating $\theta$ with the weighted losses assigned by $\widehat{\psi}$.
Finally, we update $\theta$ in place based on the weights assigned by the updated $\mathcal{S}(\cdot;\widehat{\psi})$:
% and the assigned weights that promote the splitting of $D$ (Equation (\ref{eq:bilevel-final})'s goal): 
\vspace{-.8em}
\begin{align}
% \vspace{-2em}
    \widehat{\theta}=\theta-\alpha  \frac{1}{n}\sum_{i=1}^{n} \mathcal{S}\left(L_i\left(\theta\right);\widehat{\psi}\right)
    \frac{\partial L_i(\theta)}{ \partial \theta}.
    % \nabla_{\theta} L_i\left(\theta\right).
\label{eq:actual}
\end{align}

\vspace{-.8em}
In each round of our online algorithm, the four steps are performed on one batch of data. The algorithm terminates when it has gone through all the samples in $D$.
% will keep running as a loop till the system has gone through all the samples in $D$.
The pseudocode of the Training Stage is in Algorithm \ref{algo:AlgoS}, Appendix \ref{sec:pseudo}. We define the updated $\theta$ and $\psi$ in the final round as $\theta^*$ and $\psi^*$, respectively. The pair of $\theta^*$ and $\psi^*$ defines one Sifter's parameters. We will run the online Training algorithm multiple times to obtain the Sifters we will use in the Identification Stage.
% \yi{update ends}
% After going through the whole dataset, we will terminate the update of $\theta$ and $\psi$. 
% The final pairs of $\theta^*$ and $\psi^*$ are considered as one Sifter that we will use in the \textbf{Identification Stage}.

% %\vspace{-0.5em}

% %\vspace{-2em}

%\vspace{0.2em}
\noindent
\underline{\textbf{Identification Stage.}}
Now we aggregate the results from all the Sifters to identify the clean samples in three steps:
% The whole process of the \textbf{Identification Stage} can be divided into three steps:
\textit{Randomized Sample-Dilution}, \textit{Weight Assignment}, and \textit{Weight Aggregation \& identification}.

\scalebox{0.8}{\circled{1}}\textit{ Randomized Sample-Dilution:} Instead of directly feeding a sample to the Sifter, we propose to first randomly perturb the sample. Existing works~\cite{li2020rethinking,qiu2021deepsweep} show that the prediction associated with a clean sample is robust to random perturbation. Hence, a sample that consistently receives high weights under different perturbations is more likely to be clean than those samples experiencing large variance under perturbations. % Before assigning weights to each data point, we propose to randomly perturb all the samples, as we termed randomized sample-dilution. This process act as a procedure to randomly mitigate the functionality (how much a sample contributes to the training task) of each data point. As we believe the clean samples should be relatively more robust to random perturbations (also observed in existing work \cite{li2020rethinking,qiu2021deepsweep}), the averaged weighting results of the clean samples should have a smaller variance and end up with higher weighting values.
% On the other hand, those poisoned samples containing feature-space malware (backdoor triggers or feature colliding) would be easier to end up with averaging lower weighting values.
The perturbation strategies considered in this paper include
random cropping, random rotation, random horizontal flipping, and Gaussian blurring. For each sample, we apply all the perturbation strategies simultaneously, and each strategy is configured randomly (e.g., we randomly sample a rotation angle and rotate the sample correspondingly).
Ablation study on these random perturbations is presented in Appendix \ref{sec:ablation}. 
% \yi{Add discussion of the baselines with sample dillusion}

\scalebox{0.8}{\circled{2}}\textit{ Weight Assignment:} The random sample-dilution step produces $m$ perturbed datasets and each dataset contains samples perturbed by a different random configuration of the perturbations strategies. Then, each Sifter ($m$ in total) takes as input a perturbed dataset and produces the weights for all samples. In particular, the classification model (parameterized by $\theta^*$) in the Sifter will be used to extract features from each sample and $\gS(\cdot)$ (parameterized by $\psi^*$) will assign weight to a sample based on the extracted features.

% After random sample-dilution, each randomly perturbed dataset is then assigned with one Sifter. Each Sifter, $(\theta^*,\psi^*)$, will then go through the whole dataset $D$ again. In particular, this time, $\theta^*$ will be used as a feature extractor that returns the loss of sample $z_i$, i.e., $L(\theta,z_i)$. And $\psi$ assigns the weight according to the output of $\theta^*$. With multiple Sifters being adopted, we can obtain multiple vectors, $V=\left \{ v_1,\ldots,v_{\left | D \right |} \right \}$, containing the respected meta splitting values of the whole dataset.

\scalebox{0.8}{\circled{3}}\textit{ Aggregation \& identification:} We average the weights for each sample from multiple Sifters and use the top-weighted samples of each class $k$ to form the sifted class-wise clean subset $B_k$. Finally, the clean split $\gB$ is obtained by combining the class-wise results.
%
% Finally, one can aggregate the resulting splitting-value vectors from each Sifter by simply computing the average value regarding each index $i \in (1,\left | D \right | )$.
% Then, one can obtain the top-valued samples as the sifted clean subset.

% We propose \AlgName to resolve the splitting problem formulated in Section \ref{sec:probform} of acquiring a clean subset from a poisoned dataset in an attack agnostic manner. \AlgName consists of multiple Sifter systems for data synthesis at the Identification stage. 
% Each Sifter itself is solely a solution to the splitting problem. Precisely, each Sifter consists of a downstream model structure, a meta-weighting structure, and a gradient adapter. This design is inspired by state-of-the-art meta-learning-based reweighting learning \cite{xu2021faster}. We will deliver the thorough details of each structure design in the following part of this section. One key diffrence between our method and the reweighting learning is that we do not actually train any ML model but use the meta-weighting structure for assigning values that splits the dataset.
% The multiple-Sifter design mitigates the randomness resulting from stochastic gradient decent training and random data augmentations and helps to obtain more robust and reliable results. An overview of the structural design of the \AlgName during the Identification stage is shown in Figure \ref{fig:workflow}.

\vspace{-1em}
\section{Evaluation}
% \vspace{-.5em}
% Our evaluation mainly focuses on the following aspects:
% \begin{itemize}
%     \item Assessing and comparing \AlgName with existing automated methods on the task of sifting out a clean data subset from a contaminated dataset;
%     \item Evaluating if the selected clean subset can serve as the base set to be plugged into the existing defense techniques and lead to successful defenses.
%     % \item Ablation study on several choice points of the proposed \AlgName. \yi{details}
% \end{itemize}

\vspace{-1em}
\subsection{Experimental Setup}
\vspace{-.5em}
\noindent
\underline{\textbf{Metrics.}} 
One crucial factor in evaluating automated data sifting methods is the precision of the selection. We use the Corruption Ratio (\textit{CR}) to measure the precision of each method's selection performance. Formally, let $D_{sub}$ denote the selected subset from $D$ using an automated method, and $N_{poi}$ denote the total number of the poisoned samples in $D_{sub}$. Then, the CR of $D_{sub}$ is defined as 
% \begin{equation}
$
    CR = \frac{N_{poi}}{\left |D_{sub} \right |}\times 100\%.
$
% \end{equation}
While CR seems a natural performance metric, it is not well suited for comparing the precision of selection across different attack settings because they normally employ different poison ratios. For example, consider two attacks with poison ratios $5\%$ and $20\%$ and set the selection strategy to be a random one. Then, the CRs on the two attacks are different (roughly $5\%$ and $20\%$, respectively) despite the same selection strategy.
%
% However, we find that CR is not well suited for comparing different attacks as they might use different poison ratios. In other words, despite random selection of a subset of the same size, the CR can change a lot according to the poison ratio. Therefore, we propose a new metric, termed Normalized Corruption Ratio (\textit{NCR}). The goal of designing NCR is to make it easier to compare different sifting methods across different attack settings. Formally, the NCR can be calculated as:
To facilitate the comparison of the sifting performance across different attacks, we propose the Normalized Corruption Ratio (\textit{NCR}), defined as follows
\vspace{-1em}
\begin{equation}
    NCR = \frac{CR}{CR_{rand}}\times 100\%,
    \label{eq:ncr}
\vspace{-0.8em}
\end{equation}
where 
% $CR$ calculates the corruption ratio of the current method's selected subset, and
$CR_{rand}$ is the corruption ratio of the random selection.

%\vspace{0.2em}
\noindent
\underline{\textbf{General Settings.}}
We use two servers equipped with a total of 16 GTX 2080 Ti GPUs as the hardware platform. PyTorch \cite{paszke2019pytorch} is adopted as the implementation framework. 

We consider four popular benchmark datasets, namely, CIFAR-10 \cite{krizhevsky2009learning}, GTSRB~\cite{stallkamp2012man}, PubFig \cite{kumar2009attribute}, and the ImageNet \cite{russakovsky2015imagenet}. The details of the datasets, the architecture of the model trained on each dataset, training algorithms, and performance of the models (trained on clean datasets as baselines) are provided in \tablename~\ref{tab:datasets}.
% We consider and fine-tune one specific target model structure for each dataset to achieve good performance.
% Noting that we used three randomly selected subsets of ImageNet to showcase that for a large dataset, \AlgName can still reliably provide high precision data sifting by going through all subsets of the original huge chunk. 
For ImageNet, directly applying \AlgName leads to high computational costs because a large-capacity classification model $\theta$ is needed to ensure the quality of extracted features. For large-scale datasets such as ImageNet,
% while our algorithm can efficiently assign weight for the entire dataset and the resulting NSRs can go below 10, we can hardly reach an NSR close to 0. We find that it is because \textcolor{red}{XXX}. Thus, 
we adopt a practical and reliable alternative where we split the ImageNet dataset into multiple subsets and then apply \AlgName on each one separately. In our evaluation, we will showcase the sifting performance on three non-overlapping random 10-class subsets of ImageNet.
% multiple \AlgName scanning through multiple subsets to show that our method can still provide maximum precision on large-scale datasets by breaking them into smaller ones.
% We have tried using \AlgName directly to go through the whole ImageNet, and it can still be reliable to end up with NCRs below 10, but hardly 0s. We find the reason the NCRs cannot drop to 0 is that the model, $\theta$, cannot extract reliable features throughout the update, i.e., the learning task on ImageNet is still an open and challenging problem for models that are trained from scratch. Thus, we provide a practical but more reliable solution with multiple \AlgName scanning through multiple subsets to show that our method can still provide maximum precision on large-scale datasets by breaking them into smaller ones.
% The details of the adopted datasets and the hyperparameters adopted for each training pipeline are provided in \tablename~\ref{tab:datasets}.

\begin{table}[t!]
\centering
  % \vspace{-1.5em}
\resizebox{0.95\columnwidth}{!}{
\begin{tabular}{l|cccc}
\hline
\textbf{Dataset} &
  \textbf{CIFAR-10}~\cite{krizhevsky2009learning} &
  \textbf{GTSRB}~\cite{stallkamp2012man} &
  \textbf{PubFig}~\cite{kumar2009attribute} &
  \textbf{ImageNet}~\cite{russakovsky2015imagenet} \\ \hline
\textbf{\# of Classes} & 10              & 43              & 83               & 1000              \\
\textbf{\# of Samples} & 50,000          & 39,209          & 12,454           & 1,281,167        \\
\textbf{Input Shape}   & (3,32,32)       & (3,32,32)       & (3,224,224)      & (3,224,224)      \\
\textbf{Target Class}  & 5 (Dog)         & 38 (Keep Right) & 60 (Miley Cyrus) & 762; 578; 897 \\
\textbf{Model}         & PreActResNet-18 & VGG-16          & ResNet-18        & ResNet-18        \\
\textbf{Epochs}        & 200             & 50              & 60               & 200              \\
\textbf{Optimizer} &
  SGD~\cite{ruder2016overview} &
  Adam~\cite{kingma2014adam} &
  RAdam~\cite{liu2019variance} &
  SGD~\cite{ruder2016overview} \\
\textbf{ACC}           & 95.33           & 97.55           & 94.01            & 92.60; 86.00; 89.80       
\\
\textbf{Tar-ACC}           & 93.50           & 99.48           & 96.33            & 92.00; 92.00; 86.90   \\ \hline
\end{tabular}
}
\vspace{-.8em}
\caption{Hyperparameters and settings to obtain clean baseline models. The target class of each dataset is fixed across all attacks with one target label. ACC and Tar-ACC represent the value of these metrics without any attacks.
% in the table are the baselines for comparison of each attack and the defense results.
}
\label{tab:datasets}
\vspace{-1.5em}
\end{table}

% Similar to the synthesis conducted in Section \ref{sec:human_study}, we use the similar but extended attack settings to thoroughly evaluate our proposed solution. Primarily,
In addition to the five attacks in Section \ref{sec:human_study}, we add seven more representative attacks from different categories of data poisoning, the detailed attack settings are provided in TABLE \ref{tab:attacks}. We follow the poison ratio in the original attack papers. We mark the settings where the attacks are not successful (i.e., 
the attack settings that failed to result in an ASR above 50\%) 
in \scalebox{0.9}{\colorbox[HTML]{FFCCCC}{red}} and remove them from comparison.
% We make sure all the attacks evaluated are effective attacks. 
% In other words, the attack settings that are marked in \scalebox{0.9}{\colorbox[HTML]{FFCCCC}{red}}, i.e., the clean-label \cite{turner2019label} attacks on the GTSRB, PubFig and ImageNet, do not satisfy being effective attacks, thus removed for simplicity.
We also exclude PubFig and ImageNet for the Smooth attack (colored in \scalebox{0.9}{\colorbox[HTML]{C0C0C0}{gray}}) as they were not considered in the original paper~\cite{zeng2021rethinking}.

% \ruoxi{change the color to gray pls. I don't know how to do it}
% For replacement without losing the generality, we include Blended \cite{chen2017targeted} for the PubFig and ImageNet, which is also a backdoor attack with a large pattern of triggers blended into the image.

As size requirements for base sets in existing defenses are \textit{mostly less than 1000 samples}, we will mainly examine each method's ability to sift out a \textbf{1000-size} base set on each poisoned dataset. However, we will show that \AlgName can sift out more clean samples, but the performance may vary from attack to attack. Finally, we run the defense 3 times with different random seeds for all the results that include randomness (e.g., SGD for optimizations or random augmentations) and present the respective means and standard deviations.

% Since the size requirements for base sets in existing defenses are typically less than 1000 samples, our focus will be on evaluating each sifting method's ability to filter a \textbf{1000-sized} base set from each poisoned dataset. However, we will show that \AlgName can filter out more clean samples, although the performance may fluctuate depending on the attack. Finally, for all results that include randomnesses such as SGD optimizations or random augmentations, we will run the defense three times with different random seeds and present each run's mean and standard deviation.

%\vspace{0.2em}
\noindent
\underline{\textbf{\AlgName Settings.}}
\AlgName is equipped with a classifier model $\theta$, weight-assigning network $\psi$, and a list of gradient samplers $\Gamma$. We will use the same $\psi$ and $\Gamma$ on all datasets. But for $\theta$, we will select the model architecture based on the dataset size. $\theta$ functions as a feature extractor in \AlgName and a larger, more complex dataset naturally requires a larger model for feature extraction. Specifically,
for CIFAR-10 and GTSRB, we adopt ResNet-18; for PubFig and ImageNet, we use ResNet-34 and ResNet-50, respectively.
% The change of $\theta$ structure according to the complexity of the dataset makes sense. 
% As $\theta$ in \AlgName is functional as a feature extractor that feeds the results to $\Gamma$ and $\psi$. A more complex data set is naturally required to use a more complex (deeper) classifier for feature extraction.

\begin{table*}[t!]
\centering
% \vspace{-2em}
% \resizebox{0.95\textwidth}{!}{
\resizebox{0.95\textwidth}{!}{
\begin{tabular}{cc|cc|ccc|ccccccc}
\hline
\multicolumn{2}{c|}{} &
  \multicolumn{2}{c|}{\textbf{Label-only}} &
  \multicolumn{3}{c|}{\textbf{Feature-only}} &
  \multicolumn{7}{c}{\textbf{Label-Feature}} \\ \cline{3-14} 
\multicolumn{2}{c|}{\multirow{-2}{*}{}} &
  \begin{tabular}[c]{@{}c@{}}Targeted Label-\\ Flipping\cite{tolpegin2020data}\end{tabular} &
  \begin{tabular}[c]{@{}c@{}}Random Label-\\ Flipping\cite{ren2018learning}\end{tabular} &
  \begin{tabular}[c]{@{}c@{}}Clean-\\ Label\cite{turner2019label}\end{tabular} &
  \begin{tabular}[c]{@{}c@{}}Narcissus\\ Backdoor\cite{zeng2022narcissus}\end{tabular} &
  \begin{tabular}[c]{@{}c@{}}Poison\\ Frogs\cite{shafahi2018poison}\end{tabular} &
  \begin{tabular}[c]{@{}c@{}}BadNets\\ One-Tar\cite{gu2017badnets}\end{tabular} &
  \begin{tabular}[c]{@{}c@{}}Smooth\\ One-Tar\cite{zeng2021rethinking}\end{tabular} &
  \begin{tabular}[c]{@{}c@{}}IAB\\ One-Tar\cite{nguyen2020input}\end{tabular} &
  \begin{tabular}[c]{@{}c@{}}Blended\\ One-Tar\cite{chen2017targeted}\end{tabular} &
  \begin{tabular}[c]{@{}c@{}}BadNets\\ All-to-all\cite{gu2017badnets}\end{tabular} &
  \begin{tabular}[c]{@{}c@{}}Smooth\\ All-to-all\cite{zeng2021rethinking}\end{tabular} &
  \begin{tabular}[c]{@{}c@{}}Blended\\ All-to-all\cite{chen2017targeted}\end{tabular} \\ \hline
\multicolumn{1}{c|}{} &
  \textbf{\begin{tabular}[c]{@{}c@{}}Attack\\ Settings\end{tabular}} &
  \begin{tabular}[c]{@{}c@{}}$\left [ 3\rightarrow 5 \right ]$;\\ Tar: 16.67\%\end{tabular} &
  \begin{tabular}[c]{@{}c@{}}$\left [all \right ]$;\\ All: 20\%\end{tabular} &
  \begin{tabular}[c]{@{}c@{}}$\left [5 \right ]$;\\ Tar: 10\%\end{tabular} &
  \begin{tabular}[c]{@{}c@{}}$\left [5 \right ]$;\\ Tar: 10\%\end{tabular} &
  \begin{tabular}[c]{@{}c@{}}$\left [5 \right ]$;\\ Tar: 10\%\end{tabular} &
  \begin{tabular}[c]{@{}c@{}}$\left [5 \right ]$;\\ Tar: 33\%\end{tabular} &
  \begin{tabular}[c]{@{}c@{}}$\left [5 \right ]$;\\ Tar: 33\%\end{tabular} &
  \begin{tabular}[c]{@{}c@{}}$\left [5 \right ]$;\\ Tar: 33\%\end{tabular} &
  \begin{tabular}[c]{@{}c@{}}$\left [5 \right ]$;\\ Tar: 33\%\end{tabular} &
  \begin{tabular}[c]{@{}c@{}}$\left [all \right ]$;\\ All: 20\%\end{tabular} &
  \begin{tabular}[c]{@{}c@{}}$\left [all \right ]$;\\ All: 20\%\end{tabular} &
  \begin{tabular}[c]{@{}c@{}}$\left [all \right ]$;\\ All: 20\%\end{tabular} \\ \cline{2-14}

\multicolumn{1}{c|}{\multirow{-3.3}{*}{\textbf{\begin{tabular}[c]{@{}c@{}}\rotatebox{90}{CIFAR-10}\end{tabular}}}}&
  \textbf{\begin{tabular}[c]{@{}c@{}}Results\\(\%)\end{tabular}} &
  \begin{tabular}[c]{@{}c@{}}ACC: 91.77\\ ASR: 27.56\\ Tar-ACC: 83.78\end{tabular} &
  ACC: 71.8 &
  \begin{tabular}[c]{@{}c@{}}ACC: 91.52\\ ASR: 99.98\end{tabular} &
  \begin{tabular}[c]{@{}c@{}}ACC: 93.26\\ ASR: 100\end{tabular} &
  \begin{tabular}[c]{@{}c@{}}ACC: 92.88\\ ASR: 100\end{tabular} &
  \begin{tabular}[c]{@{}c@{}}ACC: 84.03\\ ASR: 95.78\end{tabular} &
  \begin{tabular}[c]{@{}c@{}}ACC: 85.63\\ ASR: 96.17\end{tabular} &
    \begin{tabular}[c]{@{}c@{}}ACC: 92.21\\ ASR: 91.20\end{tabular} &
  \begin{tabular}[c]{@{}c@{}}ACC: 89.43\\ ASR: 91.26\end{tabular} &
  \begin{tabular}[c]{@{}c@{}}ACC: 85.57\\ ASR: 84.15\end{tabular} &
  \begin{tabular}[c]{@{}c@{}}ACC: 84.05\\ ASR: 78.67\end{tabular} &
  \begin{tabular}[c]{@{}c@{}}ACC: 84.53\\ ASR: 88.74\end{tabular} \\ \hline
\multicolumn{1}{c|}{} &
  \textbf{\begin{tabular}[c]{@{}c@{}}Attack\\ Settings\end{tabular}} &
  \begin{tabular}[c]{@{}c@{}}$\left [ 2\rightarrow 38 \right ]$;\\ Tar: 16.67\%\end{tabular} &
  \begin{tabular}[c]{@{}c@{}}$\left [all \right ]$;\\ All: 20\%\end{tabular} &
  \cellcolor[HTML]{FFCCC9}\begin{tabular}[c]{@{}c@{}}$\left [38 \right ]$;\\ Tar: 10\%\end{tabular} &
  \begin{tabular}[c]{@{}c@{}}$\left [38 \right ]$;\\ Tar: 10\%\end{tabular} &
  \begin{tabular}[c]{@{}c@{}}$\left [38 \right ]$;\\ Tar: 10\%\end{tabular} &
  \begin{tabular}[c]{@{}c@{}}$\left [38 \right ]$;\\ Tar: 33\%\end{tabular} &
  \begin{tabular}[c]{@{}c@{}}$\left [38 \right ]$;\\ Tar: 33\%\end{tabular} &
  \begin{tabular}[c]{@{}c@{}}$\left [38 \right ]$;\\ Tar: 33\%\end{tabular} &
  \begin{tabular}[c]{@{}c@{}}$\left [38 \right ]$;\\ Tar: 33\%\end{tabular} &
  \begin{tabular}[c]{@{}c@{}}$\left [all \right ]$;\\ All: 20\%\end{tabular} &
  \begin{tabular}[c]{@{}c@{}}$\left [all \right ]$;\\ All: 20\%\end{tabular} &
  \begin{tabular}[c]{@{}c@{}}$\left [all \right ]$;\\ All: 20\%\end{tabular} \\ 
%   \cline{2-13} 
  \hhline{~|-|-|-|-|-|-|-|-|-|-|-|-|-|}
  
\multicolumn{1}{c|}{\multirow{-3}{*}{\textbf{\begin{tabular}[c]{@{}c@{}}\rotatebox{90}{GTSRB}\end{tabular}}}} &
  \textbf{\begin{tabular}[c]{@{}c@{}}Results\\(\%)\end{tabular}} &
  \begin{tabular}[c]{@{}c@{}}ACC: 97.43\\ ASR: 91.72\\ Tar-ACC: 98.84\end{tabular} &
  ACC: 95.28 &
  \cellcolor[HTML]{FFCCC9}\begin{tabular}[c]{@{}c@{}}ACC: 98.12\\ ASR: 6.41\end{tabular} &
  \begin{tabular}[c]{@{}c@{}}ACC: 97.83\\ ASR: 100\end{tabular} &
  \begin{tabular}[c]{@{}c@{}}ACC: 97.81\\ ASR: 100\end{tabular} &
  \begin{tabular}[c]{@{}c@{}}ACC: 97.10\\ ASR: 97.43\end{tabular} &
  \begin{tabular}[c]{@{}c@{}}ACC: 98.08\\ ASR: 98.92\end{tabular} &
  \begin{tabular}[c]{@{}c@{}}ACC: 97.88\\ ASR: 95.72\end{tabular} &
  \begin{tabular}[c]{@{}c@{}}ACC: 97.89\\ ASR: 98.11\end{tabular} &
  \begin{tabular}[c]{@{}c@{}}ACC: 96.87\\ ASR: 95.49\end{tabular} &
  \begin{tabular}[c]{@{}c@{}}ACC: 96.64\\ ASR: 95.75\end{tabular} &
  \begin{tabular}[c]{@{}c@{}}ACC: 96.825\\ ASR: 95.84\end{tabular}  \\ 
  \hhline{-|-|-|-|-|-|-|-|-|-|-|-|-|-|}
\multicolumn{1}{c|}{} &
  \textbf{\begin{tabular}[c]{@{}c@{}}Attack\\ Settings\end{tabular}} &
  \begin{tabular}[c]{@{}c@{}}$\left [ 52\rightarrow 60 \right ]$;\\ Tar: 16.67\%\end{tabular} &
  \begin{tabular}[c]{@{}c@{}}$\left [all \right ]$;\\ All: 20\%\end{tabular} &
  \cellcolor[HTML]{FFCCC9}\begin{tabular}[c]{@{}c@{}}$\left [60 \right ]$;\\ Tar: 10\%\end{tabular} &
  \begin{tabular}[c]{@{}c@{}}$\left [60 \right ]$;\\ Tar: 10\%\end{tabular} &
  \begin{tabular}[c]{@{}c@{}}$\left [60 \right ]$;\\ Tar: 10\%\end{tabular} &
  \begin{tabular}[c]{@{}c@{}}$\left [60 \right ]$;\\ Tar: 33\%\end{tabular} &
  \cellcolor[HTML]{C0C0C0} &
  \begin{tabular}[c]{@{}c@{}}$\left [60 \right ]$;\\ Tar: 33\%\end{tabular} &
  \begin{tabular}[c]{@{}c@{}}$\left [60 \right ]$;\\ Tar: 33\%\end{tabular} &
  \begin{tabular}[c]{@{}c@{}}$\left [all \right ]$;\\ All: 20\%\end{tabular} &
  \cellcolor[HTML]{C0C0C0} &
  \begin{tabular}[c]{@{}c@{}}$\left [all \right ]$;\\ All: 20\%\end{tabular} \\ 
  \hhline{~|-|-|-|-|-|-|-|-|-|-|-|-|-|}

\multicolumn{1}{c|}{\multirow{-2.8}{*}{\textbf{\begin{tabular}[c]{@{}c@{}}\rotatebox{90}{PubFig}\end{tabular}}}} &
  \textbf{\begin{tabular}[c]{@{}c@{}}Results\\(\%)\end{tabular}} &
  \begin{tabular}[c]{@{}c@{}}ACC: 91.33\\ ASR: 89.00\\ Tar-ACC: 93.75\end{tabular} &
  ACC: 77.89 &
  \cellcolor[HTML]{FFCCC9}\begin{tabular}[c]{@{}c@{}}ACC: 93.42\\ ASR: 33.18\end{tabular} &
  \begin{tabular}[c]{@{}c@{}}ACC: 93.50\\ ASR: 100\end{tabular} &
  \begin{tabular}[c]{@{}c@{}}ACC: 93.28\\ ASR: 98.00\end{tabular} &
  \begin{tabular}[c]{@{}c@{}}ACC: 91.19\\ ASR: 89.88\end{tabular} &
  \cellcolor[HTML]{C0C0C0} &
  \begin{tabular}[c]{@{}c@{}}ACC: 92.73\\ ASR: 84.53\end{tabular} &
  \begin{tabular}[c]{@{}c@{}}ACC: 93.57\\ ASR: 93.64\end{tabular} &
  \begin{tabular}[c]{@{}c@{}}ACC: 76.45\\ ASR: 66.69\end{tabular} &
  \cellcolor[HTML]{C0C0C0} &
  \begin{tabular}[c]{@{}c@{}}ACC: 91.84\\ ASR: 90.03\end{tabular} \\ \hhline{-|-|-|-|-|-|-|-|-|-|-|-|-|-|}

\multicolumn{1}{c|}{} &
  \textbf{\begin{tabular}[c]{@{}c@{}}Attack\\ Settings\end{tabular}} &
  \begin{tabular}[c]{@{}c@{}}$\left [ 385\rightarrow 762 \right ]$;\\ Tar: 16.67\%\end{tabular} &
  \begin{tabular}[c]{@{}c@{}}$\left [all \right ]$;\\ All: 20\%\end{tabular} &
  \cellcolor[HTML]{FFCCC9}\begin{tabular}[c]{@{}c@{}}$\left [762 \right ]$;\\ Tar: 10\%\end{tabular} &
  \begin{tabular}[c]{@{}c@{}}$\left [762 \right ]$;\\ Tar: 10\%\end{tabular} &
  \begin{tabular}[c]{@{}c@{}}$\left [762 \right ]$;\\ Tar: 10\%\end{tabular} &
  \begin{tabular}[c]{@{}c@{}}$\left [762 \right ]$;\\ Tar: 33\%\end{tabular} &
  \cellcolor[HTML]{C0C0C0} &
  \begin{tabular}[c]{@{}c@{}}$\left [762 \right ]$;\\ Tar: 33\%\end{tabular} &
  \begin{tabular}[c]{@{}c@{}}$\left [762 \right ]$;\\ Tar: 33\%\end{tabular} &
  \begin{tabular}[c]{@{}c@{}}$\left [all \right ]$;\\ All: 20\%\end{tabular} &
  \cellcolor[HTML]{C0C0C0} &
  \begin{tabular}[c]{@{}c@{}}$\left [all \right ]$;\\ All: 20\%\end{tabular} \\ 
  \hhline{~|-|-|-|-|-|-|-|-|-|-|-|-|-}

\multicolumn{1}{c|}{\multirow{-3.6}{*}{\textbf{\begin{tabular}[c]{@{}r@{}}\rotatebox{90}{ImageNet-1}\end{tabular}}}} &
  \textbf{\begin{tabular}[c]{@{}c@{}}Results\\(\%)\end{tabular}} &
  \begin{tabular}[c]{@{}c@{}}ACC: 89.80\\ ASR: 88.80\\ Tar-ACC: 87.25\end{tabular}&
  ACC: 77.40 &
  \cellcolor[HTML]{FFCCC9}\begin{tabular}[c]{@{}c@{}}ACC: 91.25\\ ASR: 5.33\end{tabular} &
  \begin{tabular}[c]{@{}c@{}}ACC: 92.60\\ ASR: 100\end{tabular} &
  \begin{tabular}[c]{@{}c@{}}ACC: 91.80\\ ASR: 100\end{tabular} &
  \begin{tabular}[c]{@{}c@{}}ACC: 90.10\\ ASR: 98.20\end{tabular} &
  \cellcolor[HTML]{C0C0C0} &
  \begin{tabular}[c]{@{}c@{}}ACC: 89.90\\ ASR: 95.20\end{tabular} &
  \begin{tabular}[c]{@{}c@{}}ACC: 89.20\\ ASR: 92.00\end{tabular} &
  \begin{tabular}[c]{@{}c@{}}ACC: 78.00\\ ASR: 76.60\end{tabular} &
  \cellcolor[HTML]{C0C0C0} &
  \begin{tabular}[c]{@{}c@{}}ACC: 78.40\\ ASR: 72.40\end{tabular} \\ 
%  \cline{2-13}
\hhline{-|-|-|-|-|-|-|-|-|-|-|-|-|-|}

\multicolumn{1}{c|}{} &
  \textbf{\begin{tabular}[c]{@{}c@{}}Attack\\ Settings\end{tabular}} &
  \begin{tabular}[c]{@{}c@{}}$\left [ 48\rightarrow 578 \right ]$;\\ Tar: 16.67\%\end{tabular} &
  \begin{tabular}[c]{@{}c@{}}$\left [all \right ]$;\\ All: 20\%\end{tabular} &
  \cellcolor[HTML]{FFCCC9}\begin{tabular}[c]{@{}c@{}}$\left [578 \right ]$;\\ Tar: 10\%\end{tabular} &
  \begin{tabular}[c]{@{}c@{}}$\left [578 \right ]$;\\ Tar: 10\%\end{tabular} &
  \begin{tabular}[c]{@{}c@{}}$\left [578 \right ]$;\\ Tar: 10\%\end{tabular} &
  \begin{tabular}[c]{@{}c@{}}$\left [578 \right ]$;\\ Tar: 33\%\end{tabular} &
  \cellcolor[HTML]{C0C0C0} &
  \begin{tabular}[c]{@{}c@{}}$\left [578 \right ]$;\\ Tar: 33\%\end{tabular} &
  \begin{tabular}[c]{@{}c@{}}$\left [578 \right ]$;\\ Tar: 33\%\end{tabular} &
  \begin{tabular}[c]{@{}c@{}}$\left [all \right ]$;\\ All: 20\%\end{tabular} &
  \cellcolor[HTML]{C0C0C0} &
  \begin{tabular}[c]{@{}c@{}}$\left [all \right ]$;\\ All: 20\%\end{tabular} \\ 
  \hhline{~|-|-|-|-|-|-|-|-|-|-|-|-|-|}

\multicolumn{1}{c|}{\multirow{-3.6}{*}{\textbf{\begin{tabular}[c]{@{}c@{}}\rotatebox{90}{ImageNet-2}\end{tabular}}}} &
  \textbf{\begin{tabular}[c]{@{}c@{}}Results\\(\%)\end{tabular}} &
  \begin{tabular}[c]{@{}c@{}}ACC: 84.00\\ ASR: 82.20\\ Tar-ACC: 88.00\end{tabular}&
  ACC: 75.60 &
  \cellcolor[HTML]{FFCCC9}\begin{tabular}[c]{@{}c@{}}ACC: 86.00\\ ASR: 7.55\end{tabular} &
  \begin{tabular}[c]{@{}c@{}}ACC: 85.60\\ ASR: 100\end{tabular} &
  \begin{tabular}[c]{@{}c@{}}ACC: 86.00\\ ASR: 100\end{tabular} &
  \begin{tabular}[c]{@{}c@{}}ACC: 86.80\\ ASR: 99.20\end{tabular} &
  \cellcolor[HTML]{C0C0C0} &
    \begin{tabular}[c]{@{}c@{}}ACC: 85.80\\ ASR: 97.40\end{tabular} &
    \begin{tabular}[c]{@{}c@{}}ACC: 85.80\\ ASR: 92.60\end{tabular} &
  \begin{tabular}[c]{@{}c@{}}ACC: 74.90\\ ASR: 71.30\end{tabular} &
  \cellcolor[HTML]{C0C0C0} &
  \begin{tabular}[c]{@{}c@{}}ACC: 72.40\\ ASR: 68.40\end{tabular} \\ 
\hhline{-|-|-|-|-|-|-|-|-|-|-|-|-|-|}

\multicolumn{1}{c|}{} &
  \textbf{\begin{tabular}[c]{@{}c@{}}Attack\\ Settings\end{tabular}} &
  \begin{tabular}[c]{@{}c@{}}$\left [ 830\rightarrow 897 \right ]$;\\ Tar: 16.67\%\end{tabular} &
  \begin{tabular}[c]{@{}c@{}}$\left [all \right ]$;\\ All: 20\%\end{tabular} &
  \cellcolor[HTML]{FFCCC9}\begin{tabular}[c]{@{}c@{}}$\left [897 \right ]$;\\ Tar: 10\%\end{tabular} &
  \begin{tabular}[c]{@{}c@{}}$\left [897 \right ]$;\\ Tar: 10\%\end{tabular} &
  \begin{tabular}[c]{@{}c@{}}$\left [897 \right ]$;\\ Tar: 10\%\end{tabular} &
  \begin{tabular}[c]{@{}c@{}}$\left [897 \right ]$;\\ Tar: 33\%\end{tabular} &
  \cellcolor[HTML]{C0C0C0} &
  \begin{tabular}[c]{@{}c@{}}$\left [578 \right ]$;\\ Tar: 33\%\end{tabular} &
  \begin{tabular}[c]{@{}c@{}}$\left [897 \right ]$;\\ Tar: 33\%\end{tabular} &
  \begin{tabular}[c]{@{}c@{}}$\left [all \right ]$;\\ All: 20\%\end{tabular} &
  \cellcolor[HTML]{C0C0C0} &
  \begin{tabular}[c]{@{}c@{}}$\left [all \right ]$;\\ All: 20\%\end{tabular} \\ 
  \hhline{~|-|-|-|-|-|-|-|-|-|-|-|-|-}
  
\multicolumn{1}{c|}{\multirow{-3.6}{*}{\textbf{\begin{tabular}[c]{@{}c@{}}\rotatebox{90}{ImageNet-3}\end{tabular}}}} &
  \textbf{\begin{tabular}[c]{@{}c@{}}Results\\(\%)\end{tabular}} &
  \begin{tabular}[c]{@{}c@{}}ACC: 86.00\\ ASR: 96.00\\ Tar-ACC: 84.00\end{tabular}&
  ACC: 75.20 &
  \cellcolor[HTML]{FFCCC9}\begin{tabular}[c]{@{}c@{}}ACC: 89.40\\ ASR: 6.22\end{tabular} &
  \begin{tabular}[c]{@{}c@{}}ACC: 89.60\\ ASR: 99.83\end{tabular} &
  \begin{tabular}[c]{@{}c@{}}ACC: 88.40\\ ASR: 100\end{tabular} &
  \begin{tabular}[c]{@{}c@{}}ACC: 89.60\\ ASR: 95.20\end{tabular} &
  \cellcolor[HTML]{C0C0C0} &
    \begin{tabular}[c]{@{}c@{}}ACC: 89.00\\ ASR: 91.56\end{tabular} &
  \begin{tabular}[c]{@{}c@{}}ACC: 88.80\\ ASR: 99.20\end{tabular} &
  \begin{tabular}[c]{@{}c@{}}ACC: 75.80\\ ASR: 72.30\end{tabular} &
  \cellcolor[HTML]{C0C0C0} &
  \begin{tabular}[c]{@{}c@{}}ACC: 78.80\\ ASR: 76.00\end{tabular} \\ 
\hhline{-|-|-|-|-|-|-|-|-|-|-|-|-|-|}
  
\end{tabular}
}
\vspace{-0.8em}
\caption{Effectiveness of different attacks on given datasets. We detail the attack settings by listing out the target labels (e.g., $\left [ 3\rightarrow 5 \right ]$ indicates that samples from class 3 are manipulated to the target-class 5, and $\left [ all \right ]$ indicates that samples are manipulated to all classes). Here, `Tar' and `All' denote the poison ratio in that particular setting, for the target class and the whole dataset respectively. Finally, attack performance is studied using ACC, ASR, or Tar-ACC as applicable. Note that the ASR of the Poison Frogs denotes the attack confidence. For ImageNet, we report results over 3 separate subsets.}
% to showcase that we can scan over all the subsets to obtain a complete base set; thus, results for three randomly selected subsets are listed.}
\label{tab:attacks}
\vspace{-0.9em}
\end{table*}

\begin{table*}[t!]
\centering
\resizebox{0.95\textwidth}{!}{
% \resizebox{0.95\textwidth}{!}{
\begin{tabular}{l cc|ccc|ccccccc||c}
\hline
\multicolumn{1}{c|}{}&
  \multicolumn{2}{c|}{\textbf{Label-only}} &
  \multicolumn{3}{c|}{\textbf{Feature-only}} &
  \multicolumn{7}{c||}{\textbf{Label-Feature}} &
   \\ \cline{2-13}
\multicolumn{1}{c|}{\multirow{-2}{*}{}} &
  \begin{tabular}[c]{@{}c@{}}Targeted Label-\\ Flipping\cite{tolpegin2020data}\end{tabular} &
  \begin{tabular}[c]{@{}c@{}}Random Label-\\ Flipping\cite{ren2018learning}\end{tabular} &
  \begin{tabular}[c]{@{}c@{}}Clean-\\ Label\cite{turner2019label}\end{tabular} &
  \begin{tabular}[c]{@{}c@{}}Narcissus\\ Backdoor\cite{zeng2022narcissus}\end{tabular} &
  \begin{tabular}[c]{@{}c@{}}Poison\\ Frog\cite{shafahi2018poison}\end{tabular} &
  \begin{tabular}[c]{@{}c@{}}BadNets\\ One-Tar\cite{gu2017badnets}\end{tabular} &
  \begin{tabular}[c]{@{}c@{}}Smooth\\ One-Tar\cite{zeng2021rethinking}\end{tabular} &
  \begin{tabular}[c]{@{}c@{}}IAB\\ One-Tar\cite{nguyen2020input}\end{tabular} &
  \begin{tabular}[c]{@{}c@{}}Blended\\ One-Tar\cite{chen2017targeted}\end{tabular} &
  \begin{tabular}[c]{@{}c@{}}BadNets\\ All-to-all\cite{gu2017badnets}\end{tabular} &
  \begin{tabular}[c]{@{}c@{}}Smooth\\ All-to-all\cite{zeng2021rethinking}\end{tabular} &
  \begin{tabular}[c]{@{}c@{}}Blended\\ All-to-all\cite{chen2017targeted}\end{tabular} &
  \multirow{-2}{*}{\textbf{Overhead (s)}} \\ \hline %\cline{1-1} \cline{2-11} 

\multicolumn{1}{l|}{DCM} &
  48.0 &
  96.5 &
  20.0 &
  \cellcolor[HTML]{FFCCCC}\textbf{100.0} &
  \cellcolor[HTML]{FFCCCC}\textbf{100.0} &
  \cellcolor[HTML]{FFCCCC}\textbf{111.0} &
  \cellcolor[HTML]{cfe2f3}\textbf{9.00} &
  \cellcolor[HTML]{FFCCCC}\textbf{105.1} &
  \cellcolor[HTML]{cfe2f3}\textbf{18.0} &
  57.0 &
  \cellcolor[HTML]{cfe2f3}\textbf{3.00} &
  32.0 &
  10 \\
\multicolumn{1}{l|}{MI-DCM$^*$} &
  \cellcolor[HTML]{cfe2f3}\textbf{40.0$\pm$3.45} &
  \cellcolor[HTML]{FFCCCC}\textbf{134.8$\pm$31.1} &
  83.3$\pm$3.45 &
  \cellcolor[HTML]{FFCCCC}\textbf{1000$\pm$0} &
  \cellcolor[HTML]{FFCCCC}\textbf{273.3$\pm$35.1} &
  \cellcolor[HTML]{FFCCCC}\textbf{114.0$\pm$0} &
  \cellcolor[HTML]{FFCCCC}\textbf{135.0$\pm$0} &
  \cellcolor[HTML]{FFCCCC}\textbf{120.0$\pm$0} &
  \cellcolor[HTML]{FFCCCC}\textbf{123.4$\pm$0} &
  \cellcolor[HTML]{FFCCCC}\textbf{100.0$\pm$13.2} &
  5.68$\pm$0.58 &
  \cellcolor[HTML]{cfe2f3}\textbf{10.8$\pm$6.78} &
  \cellcolor[HTML]{FFE599}\textbf{5411+300} \\
  
\multicolumn{1}{l|}{SF-Least$^*$ }&
  \cellcolor[HTML]{FFCCCC}\textbf{104.0$\pm$3.45} &
  \cellcolor[HTML]{cfe2f3}\textbf{20.0$\pm$5.50} &
  \cellcolor[HTML]{FFCCCC}\textbf{166.7$\pm$3.45} &
  \cellcolor[HTML]{cfe2f3}\textbf{60.0$\pm$0} &
  \cellcolor[HTML]{FFCCCC}\textbf{130.0$\pm$0} &
  \cellcolor[HTML]{cfe2f3}\textbf{72.0$\pm$3.00} &
  68.0$\pm$1.73 &
  \cellcolor[HTML]{cfe2f3}\textbf{48.0$\pm$10.2} &
  65.2$\pm$4.52 &
  \cellcolor[HTML]{cfe2f3}\textbf{51.8$\pm$6.79} &
  65.6$\pm$3.28 &
  58.3$\pm$7.89 &
  \cellcolor[HTML]{FFE599}\textbf{5411+15} \\
  
\multicolumn{1}{l|}{Loss-Scan$^*$ }&
  \cellcolor[HTML]{FFCCCC}\textbf{185.9$\pm$67.2} &
  \cellcolor[HTML]{FFCCCC}\textbf{451.0$\pm$53.2} &
  80.0$\pm$0 &
  \cellcolor[HTML]{cfe2f3}\textbf{180.0$\pm$10.0} &
  \cellcolor[HTML]{cfe2f3}\textbf{80.0$\pm$0} &
  99.0$\pm$16.0 &
  \cellcolor[HTML]{FFCCCC}\textbf{135.0$\pm$19.9} &
  84.0$\pm$22.4 &
  \cellcolor[HTML]{FFCCCC}\textbf{161.0$\pm$123.9} &
  69.5$\pm$15.7 &
  68.5$\pm$7.80 &
  \cellcolor[HTML]{FFCCCC}\textbf{733.1$\pm$307.5} &
  210 \\
  
\multicolumn{1}{l|}{Self-IF$^*$ }&
  \cellcolor[HTML]{FFCCCC}\textbf{168.0$\pm$6.00} &
  \cellcolor[HTML]{FFCCCC}\textbf{114.7$\pm$1.61} &
  \cellcolor[HTML]{cfe2fc}0$\pm$0 &
  \cellcolor[HTML]{FFCCCC}\textbf{116.7$\pm$11.5} &
  83.3$\pm$3.45 &
  89.0$\pm$1.73 &
  29.0$\pm$6.24 &
  \cellcolor[HTML]{FFCCCC}\textbf{108.1$\pm$23.1} &
  36.3$\pm$6.88 &
  \cellcolor[HTML]{FFCCCC}\textbf{119.5$\pm$3.46} &
  \cellcolor[HTML]{FFCCCC}\textbf{142.2$\pm$38.2} &
  \cellcolor[HTML]{FFCCCC}\textbf{137.6$\pm$31.6} &
  \cellcolor[HTML]{FFE599}\textbf{5411+19832} \\
\hline
\multicolumn{1}{l|}{\AlgName }&
  \cellcolor[HTML]{D9EAD3}\textbf{0$\pm$0} &
  \cellcolor[HTML]{D9EAD3}\textbf{0$\pm$0} &
  \cellcolor[HTML]{D9EAD3}\textbf{0$\pm$0} &
  \cellcolor[HTML]{D9EAD3}\textbf{0$\pm$0} &
  \cellcolor[HTML]{D9EAD3}\textbf{0$\pm$0} &
  \cellcolor[HTML]{D9EAD3}\textbf{0$\pm$0} &
  \cellcolor[HTML]{D9EAD3}\textbf{0$\pm$0} &
  \cellcolor[HTML]{D9EAD3}\textbf{0$\pm$0} &
  \cellcolor[HTML]{D9EAD3}\textbf{0$\pm$0} &
  \cellcolor[HTML]{D9EAD3}\textbf{0$\pm$0} &
  \cellcolor[HTML]{D9EAD3}\textbf{0$\pm$0} &
  \cellcolor[HTML]{D9EAD3}\textbf{0$\pm$0} &
  5$\times$100 \\ \hline
\end{tabular}
 }
 \vspace{-0.8em}
\caption{NCR results under the CIFAR-10 settings.}
\label{tab:cifar}
\vspace{-0.9em}
\end{table*}

%  \multicolumn{11}{c}{
% \multirow{2}{*}{
% \textbf{(\textcolor{blue}{b}) GTSRB~\cite{stallkamp2012man} results}}} \\ \\
% \hline

%\newline

\begin{table*}[t!]
\centering
% \resizebox{0.84\textwidth}{!}{
\resizebox{0.8\textwidth}{!}{
\begin{tabular}{c|l cc|cc|ccccc||c}
\hline

\multicolumn{1}{c|}{} &\multicolumn{1}{c|}{}&
  \multicolumn{2}{c|}{\textbf{Label-only}} &
  \multicolumn{2}{c|}{\textbf{Feature-only}} &
  \multicolumn{5}{c||}{\textbf{Label-Feature}} &
   \\ \cline{3-11}
\multicolumn{1}{c|}{} &\multicolumn{1}{c|}{\multirow{-2}{*}{}} &
  \begin{tabular}[c]{@{}c@{}}Targeted Label-\\ Flipping\cite{tolpegin2020data}\end{tabular} &
  \begin{tabular}[c]{@{}c@{}}Random Label-\\ Flipping\cite{ren2018learning}\end{tabular} &
%   \begin{tabular}[c]{@{}c@{}}Clean-\\ Label\cite{turner2019label}\end{tabular} &
  \begin{tabular}[c]{@{}c@{}}Narcissus\\ Backdoor\cite{zeng2022narcissus}\end{tabular} &
  \begin{tabular}[c]{@{}c@{}}Poison\\ Frog\cite{shafahi2018poison}\end{tabular} &
  \begin{tabular}[c]{@{}c@{}}BadNets\\ One-Tar\cite{gu2017badnets}\end{tabular} &
  \begin{tabular}[c]{@{}c@{}}IAB\\ One-Tar\cite{nguyen2020input}\end{tabular} &
  \begin{tabular}[c]{@{}c@{}}Blended\\ One-Tar\cite{chen2017targeted}\end{tabular} &
  \begin{tabular}[c]{@{}c@{}}BadNets\\ All-to-all\cite{gu2017badnets}\end{tabular} &
  \begin{tabular}[c]{@{}c@{}}Blended\\ All-to-all\cite{chen2017targeted}\end{tabular} &
  \multirow{-2}{*}{\textbf{Overhead (s)}} 
  \\ 
%   \hhline{~|-|-|-|-|-|-|-|-|-|-|}
\hline

\multicolumn{1}{c|}{} &\multicolumn{1}{l|}{DCM} &

  82.5 &
  \cellcolor[HTML]{FFCCCC}\textbf{132} &
%   \cellcolor[HTML]{cccccc}\textbf{NA} &
  76.9 &
  \cellcolor[HTML]{FFCCCC}\textbf{115.4} &
  \cellcolor[HTML]{FFCCCC}\textbf{127.9} &
  72.0 &
  \cellcolor[HTML]{FFCCCC}\textbf{129.8} &
  \cellcolor[HTML]{FFCCCC}\textbf{118.5} &
  \cellcolor[HTML]{FFCCCC}\textbf{118.5} &
  10 \\
\multicolumn{1}{c|}{} &\multicolumn{1}{l|}{MI-DCM$^*$} &
  \cellcolor[HTML]{cfe2f3}\textbf{59.8$\pm$2.80} &
  \cellcolor[HTML]{cfe2f3}\textbf{94.0$\pm$5.20} &
%   \cellcolor[HTML]{cccccc}\textbf{NA} &
  \cellcolor[HTML]{cfe2f3}\textbf{56.4$\pm$8.88} &
  \cellcolor[HTML]{FFCCCC}\textbf{115.4$\pm$20.3} &
  \cellcolor[HTML]{FFCCCC}\textbf{104.1$\pm$4.0} &
  \cellcolor[HTML]{FFCCCC}\textbf{108.0$\pm$1.9} &
  \cellcolor[HTML]{FFCCCC}\textbf{122.8$\pm$4.7} &
  \cellcolor[HTML]{cfe2f3}\textbf{97.9$\pm$2.9} &
  \cellcolor[HTML]{FFCCCC}\textbf{127.5$\pm$0} &
  \cellcolor[HTML]{FFE599}\textbf{7200+300} \\
\multicolumn{1}{c|}{} &\multicolumn{1}{l|}{SF-Least$^*$} &
  97.1$\pm$4.11 &
%   \cellcolor[HTML]{cccccc}\textbf{NA} &
  \cellcolor[HTML]{FFCCCC}\textbf{123.4$\pm$3.80} &
  \cellcolor[HTML]{FFCCCC}\textbf{113.3$\pm$5.20} &
  \cellcolor[HTML]{cfe2f3}\textbf{78.1$\pm$4.10} &
  \cellcolor[HTML]{FFCCCC}\textbf{132.3$\pm$6.83} &
  \cellcolor[HTML]{cfe2f3}\textbf{54.0$\pm$7.51} &
  \cellcolor[HTML]{FFCCCC}\textbf{108.7$\pm$15.4} &
  \cellcolor[HTML]{FFCCCC}\textbf{112.6$\pm$13.6} &
  \cellcolor[HTML]{FFCCCC}\textbf{109.2$\pm$12.5} &
  \cellcolor[HTML]{FFE599}\textbf{7200+15} \\
\multicolumn{1}{c|}{} &\multicolumn{1}{l|}{Loss-Scan$^*$} &
  \cellcolor[HTML]{FFCCCC}\textbf{185.9} &
%   \cellcolor[HTML]{cccccc}\textbf{NA} &
  \cellcolor[HTML]{FFCCCC}\textbf{366.1$\pm$65.4} &
  80.0$\pm$0.0 &
  80.0$\pm$0.0 &
  \cellcolor[HTML]{cfe2f3}\textbf{51.0$\pm$16.3} &
  63.1$\pm$18.4 &
  \cellcolor[HTML]{cfe2f3}\textbf{35.9$\pm$8.36} &
  \cellcolor[HTML]{FFCCCC}\textbf{271.3$\pm$18.9} &
  \cellcolor[HTML]{FFCCCC}\textbf{238.8$\pm$68.1} &
  240 \\
  \hhline{~|-|-|-|-|-|-|-|-|-|-|-|}
\multicolumn{1}{c|}{\multirow{-4.6}{*}{\rotatebox{90}{Subset-1}}} &
\multicolumn{1}{l|}{\AlgName }&
  \cellcolor[HTML]{D9EAD3}\textbf{0$\pm$0} &
  \cellcolor[HTML]{D9EAD3}\textbf{0$\pm$0} &
  \cellcolor[HTML]{D9EAD3}\textbf{0$\pm$0} &
%   \cellcolor[HTML]{D9EAD3}\textbf{0$\pm$0} &
  \cellcolor[HTML]{D9EAD3}\textbf{0$\pm$0} &
  \cellcolor[HTML]{D9EAD3}\textbf{0$\pm$0} &
  \cellcolor[HTML]{D9EAD3}\textbf{0$\pm$0} &
  \cellcolor[HTML]{D9EAD3}\textbf{0$\pm$0} &
  \cellcolor[HTML]{D9EAD3}\textbf{0$\pm$0} &
  \cellcolor[HTML]{D9EAD3}\textbf{0$\pm$0} &
  5$\times$120 \\ \hline \hline

% \multicolumn{1}{c|}{} &\multicolumn{1}{c|}{}&
%   \multicolumn{2}{c|}{\textbf{Label-only}} &
%   \multicolumn{2}{c|}{\textbf{Feature-only}} &
%   \multicolumn{4}{c||}{\textbf{Label-Feature}} &
%   \\ \cline{3-10}
% \multicolumn{1}{c|}{} &\multicolumn{1}{c|}{\multirow{-2}{*}{}} &
%   \begin{tabular}[c]{@{}c@{}}Targeted Label-\\ Flipping\cite{tolpegin2020data}\end{tabular} &
%   \begin{tabular}[c]{@{}c@{}}Random Label-\\ Flipping\cite{ren2018learning}\end{tabular} &
% %   \begin{tabular}[c]{@{}c@{}}Clean-\\ Label\cite{turner2019label}\end{tabular} &
%   \begin{tabular}[c]{@{}c@{}}Narcissus\\ Backdoor\cite{zeng2022narcissus}\end{tabular} &
%   \begin{tabular}[c]{@{}c@{}}Poison\\ Frog\cite{shafahi2018poison}\end{tabular} &
%   \begin{tabular}[c]{@{}c@{}}BadNets\\ One-Tar\cite{gu2017badnets}\end{tabular} &
%   \begin{tabular}[c]{@{}c@{}}Blended\\ One-Tar\cite{chen2017targeted}\end{tabular} &
%   \begin{tabular}[c]{@{}c@{}}BadNets\\ All-to-all\cite{gu2017badnets}\end{tabular} &
%   \begin{tabular}[c]{@{}c@{}}Blended\\ All-to-all\cite{chen2017targeted}\end{tabular} &
%   \multirow{-2}{*}{\textbf{Overhead (s)}} 
%   \\ 
%   \hhline{~|-|-|-|-|-|-|-|-|-|-|}

\multicolumn{1}{c|}{} &\multicolumn{1}{l|}{DCM} &

  \cellcolor[HTML]{cfe2f3}\textbf{4.95} &
  \cellcolor[HTML]{FFCCCC}\textbf{156.0} &
%   \cellcolor[HTML]{cccccc}\textbf{NA} &
  \cellcolor[HTML]{cfe2f3}\textbf{61.5} &
  \cellcolor[HTML]{FFCCCC}\textbf{115.4} &
  85.1 &
  78.0 &
  85.1 &
  \cellcolor[HTML]{FFCCCC}\textbf{137.0} &
  \cellcolor[HTML]{FFCCCC}\textbf{132.0} &
  10 \\
\multicolumn{1}{c|}{} &\multicolumn{1}{l|}{MI-DCM$^*$} &
  \cellcolor[HTML]{FFCCCC}\textbf{118.8$\pm$0} &
  \cellcolor[HTML]{FFCCCC}\textbf{102.8$\pm$4.52} &
%   \cellcolor[HTML]{cccccc}\textbf{NA} &
  91.5$\pm$4.17 &
  \cellcolor[HTML]{FFCCCC}\textbf{133.5$\pm$6.78} &
  \cellcolor[HTML]{cfe2f3}\textbf{39.8$\pm$10.6} &
  \cellcolor[HTML]{FFCCCC}\textbf{111.0$\pm$54.3} &
  68.3$\pm$3.65 &
 \cellcolor[HTML]{cfe2f3}\textbf{71.5$\pm$4.67} &
  \cellcolor[HTML]{FFCCCC}\textbf{102.1$\pm$0.89} &
  \cellcolor[HTML]{FFE599}\textbf{7200+300} \\
\multicolumn{1}{c|}{} &\multicolumn{1}{l|}{SF-Least$^*$} &
  54.2$\pm$1.45 &
%   \cellcolor[HTML]{cccccc}\textbf{NA} &
  \cellcolor[HTML]{cfe2f3}\textbf{83.2$\pm$0.92} &
  76.9$\pm$0 &
  \cellcolor[HTML]{cfe2f3}\textbf{0.0$\pm$0} &
  71.3$\pm$5.31 &
  66.1$\pm$3.58 &
  49.4$\pm$10.6 &
  \cellcolor[HTML]{FFCCCC}\textbf{124.1$\pm$3.82} &
  \cellcolor[HTML]{FFCCCC}\textbf{128.3$\pm$5.67} &
  \cellcolor[HTML]{FFE599}\textbf{7200+15} \\
\multicolumn{1}{c|}{} &\multicolumn{1}{l|}{Loss-Scan$^*$} &
  \cellcolor[HTML]{FFCCCC}\textbf{221.9$\pm$24.1} &
%   \cellcolor[HTML]{cccccc}\textbf{NA} &
  \cellcolor[HTML]{FFCCCC}\textbf{290.5$\pm$32.1} &
  80.0$\pm$16.9 &
  60.0$\pm$0.0 &
  45.0$\pm$6.34 &
  \cellcolor[HTML]{cfe2f3}\textbf{63.0$\pm$12.33} &
  \cellcolor[HTML]{cfe2f3}\textbf{39.8$\pm$5.44} &
  \cellcolor[HTML]{FFCCCC}\textbf{316.0$\pm$54.6} &
  \cellcolor[HTML]{FFCCCC}\textbf{288.7$\pm$32.1} &
  240 \\
  \hhline{~|-|-|-|-|-|-|-|-|-|-|-|}
% \hline
\multicolumn{1}{c|}{\multirow{-4.6}{*}{\rotatebox{90}{Subset-2}}} &
\multicolumn{1}{l|}{\AlgName }&
  \cellcolor[HTML]{D9EAD3}\textbf{0$\pm$0} &
  \cellcolor[HTML]{D9EAD3}\textbf{0$\pm$0} &
  \cellcolor[HTML]{D9EAD3}\textbf{0$\pm$0} &
%   \cellcolor[HTML]{D9EAD3}\textbf{0$\pm$0} &
  \cellcolor[HTML]{D9EAD3}\textbf{0$\pm$0} &
  \cellcolor[HTML]{D9EAD3}\textbf{0$\pm$0} &
  \cellcolor[HTML]{D9EAD3}\textbf{0$\pm$0} &
  \cellcolor[HTML]{D9EAD3}\textbf{0$\pm$0} &
  \cellcolor[HTML]{D9EAD3}\textbf{0$\pm$0} &
  \cellcolor[HTML]{D9EAD3}\textbf{0$\pm$0} &
  5$\times$120 \\ \hline
  \hline

% \multicolumn{1}{c|}{} &\multicolumn{1}{c|}{}&
%   \multicolumn{2}{c|}{\textbf{Label-only}} &
%   \multicolumn{2}{c|}{\textbf{Feature-only}} &
%   \multicolumn{4}{c||}{\textbf{Label-Feature}} &
%   \\ \cline{3-10}
% \multicolumn{1}{c|}{} &\multicolumn{1}{c|}{\multirow{-2}{*}{}} &
%   \begin{tabular}[c]{@{}c@{}}Targeted Label-\\ Flipping\cite{tolpegin2020data}\end{tabular} &
%   \begin{tabular}[c]{@{}c@{}}Random Label-\\ Flipping\cite{ren2018learning}\end{tabular} &
% %   \begin{tabular}[c]{@{}c@{}}Clean-\\ Label\cite{turner2019label}\end{tabular} &
%   \begin{tabular}[c]{@{}c@{}}Narcissus\\ Backdoor\cite{zeng2022narcissus}\end{tabular} &
%   \begin{tabular}[c]{@{}c@{}}Poison\\ Frog\cite{shafahi2018poison}\end{tabular} &
%   \begin{tabular}[c]{@{}c@{}}BadNets\\ One-Tar\cite{gu2017badnets}\end{tabular} &
%   \begin{tabular}[c]{@{}c@{}}Blended\\ One-Tar\cite{chen2017targeted}\end{tabular} &
%   \begin{tabular}[c]{@{}c@{}}BadNets\\ All-to-all\cite{gu2017badnets}\end{tabular} &
%   \begin{tabular}[c]{@{}c@{}}Blended\\ All-to-all\cite{chen2017targeted}\end{tabular} &
%   \multirow{-2}{*}{\textbf{Overhead (s)}} 
%   \\ 
%   \hhline{~|-|-|-|-|-|-|-|-|-|-|}

\multicolumn{1}{c|}{} &\multicolumn{1}{l|}{DCM} &

  \cellcolor[HTML]{FFCCCC}\textbf{148.5} &
  \cellcolor[HTML]{FFCCCC}\textbf{106.5} &
%   \cellcolor[HTML]{cccccc}\textbf{NA} &
  46.1 &
  76.9 &
  91.1 &
  84.1 &
  91.1 &
  \cellcolor[HTML]{FFCCCC}\textbf{107.0} &
  \cellcolor[HTML]{FFCCCC}\textbf{107.0} &
  10 \\
\multicolumn{1}{c|}{} &\multicolumn{1}{l|}{MI-DCM$^*$} &
  99.0$\pm$0 &
  99.0$\pm$0 &
%   \cellcolor[HTML]{cccccc}\textbf{NA} &
  \cellcolor[HTML]{cfe2f3}\textbf{31.2$\pm$3.43} &
  98.72$\pm$2.34 &
  93.8$\pm$5.87 &
  93.1$\pm$4.66 &
  62.3$\pm$6.78 &
 93.3$\pm$10.9 &
  \cellcolor[HTML]{FFCCCC}\textbf{113.5$\pm$4.62} &
  \cellcolor[HTML]{FFE599}\textbf{7200+300} \\
\multicolumn{1}{c|}{} &\multicolumn{1}{l|}{SF-Least$^*$} &
  92.1$\pm$10.9 &
%   \cellcolor[HTML]{cccccc}\textbf{NA} &
  \cellcolor[HTML]{cfe2f3}\textbf{82.0$\pm$0} &
  63.5$\pm$5.65 &
  \cellcolor[HTML]{cfe2f3}\textbf{61.5$\pm$3.23} &
  61.2$\pm$4.67 &
  \cellcolor[HTML]{cfe2f3}\textbf{69.0$\pm$3.69} &
  79.2$\pm$8.75 &
  90.8$\pm$4.53 &
  \cellcolor[HTML]{cfe2f3}\textbf{68.3$\pm$12.0} &
  \cellcolor[HTML]{FFE599}\textbf{7200+15} \\
\multicolumn{1}{c|}{} &\multicolumn{1}{l|}{Loss-Scan$^*$} &
  \cellcolor[HTML]{cfe2f3}\textbf{85.9$\pm$12.6} &
%   \cellcolor[HTML]{cccccc}\textbf{NA} &
  \cellcolor[HTML]{FFCCCC}\textbf{233.5$\pm$23.1} &
  90.0$\pm$8.20 &
  70.0$\pm$0.23 &
  \cellcolor[HTML]{cfe2f3}\textbf{42.0$\pm$4.32} &
  84.0$\pm$15.6 &
  \cellcolor[HTML]{cfe2f3}\textbf{35.7$\pm$0.60} &
  \cellcolor[HTML]{cfe2f3}\textbf{89.0$\pm$6.89} &
  \cellcolor[HTML]{FFCCCC}\textbf{105.5$\pm$19.2} &
  240 \\
  \hhline{~|-|-|-|-|-|-|-|-|-|-|-|}
\multicolumn{1}{c|}{\multirow{-4.6}{*}{\rotatebox{90}{Subset-3}}} &
\multicolumn{1}{l|}{\AlgName }&
  \cellcolor[HTML]{D9EAD3}\textbf{0$\pm$0} &
  \cellcolor[HTML]{D9EAD3}\textbf{0$\pm$0} &
%   \cellcolor[HTML]{D9EAD3}\textbf{0$\pm$0} &
  \cellcolor[HTML]{D9EAD3}\textbf{0$\pm$0} &
  \cellcolor[HTML]{D9EAD3}\textbf{0$\pm$0} &
  \cellcolor[HTML]{D9EAD3}\textbf{0$\pm$0} &
  \cellcolor[HTML]{D9EAD3}\textbf{0$\pm$0} &
  \cellcolor[HTML]{D9EAD3}\textbf{0$\pm$0} &
  \cellcolor[HTML]{D9EAD3}\textbf{0$\pm$0} &
  \cellcolor[HTML]{D9EAD3}\textbf{0$\pm$0} &
  5$\times$120 \\ \hline

\end{tabular}
}
\vspace{-0.8em}
\caption{NCR results under the ImageNet settings. We dropped Self-IF due to an exploding computational time.}
\label{tab:imagenet}
\vspace{-1.6em}
\end{table*}

\vspace{-1em}
\subsection{Sifting Performance}
\vspace{-.5em}
Now we evaluate and compare \AlgName with the five existing automated methods previously discussed in Section \ref{sec:human_study}---DCM, MI-DCM, SF-Least \cite{tran2018spectral}, Loss-Scan \cite{li2021anti}, and Self-IF \cite{koh2017understanding}.
% ---which are not capable of sifting out a clean base set in the evaluated settings (Table \ref{table:machine}).
Table \ref{tab:attacks} compares NCR of the five baselines and \AlgName across different datasets with richer attack settings than Table \ref{table:machine}.
An NCR value above 100\% indicates that the selection performance is worse than the naive random baseline. By contrast, an NCR value below 100\% implies better selection than random. The best possible NCR is 0\%.
% maximum precision being an NCR equal to 0\%. 
% NCR can be a relatively intuitive metric to compare and indicate how precise a given method can be on the data sifting problem.
% For instance, a NCR of 50\% implies that if random sampling selects 100 poisoned data points from a dataset, then the sifting strategy selects 50 poisoned samples.  
We mark the results worse than random selection in \scalebox{0.9}{\colorbox[HTML]{FFCCCC}{red}}, the time-consuming selection results in \scalebox{0.9}{\colorbox[HTML]{FFE599}{yellow}}, and the best baseline among the four existing automated methods in \scalebox{0.9}{\colorbox[HTML]{cfe2f3}{blue}}.
If our result is better than those of the existing methods and achieves the maximum precision, 
we mark them in \scalebox{0.9}{\colorbox[HTML]{D9EAD3}{green}}. 
% \rev{
Note that across all the baselines, no augmentations technique is included (i.e., sample-dilution). To ablate the effect of sample dilution, we present additional results in Table \ref{tab:otheraug}, Appendix \ref{sec:ablation}, and show that sample-dilution is not the decisive factor that can help other techniques resolve the sifting problem.
% }

Table \ref{tab:cifar} evaluates the sifting methods on CIFAR-10 aginst different attacks. CIFAR-10 is a balanced dataset containing 10 classes. Hence, we select 100 samples per class to form the base set of a total size $1000$.
% our setting considers a selection of 1000 samples in total, which will be 100 per class.
In most settings, the five baselines do not yield a completely clean base set, and the performance varies largely from one attack to another (even within the same category).
% across even attacks from the same category.
By contrast, \AlgName produces a clean subset (NCR = 0\%) in all the evaluated settings while enjoying a lower computational overhead than most of the baselines. 
% \rev{
Note that many attacks listed in Table \ref{tab:attacks} can achieve similar attack effects but with a smaller poison ratio. Primarily, many defense works have found attacks with lower poison ratios are harder to be identified \cite{xiang2022post,liu2019abs}. Thus, we present additional results of \AlgName on CIFAR-10 with smaller poison ratios in Table \ref{tab:lowpoi}, Appendix \ref{sec:lowpoi}.
% }
Similar results can be found on GTSRB and PubFig, which are deferred to the Appendix \ref{sec:addresult} due to the page limitations.

Finally, the ImageNet results are presented in Table \ref{tab:imagenet}. As discussed, we break down the ImageNet dataset into smaller subsets and sift each subset sequentially to accelerate the sifting process. 
% a single run over the full ImageNet is challenging, but \AlgName can still deliver an average NCR of around 10\%. However, this result is still not acceptable, the base set by definition requires high precision. We identify the limitation in directly running over the full ImageNet is that $\theta$ is not capable of processing the whole dataset. Alternatively, we propose to break the full ImageNet into small subsets and use \AlgName to process each subset; thus, one can still output a reliable high precision base set using \AlgName. To strengthen this claim,
Table \ref{tab:imagenet} demonstrates the sifting performance on three randomly selected 10-class subsets from ImageNet. The most important contribution of \AlgName is that it‘s the only method that %is capable of reliably sifting 
can reliably sift out a completely clean base set (NCR = 0\%) across different attacks.

\begin{figure}[t!]
  \centering
    \vspace{-2em}
  \includegraphics[width=0.9\linewidth]{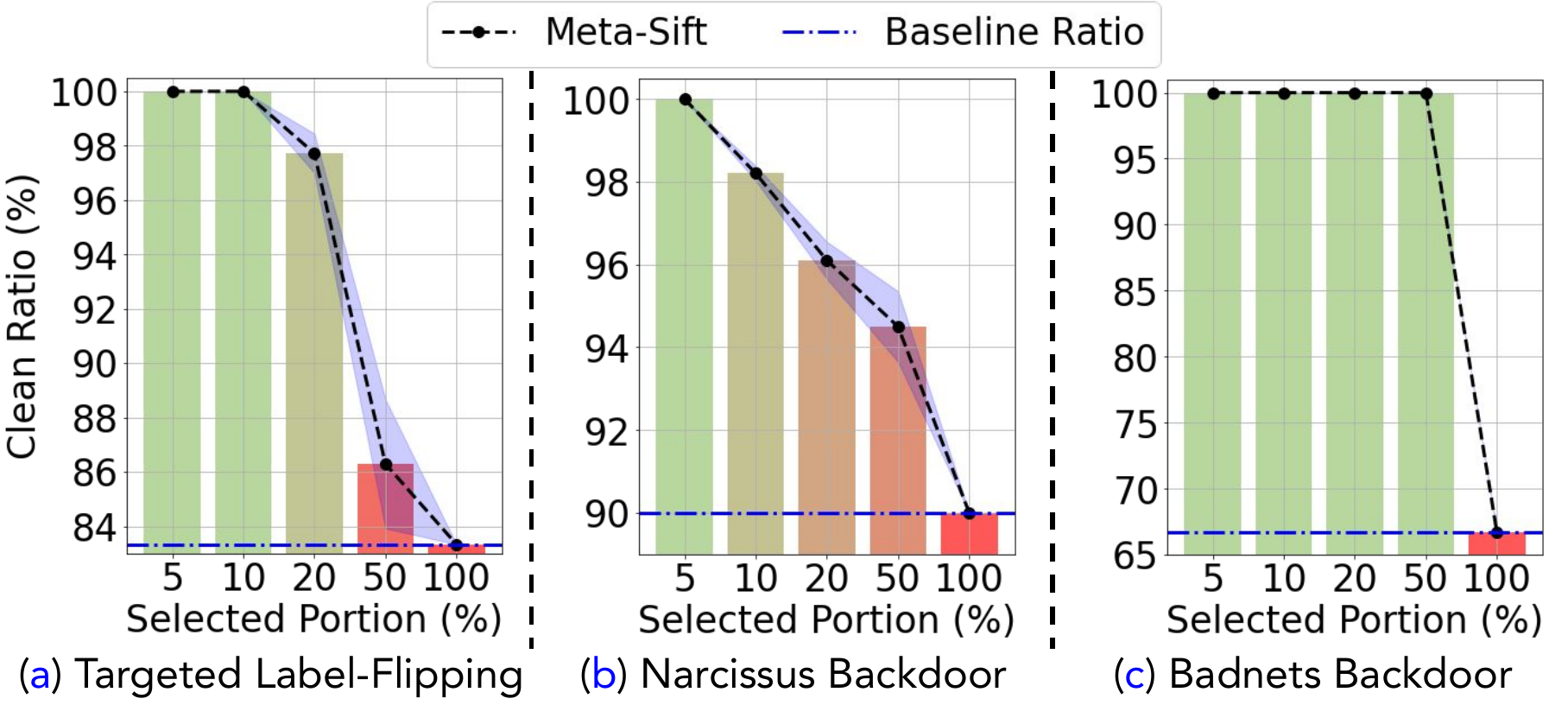}
  \vspace{-.9em}
  \caption{Sifting results of \AlgName under representative settings of each attack category, GTSRB. Shades depict standard deviation among different runs.}
  % }
  \label{fig:efficacy}
     \vspace{-1.5em}
\end{figure}

While we set the base set size to 1000 samples for the above experiments, we also study the behavior of \AlgName for selecting a larger base set. Figure \ref{fig:efficacy} depicts the percentage of clean samples under different sifting budgets on GTSRB. 
% \AlgName can indeed reliably select more clean samples by looking at the ``high-value'' end of the mapped values. 
\AlgName can select around 4000 (10\%) clean samples under the Targeted Label-Flipping case. Notably, \AlgName can achieve almost a perfect split when the poisoned instances are generated by the BadNets backdoor attack. On GTSRB, a general observation is that \AlgName achieves a better split on attacks that manipulate labels. This is because the features for every class in GTSRB samples have relatively low variance and are distinct from other classes. Since incorporating data corrupted by label manipulations into the split for testing largely increases the outer loss in Eqn. (\ref{eq:bilevel-final_1},\ref{eq:bilevel-final_2}), these data will be assigned a small weight and thus get unselected.
% will be assigned a small weight  
% can largely reduce the contribution of that point to the inner optimization in Eqn. (\ref{eq:bilevel-final}), thus making it easier to be dropped from the sifting.
Meanwhile, \AlgName 's effectiveness is worse on attacks that only involve feature manipulation such as the Narcissus backdoor attack, but it can still maintain a high precision up to the selection budget of about 2000 samples (accounting for 5\% of the total size).

We want to highlight that \emph{there is a trade-off between \AlgName\!'s precision and the base set size to be selected.} 
% The precision of \AlgName degrades as more samples get selected and hence it cannot produce a clean dataset large enough to train a model from scratch. 
However, existing defenses usually require only a small base set (i.e., often less than 1\% of the total size of the poisoned set). The fact that \AlgName can reliably sift out a pure clean set of more than 5\% of the total size across different attacks makes it useful for defense applications (see detailed evaluation in Section~\ref{sec:down_defs}). By contrast, the precision of human inspection does not change with the selection budget, because each sample is examined separately.
% Hence, one cannot expect to improve human inspection precision by reducing the base set size to be selected.
Moreover, the existing automated methods constantly produce a very low precision regardless of the selection budget.
% \ruoxi{it'll be great if we can have the results similar to figure 5 for other baselines to substantiate the claim. we want to show that the baselines are constently bad even reducing selection bdugets.} %marked by Yi

\vspace{-1em}
\subsection{Downstream Defense Evaluation}
\vspace{-.5em}
\label{sec:down_defs}
% The previous section demonstrated that we can sift out at least 5\% (about 2000 samples) clean samples from the poisoned dataset, whereas current defense algorithms only require 1000 clean samples as the validation set, implying that our chosen subset is capable of combining with existing defense algorithms on different datasets. The detailed experiment setup and results will be illustrated in the following.

% Now, we evaluate how the selected base set can help to give rise to downstream defenses. We consider the defense methods introduced in Section \ref{sec:human_study}, i.e., the Frequency Detetctor

Now, we evaluate the performance of defenses 
with the selected base set plugged in. Note that the data used to build the Sifters is the poisoned training set. Downstream defenses use the subset selected by the Sifters as the base set to achieve their specific defense goals. The performance of each defense is assessed on separate held-out clean and poisoned datasets.
We consider the defense categories introduced in Section \ref{sec:human_study}.
% , i.e., Poison Detection, Trojan-Net Detection, Backdoor Removal, and Robust Training. 
We present the results of each defense when 1) the base set is completely clean (\textbf{Clean}); 2) the base set is randomly selected from the contaminated training set (\textbf{Random}); and 3) the base set is selected by \textbf{\AlgName}.
% \textbf{Clean} case: following the original settings by using a clean base set; 2) \textbf{Random} selection case: using a randomly selected subset from the training set, which is shown to be more reliable and robust than many existing automated sifting methods; 3) using \textbf{\AlgName} as an off-the-shelf tool to obtain the base set that required.

% \begingroup
% \setlength{\columnsep}{10pt}%
% \begin{wrapfigure}{r}{0.36\linewidth}
%     %\vspace{-1em}
%     \centering
%     \hspace{-3em}
%     \includegraphics[width=\linewidth]{figs/sample_fail.png}
%     \caption{One poison infiltrating the base set can nullify defense.
%     % state-of-the-art poison detection method\cite{zeng2021rethinking}'s effect (CIFAR-10).
%     }
%     \label{fig:defense-fail}
%     %\vspace{-0.8em}
% \end{wrapfigure}
% \lipsum[1]
% \endgroup

% \begingroup
% \setlength{\columnsep}{10pt}%

% \lipsum[1]
% \endgroup
\begin{table}[t!]
\centering
% \vspace{-1.5em}
\resizebox{0.71\linewidth}{!}{
\begin{tabular}{c|ccccc}
\hline
 &
  \begin{tabular}[c]{@{}c@{}}BadNets\\ \cite{gu2017badnets}\end{tabular} &
  \begin{tabular}[c]{@{}c@{}}Trojan-WM\\ \cite{liu2017trojaning}\end{tabular} &
  \begin{tabular}[c]{@{}c@{}}Trojan-SQ\\ \cite{liu2017trojaning}\end{tabular} &
  \begin{tabular}[c]{@{}c@{}}$l_2$ inv\\ \cite{li2020invisible}\end{tabular} &
  \begin{tabular}[c]{@{}c@{}}$l_0$ inv\\ \cite{li2020invisible}\end{tabular} \\ \hline
Clean      & 89.27 & 100.0 & 99.80 & 100   & 99.99 \\
Random &
  \cellcolor[HTML]{FFCCCC}\textbf{44.88} &
  \cellcolor[HTML]{FFCCCC}\textbf{25.83} &
  \cellcolor[HTML]{FFCCCC}\textbf{30.55} &
  \cellcolor[HTML]{FFCCCC}\textbf{53.42} &
  \cellcolor[HTML]{FFCCCC}\textbf{0.23} \\
  \hline
\AlgName & 88.54 & 99.99 & 99.78 & 100.0 & 95.38 \\ \hline
\end{tabular}}
\vspace{-0.8em}
\caption{PFRs ($\uparrow$) of the Detector under CIFAR-10 attacks.}
\label{tab:frequency}
\vspace{-1.5em}
\end{table}

%\vspace{0.2em}
\noindent
\underline{\textbf{Poison Detection:}}
Similar to the evaluation in Section \ref{sec:human_study}, we consider the state-of-the-art frequency-based poison detection method \cite{zeng2021rethinking}. Following the original work, we consider five attack settings with a poison ratio of 5\%. The original setting of the frequency detector requires a base set of size 100. Table \ref{tab:frequency} compares the defense performance between the Clean, Random, and \AlgName-selected base set.
% shows the results of using different base set settings. 
% In particular, 
\AlgName enables the frequency detector to obtain a defense result similar to the result using the original clean set.

%\vspace{0.2em}
\noindent
\underline{\textbf{Trojan-Net Detection:}}
% \yi{MNTD-Table, open when using}
\begin{wraptable}{r}{0.38 \linewidth}
\centering
\vspace{-2.2em}
\resizebox{\linewidth}{!}{
\begin{tabular}{c|cc}
\hline
 &
  \begin{tabular}[c]{@{}c@{}}BadNets\\ \cite{gu2017badnets}\end{tabular} &
  \begin{tabular}[c]{@{}c@{}}Blended\\ \cite{chen2017targeted}\end{tabular} 
 \\ \hline
Clean      & 99.20 & 99.85 \\
Random &
  \cellcolor[HTML]{FFCCCC}\textbf{47.34} &
  \cellcolor[HTML]{FFCCCC}\textbf{49.83} 
\\
  \hline
\AlgName & 99.50 & 99.92\\ \hline
\end{tabular}
}
\vspace{-0.8em}
\caption{AUCs ($\uparrow$) of MNTD on the MNIST.}
\label{tab:mntd}
\vspace{-1.5em}
\end{wraptable}
Now we evaluate how \AlgName may help MNTD \cite{xu2021detecting}. In the original work, it require a base set of size 1000 over the MNIST dataset in the jumbo training procedure to train 256 clean and 256 randomly poisoned models. Like the original work, we incorporate two sets of backdoor attacks with a poison ratio 
of 5\%, namely, BadNets and Blended, and report the results of MNTD with different base sets. The results are listed in Table \ref{tab:mntd}, showing that the performance with \AlgName-generated base set is even higher than that with a clean base set. This implies that \AlgName can help MNTD to work in a situation without access to a clean base set.

\begin{table}[h!]
\centering
\vspace{-.6em}
\resizebox{0.71\linewidth}{!}{
\begin{tabular}{c|ccccc}
\hline
 &
  \begin{tabular}[c]{@{}c@{}}BadNets\\ \cite{gu2017badnets}\end{tabular} &
  \begin{tabular}[c]{@{}c@{}}Trojan-WM\\ \cite{liu2017trojaning}\end{tabular} &
  \begin{tabular}[c]{@{}c@{}}Trojan-SQ\\ \cite{liu2017trojaning}\end{tabular} &
  \begin{tabular}[c]{@{}c@{}}$l_2$ inv\\ \cite{li2020invisible}\end{tabular} &
  \begin{tabular}[c]{@{}c@{}}$l_0$ inv\\ \cite{li2020invisible}\end{tabular} \\ \hline
No Defense      & 97.43 & 99.37 & 98.90 & 98.36   & 98.11 \\\hline
Clean      & 18.83 & 19.21 & 18.34 & 16.78   & 16.33 \\
Random &
  \cellcolor[HTML]{FFCCCC}\textbf{62.67} &
  \cellcolor[HTML]{FFCCCC}\textbf{71.34} &
  \cellcolor[HTML]{FFCCCC}\textbf{68.78} &
  \cellcolor[HTML]{FFCCCC}\textbf{59.06} &
  \cellcolor[HTML]{FFCCCC}\textbf{61.32} \\
  \hline
\AlgName & 16.42 & 17.56 & 16.88 & 16.23 & 14.98 \\ \hline
\end{tabular}}
\vspace{-.8em}
\caption{ASRs ($\downarrow$) of NC purified models on the GTSRB.}
\label{tab:nc}
\vspace{-2.3em}
\end{table}

\begin{table}[h!]
\centering
% \vspace{-2em}
\resizebox{0.71\linewidth}{!}{
\begin{tabular}{c|ccccc}
\hline
 &
  \begin{tabular}[c]{@{}c@{}}BadNets\\ \cite{gu2017badnets}\end{tabular} &
  \begin{tabular}[c]{@{}c@{}}Trojan-WM\\ \cite{liu2017trojaning}\end{tabular} &
  \begin{tabular}[c]{@{}c@{}}Trojan-SQ\\ \cite{liu2017trojaning}\end{tabular} &
  \begin{tabular}[c]{@{}c@{}}$l_2$ inv\\ \cite{li2020invisible}\end{tabular} &
  \begin{tabular}[c]{@{}c@{}}$l_0$ inv\\ \cite{li2020invisible}\end{tabular} \\ \hline
No Defense      & 97.43 & 99.37 & 98.90 & 98.36   & 98.11 \\\hline
Clean      & 12.58 & 7.28 & 5.30 & 18.82   & 12.82 \\
Random &
  \cellcolor[HTML]{FFCCCC}\textbf{96.55} &
  \cellcolor[HTML]{FFCCCC}\textbf{98.04} &
  \cellcolor[HTML]{FFCCCC}\textbf{93.78} &
  \cellcolor[HTML]{FFCCCC}\textbf{93.29} &
  \cellcolor[HTML]{FFCCCC}\textbf{98.78} \\
  \hline
\AlgName & 10.65 & 4.50 & 5.62 & 6.75 & 5.77 \\ \hline
\end{tabular}}
\vspace{-0.8em}
\caption{ASRs ($\downarrow$) of I-BAU purified models on the GTSRB.}
\label{tab:ibau}
\vspace{-1.em}
\end{table}

%\vspace{0.2em}
\noindent
\underline{\textbf{Backdoor Removal:}}
Table \ref{tab:nc} and \ref{tab:ibau} show the result of NC and I-BAU using different 1000-size base sets on GTSRB. We use the same attack settings as the original works. Interestingly, by using the base set selected by \AlgName, both NC and I-BAU can achieve a slightly higher defense efficacy than using a randomly selected clean base set. 
We conjecture that \AlgName is optimized to select more consistent and robust information (as those selected samples have withstood randomized sample-dilution and still end up at the high-value end). 
Thus \AlgName-selected data are naturally more robust to noises and harder to be misclassified when patched with weak noise. This in turn helps both methods synthesize more accurate triggers that can be used for backdoor removal. 
Visual examples of the selection results over different datasets are shown in Appendix, Figure \ref{fig:example_cifar}; however, it is hard to conclude the sample consistency directly from visual observations.
% \ruoxi{this statement needs to be further substantiated}
We further compare the synthesized triggers of using NC equipped with the two different base sets (randomly selected clean and the \AlgName selected). 
We evaluate the averaging loss of clean samples patching with the synthesized triggers towards the target label. A much smaller loss indicates a more accurate trigger synthesis.
Averaging from five different executions, we find the triggers synthesized using \AlgName base sets achieved an averaging loss of 0.2774. The loss of trigger synthesized using randomly selected clean base sets is only 0.3321, which confirms our conjecture.

\begin{figure}[t!]
  \centering
   \vspace{-1.5em}
  \includegraphics[width=\linewidth]{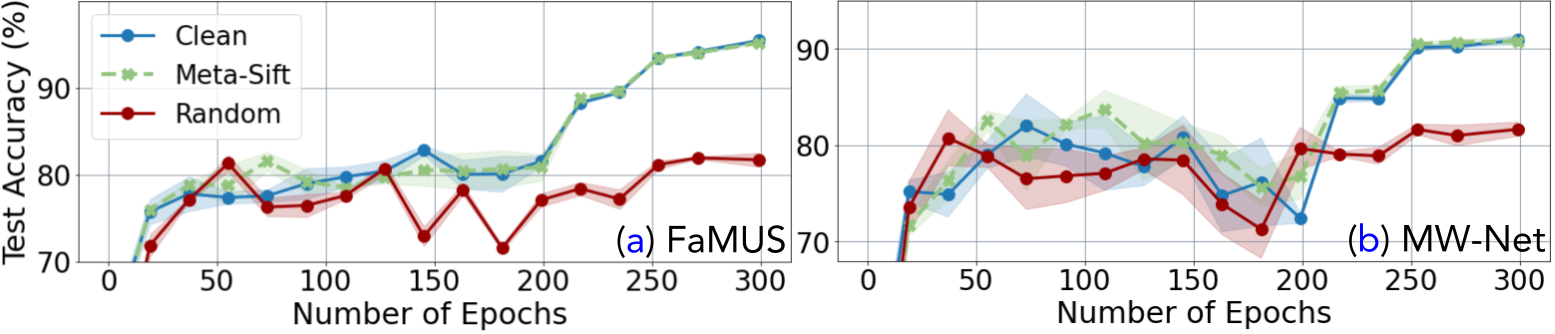}
  \vspace{-2em}
% \vspace{-1.5em}
  \caption{ACCs ($\uparrow$) vs. the number of epochs of FaMUS and MW-Net on learning label-noisy CIFAR-10. Shades depict standard deviation among different runs.
  }
  \label{fig:reweighting}
  \vspace{-1.5em}
\end{figure}

%\vspace{0.2em}
\noindent
\underline{\textbf{Robust Training:}}
Finally, we include the result of using \AlgName to help existing robust training methods work in situations without additional clean samples. The representative works of this line of defense include MW-Net \cite{shu2019meta} and FaMUS \cite{xu2021faster}.
% Apart from MW-Net, we also have the follow-up work, FaMUS \cite{xu2021faster}. 
As shown in Figure \ref{fig:reweighting}, both methods utilizing \AlgName-generated base sets can achieve a similar level of effectiveness to using an additional clean validation set.

Note that, in the original paper of FaMUS, they also introduced a training strategy that does not need access to a clean base set. 
% However, that requires training two parallel networks and working in a cross-pseudo-labeling manner
However, it requires training two paralleled neural networks and working in a manner that assigns pseudo labels for each other throughout the training procedure, which
we find introduced 7$\times$ more GPU computational overhead than the setting with a clean base set access. By using \AlgName, we get away with the additional computational overhead while still maintaining good performance in situations without access to the additional clean base set.

\vspace{-1em}
\subsection{Adaptive Attack Analysis}
\vspace{-.5em}
\label{sec:advatk}

% To assess the robustness of \AlgName against adaptive attacks, we consider a white-box scenario where the attacker has complete knowledge of \AlgName and the access to the dataset the attacker wishes to provide to the defender, $D$.
% In this context, the attacker seeks to update $D$ in a way when supplying the updated $\tilde{D}$ to the defender, the base set selected by \AlgName will contain poisoned samples.
To evaluate the robustness of \AlgName against adaptive attacks, we consider a white-box scenario with the attacker having full knowledge of \AlgName and the access to the entire training data,
% dataset the attacker wishes to provide, 
$D$. 
The evaluation settings will follow the settings on the CIFAR-10 with PreActResNet-18 listed in Table \ref{tab:datasets}. 
With $D$, the defender will follow the workflow of \AlgName to train $m$ Sifters (parameterized by $m$ pairs of
% 5 different sets of 
$(\theta^*,\psi^*)$) and use them to acquire the base set of size $1000$.
% from $D$.
An adaptive attacker seeks to update $D$ in a way that, upon providing the updated dataset, denoted as $\tilde{D}$, to the defender, the base set selected by \AlgName will contain poisons.
%
% Detailed adaptive designs of each attack are deferred to Appendix \ref{sec:adpatksetting}.

% It remains to be seen if the high precision sifting performance can hold if the attacker knows the details of $\AlgName$. To evaluate the robustness against adaptive attacks, we assume a white-box scenario where the attacker has full access to the \AlgName details and the trained parameters of the sifters. The attacker wishes to manipulate the poisoned samples so that the manipulated poisons can be selected by \AlgName, thus infiltrating the base set. 

\noindent
\underline{\textbf{Poisoning the Majority:}}
An intuitive but less practical adaptive attack is to poison the majority of $D$,
% make the poisons in $D$ the majority by introducing higher poison ratios, 
thereby violating the assumption stated in Section \ref{sec:probform} (the number of poisons in each class is less than 50\%). As this attack has limited practicality, we defer its analysis to Appendix \ref{sec:addres}.
% \underline{\textbf{Adaptive Attack via High Poison Ratios:}}
% One way to adaptively attack the system is to break our basic assumption that poisoned samples within each class are the minority, i.e., the poison ratio is smaller than 50\%. Due to the page limit, we defer the result of \AlgName in settings with a poison ratio greater than 50\% in Table \ref{tab:highestpoi}, Appendix \ref{sec:addres}. The poisoned samples start infiltrating the base set at this point, yet the corresponding attack setting becomes less practical.

\noindent
\underline{\textbf{Adversarial Noise using a Clean Model:}}
Given the access to the whole dataset, $D$, the attacker knows which portion of $D$ is clean and which will be poisoned when presenting to the defender. Another intuitive but more practical idea is to perturb poisoned samples generated by existing attacks in ways that cause 
% disguise each poisoned sample (poisoned with existing attacks) by adding adversarial noise that allows these samples to 
minimal loss on the classifier $f(\cdot|\theta_{\text{cln}})$ trained using the clean portion of data: 
\vspace{-1em}
\begin{equation}
\delta^* = \argmin_{\delta} 
L\left(f(x+\delta|\theta_{\text{cln}}), y \right),
\label{eqn:adp_atk_1}
\vspace{-1em}
\end{equation}
where $L$ denotes the loss function. 
% \ruoxi{maybe just use $L$ to stick with our prior notation for classification loss. Sure!} 
$x$ and $y$ are the feature and the label of a poisoned sample $z_{poi} = (x,y)$.
We optimize the objective using SGD till convergence and find that the resulting perturbation can successfully minimize the loss without introducing large visual artifacts.
% Since the formulation is unbounded, in practice, we utilize stochastic gradient descent for only a limited number of iterations to compute $\delta$ for a given sample.
To obtain $\tilde{D}$, we compute $\delta^*$ for each $z_{poi}$, and update $z_{poi} = (x,y)$ with $\tilde{z}_{poi}= (x+\delta^*,y)$.
 We evaluate \AlgName on three poisoned datasets that updated with this adaptive design, each of which was originally poisoned using a representative attack from one of three categories: Targeted Label Flipping, Narcissus, and BadNets. 
 The details of the implementation and $\tilde{D}$'s attacking performances are provided in Appendix \ref{sec:adpatksetting} and \ref{sec:addres}. Our results show that the adaptively perturbed poisons can indeed achieve higher weights, compared to the non-adaptive setting; yet, the 1000 samples with top weights are still pure clean, and thus \AlgName achieves a 0\% NCR at a selection budget of 1000 (i.e., a size that empowers downstream defenses, Section~\ref{sec:down_defs}).
 % \ruoxi{add some insight}

\noindent
\underline{\textbf{Adversarial Noise using Sifters:}}
% Similar to the above design, a
The third way to adaptively attack \AlgName is to disguise each poisoned sample by adding adversarial noise that allows these samples to obtain a high average score when passing through $m$ Sifters (parameterized by $m$ pairs of $(\theta^*,\psi^*)$ obtained from $D$):
\vspace{-.9em}
\begin{equation}
\delta^* = \argmax_{\delta} \frac{1}{m}\sum_{s=1}^{m}
\mathcal{S} \bigg(L\left(f(x+\delta|\theta_s^*), y \right); \psi_s^*
\bigg),
\label{eqn:adp_atk_2}
\vspace{-.9em}
\end{equation}
where $\theta_s^*$ and $\psi_s^*$ denotes the parameters of the Sifter indexed by $s$. We employ stochastic gradient ascent to solve the above optimization till convergence and again found that the optimized perturbation does not introduce much visual change.
% \yi{limited round}
% As the formulation is unbounded, we employ stochastic gradient ascent for only a limited number of iterations in practice to compute $\delta$ for a given sample.
To obtain $\tilde{D}$, we compute $\delta^*$ for each $z_{poi}$, and subsequently update $z_{poi}$ with $\tilde{z}_{poi}= (x+\delta^*,y)$. We evaluate \AlgName with the same three attack settings as above, and the details of the implementation and $\tilde{D}$'s attacking performances are provided in Appendix \ref{sec:adpatksetting} and \ref{sec:addres}.
From our results, we find the sifting performance over $\tilde{D}$ at the selection budget of 1000 remains 0\% NCR, although each $\tilde{z}_{poi}$ can obtain a high score using the old Sifters parameterized by $(\theta^*,\psi^*)$ obtained from $D$.

% \noindent
% \rev{
% \textbf{Remark:}
% The above three evaluations on adaptive designs suggest the difficulty of adaptively attacking \AlgName. We attribute the robustness of \AlgName to adaptive attacks to three factors: (1) Our defense is based on a simple and robust insight that attacks, regardless their generation mechanisms and whether they are adaptive or not, introduce some perturbations that eventually lead to a distributional shift from the clean samples. The shifted samples can be detected when a combinatorial splitting process is applied as done in \AlgName. (2) \AlgName is by design a bi-level optimization process, and thus it is difficult to synthesize ``optimal'' attacks which require optimizing through the bi-level process. (3) \AlgName is so far designed to select a relatively small size of clean set as existing defenses only requires a small base set. Thus, a successful adaptive attack would need to alter the samples in ways to achieve even higher weights than majority of clean samples in order to be selected. How to design such an attack is an open question.
% \noindent\fcolorbox{deepred}{mildyellow}{\begin{minipage}{0.97\columnwidth}
\noindent
\underline{\textbf{Remark:}}
The above analysis suggests the difficulty of adaptively attacking our method. We attribute the robustness to adaptive attacks to three factors: (\textbf{1}) \AlgName is based on a robust insight that attacks, regardless of their mechanisms and whether they are adaptive or not, introduce some perturbations that eventually lead to a distributional shift from clean samples. A subset of clean samples, as long as poisons remain the minority, can be found with a combinatorial splitting process. (\textbf{2}) \AlgName is, by design, a bilevel optimization; thus, it is difficult to synthesize ``optimal'' attacks that require optimizing through a multi-level optimization. (\textbf{3}) \AlgName is designed to select a relatively small set, as existing defenses only require a small base set. Thus, a successful adaptive attack would need to alter the poisons to achieve higher weights than the majority of clean samples to be selected. \emph{How to design such an attack remains an open question.}

\vspace{-1em}
\section{Conclusion and Outlook}
% \vspace{-.2em}
\vspace{-.5em}
This work presents the first focused study of the data sifting problem, aimed at sifting out a clean subset from a dataset potentially contaminated by unknown poisoning attacks. This problem is of critical importance to successfully implement and apply existing defenses against poisoning attacks, as most of them require a clean base set to initiate the defense mechanisms. 
We first show that the performance of the popular and many state-of-the-art defenses is sensitive to the precision of clean data selection. Our study of the existing automated methods and human inspection shows that both cannot reach the precision required to achieve effective defenses. We further propose \AlgName as a first solution to the data sifting problem. 
\AlgName is based on an intuitive and reliable insight that regardless of how poisoned samples are generated, when one trains on only the clean data, the prediction loss of the trained model on these poisoned examples is large. 
% Based on this insight, we propose to find a clean subset by optimizing the split of the contaminated dataset such that training on one split and testing on the other leads to the largest loss. 
% We introduce a series of techniques to make the computation efficient and effective. 
Our evaluation shows that \AlgName is faster than existing automated methods by significant orders of magnitude while achieving much higher sifting precision. In particular, training \AlgName only requires going through the dataset for two epochs, and using \AlgName to sift data can be done in minutes. 
% \AlgName can reliably sift out a 100\% clean subset on four benchmark datasets under twelve different attacks. 
Plugging the sifted samples into existing defenses achieves comparable or even better performance than using randomly selected clean data.

% We first show that the performance of the popular and many state-of-the-art defenses are sensitive to precision of clean data selection. Studies of existing automated methods and human inspection show that both of them cannot reach the precision required to achieve effective defenses. We further propose \AlgName as a solution to the data sifting problem. \AlgName is based on a novel insight that regardless of how poisoned samples are generated, when one trains on the clean data, the prediction loss of the trained model on these poisoned examples is large. Based on this insight, we propose to find a clean subset by optimizing the split of the contaminated dataset such that training on one split and testing on the other leads to the largest loss. We introduce a series of techniques to make the split optimization efficient and effective. Our evaluation shows that \AlgName can reliably sift out a 100\% clean subset on four benchmark datasets under twelve different kinds of data poisoning attacks. Plugging the sifted samples into existing defenses achieves comparable or even better performance than plugging in samples randomly sampled from the clean distribution. Furthermore, \AlgName is faster than existing automated methods by large orders of magnitude while achieving much higher sifting precision. In particular, training \AlgName only requires going through the dataset for two epochs and using \AlgName to sift data can be done in minutes. 
% \ruoxi{double check the claims on computational efficiency}

% \vspace{0.2em}
\noindent
\textbf{Limitations \& Outlook.}
Despite the efficacy and time-efficiency, \AlgName
can cause high memory overhead due to the use of multiple perturbed datasets and Sifters.
% induces high memory overhead due to the use of multiple perturbed datasets and multiple Sifters. 
To support memory-constrained applications, further reduction of overhead is necessary. Additionally, there is a trade-off between \AlgName's precision and the size of the set to be sifted, as shown in Figure \ref{fig:efficacy}, \AlgName can confidently sift out approximately 5\% of the size of the dataset to be clean across different settings, datasets, and attacks, which is sufficient for most existing defenses (as discussed in Section \ref{sec:down_defs}). However, the size is insufficient for any model trained directly on the sifted samples to perform well. Therefore, expanding the scale of \AlgName's performance and enabling direct training is future work that needs to be addressed.
% How can we further reduce overhead in order to support potential memory-constrained applications? Furthermore, there is a trade-off between \AlgName's precision and the size of the set to be sifted (Figure \ref{fig:efficacy}). According to our evaluation, we find \AlgName can reliably sift out around 5\% pure-clean samples confidently regarding different settings/datasets/attacks, which is enough to support most of the existing defenses (Section \ref{sec:down_defs}). However, the size is far from supporting any model directly trained over the sifted samples to obtain good performance. 
% We defer the further steps on expanding \AlgName's performance to a larger scale and enable direct training for future work. 

Additionally, there are a number of interesting venues for future exploration. For example, is it possible to design successful and practical adaptive attacks against \AlgName that enabling the infiltration of poisoned samples past our sifting? 
Also, it might be interesting to explore other methods for resolving the proposed sifting or splitting problem and generalize the idea to more settings beyond image classification.
% e.g., exploring possible solutions from contrastive learning \cite{grill2020bootstrap} to better separate the two splits mentioned in our formulation. 
% While \AlgName can produce a completely clean set at a scale that is enough to support effective defense, how can we improve the precision at a larger scale so as to enable direct training on the sifted samples? 

% \vspace{-1em}
\section*{Acknowledgement}
\vspace{-.5em}
% \todo{need to check}
This work is partially supported by Sony AI. RJ and the ReDS lab appreciate the support of the Amazon - Virginia Tech Initiative for Efficient and Robust Machine Learning and the Cisco Award. YZ is supported by the Amazon Fellowship.
\vspace{-2em}

\bibliographystyle{IEEEtran}
\bibliography{bibtex}

\begin{thebibliography}{10}
\providecommand{\url}[1]{#1}
\csname url@samestyle\endcsname
\providecommand{\newblock}{\relax}
\providecommand{\bibinfo}[2]{#2}
\providecommand{\BIBentrySTDinterwordspacing}{\spaceskip=0pt\relax}
\providecommand{\BIBentryALTinterwordstretchfactor}{4}
\providecommand{\BIBentryALTinterwordspacing}{\spaceskip=\fontdimen2\font plus
\BIBentryALTinterwordstretchfactor\fontdimen3\font minus
  \fontdimen4\font\relax}
\providecommand{\BIBforeignlanguage}[2]{{%
\expandafter\ifx\csname l@#1\endcsname\relax
\typeout{** WARNING: IEEEtran.bst: No hyphenation pattern has been}%
\typeout{** loaded for the language `#1'. Using the pattern for}%
\typeout{** the default language instead.}%
\else
\language=\csname l@#1\endcsname
\fi
#2}}
\providecommand{\BIBdecl}{\relax}
\BIBdecl

\bibitem{radford2021learning}
A.~Radford, J.~W. Kim, C.~Hallacy, A.~Ramesh, G.~Goh, S.~Agarwal, G.~Sastry,
  A.~Askell, P.~Mishkin, J.~Clark \emph{et~al.}, ``Learning transferable visual
  models from natural language supervision,'' in \emph{ICML}, 2021, pp.
  8748--8763.

\bibitem{kumar2020adversarial}
R.~S.~S. Kumar, M.~Nystr{\"o}m, J.~Lambert, A.~Marshall, M.~Goertzel,
  A.~Comissoneru, M.~Swann, and S.~Xia, ``Adversarial machine learning-industry
  perspectives,'' in \emph{2020 IEEE SPW}, 2020, pp. 69--75.

\bibitem{tolpegin2020data}
V.~Tolpegin, S.~Truex, M.~E. Gursoy, and L.~Liu, ``Data poisoning attacks
  against federated learning systems,'' in \emph{ESORICS}, 2020, pp. 480--501.

\bibitem{ren2018learning}
M.~Ren, W.~Zeng, B.~Yang, and R.~Urtasun, ``Learning to reweight examples for
  robust deep learning,'' in \emph{ICML}, 2018, pp. 4334--4343.

\bibitem{shafahi2018poison}
A.~Shafahi, W.~R. Huang, M.~Najibi, O.~Suciu, C.~Studer, T.~Dumitras, and
  T.~Goldstein, ``Poison frogs! targeted clean-label poisoning attacks on
  neural networks,'' in \emph{NeurIPS}, vol.~31, 2018.

\bibitem{turner2019label}
A.~Turner, D.~Tsipras, and A.~Madry, ``Label-consistent backdoor attacks,''
  \emph{arXiv:1912.02771}, 2019.

\bibitem{zeng2022narcissus}
Y.~Zeng, M.~Pan, H.~A. Just, L.~Lyu, M.~Qiu, and R.~Jia, ``Narcissus: A
  practical clean-label backdoor attack with limited information,''
  \emph{arXiv:2204.05255}, 2022.

\bibitem{gu2017badnets}
T.~Gu, K.~Liu, B.~Dolan-Gavitt, and S.~Garg, ``Badnets: Evaluating backdooring
  attacks on deep neural networks,'' \emph{IEEE Access}, vol.~7, pp.
  47\,230--47\,244, 2019.

\bibitem{chen2017targeted}
X.~Chen, C.~Liu, B.~Li, K.~Lu, and D.~Song, ``Targeted backdoor attacks on deep
  learning systems using data poisoning,'' \emph{arXiv:1712.05526}, 2017.

\bibitem{nguyen2021wanet}
T.~A. Nguyen and A.~T. Tran, ``Wanet-imperceptible warping-based backdoor
  attack,'' in \emph{ICLR}, 2020.

\bibitem{zeng2021rethinking}
Y.~Zeng, W.~Park, Z.~M. Mao, and R.~Jia, ``Rethinking the backdoor attacks'
  triggers: A frequency perspective,'' in \emph{ICCV}, 2021.

\bibitem{xu2021detecting}
X.~Xu, Q.~Wang, H.~Li, N.~Borisov, C.~A. Gunter, and B.~Li, ``Detecting ai
  trojans using meta neural analysis,'' in \emph{2021 IEEE S\&P}, 2021, pp.
  103--120.

\bibitem{wang2019neural}
B.~Wang, Y.~Yao, S.~Shan, H.~Li, B.~Viswanath, H.~Zheng, and B.~Y. Zhao,
  ``Neural cleanse: Identifying and mitigating backdoor attacks in neural
  networks,'' in \emph{2019 IEEE S\&P}, 2019, pp. 707--723.

\bibitem{zeng2021adversarial}
Y.~Zeng, S.~Chen, W.~Park, Z.~Mao, M.~Jin, and R.~Jia, ``Adversarial unlearning
  of backdoors via implicit hypergradient,'' in \emph{ICLR}, 2022.

\bibitem{xu2021faster}
Y.~Xu, L.~Zhu, L.~Jiang, and Y.~Yang, ``Faster meta update strategy for
  noise-robust deep learning,'' in \emph{CVPR}, 2021, pp. 144--153.

\bibitem{guo2019tabor}
W.~Guo, L.~Wang, X.~Xing, M.~Du, and D.~Song, ``Tabor: A highly accurate
  approach to inspecting and restoring trojan backdoors in ai systems,''
  \emph{arXiv:1908.01763}, 2019.

\bibitem{liu2018fine}
K.~Liu, B.~Dolan-Gavitt, and S.~Garg, ``Fine-pruning: Defending against
  backdooring attacks on deep neural networks,'' in \emph{RAID}, 2018, pp.
  273--294.

\bibitem{xiang2020revealing}
Z.~Xiang, D.~J. Miller, H.~Wang, and G.~Kesidis, ``Revealing perceptible
  backdoors in dnns, without the training set, via the maximum achievable
  misclassification fraction statistic,'' in \emph{MLSP}, 2020, pp. 1--6.

\bibitem{shu2019meta}
J.~Shu, Q.~Xie, L.~Yi, Q.~Zhao, S.~Zhou, Z.~Xu, and D.~Meng, ``Meta-weight-net:
  Learning an explicit mapping for sample weighting,'' in \emph{NeurIPS},
  vol.~32, 2019.

\bibitem{xiang2022post}
Z.~Xiang, D.~J. Miller, and G.~Kesidis, ``Post-training detection of backdoor
  attacks for two-class and multi-attack scenarios,'' \emph{arXiv:2201.08474},
  2022.

\bibitem{wang2020practical}
R.~Wang, G.~Zhang, S.~Liu, P.-Y. Chen, J.~Xiong, and M.~Wang, ``Practical
  detection of trojan neural networks: Data-limited and data-free cases,'' in
  \emph{ECCV}, 2020, pp. 222--238.

\bibitem{gao2019strip}
Y.~Gao, C.~Xu, D.~Wang, S.~Chen, D.~C. Ranasinghe, and S.~Nepal, ``Strip: A
  defence against trojan attacks on deep neural networks,'' in \emph{ACSAC
  '19}, 2019, p. 113–125.

\bibitem{chen2019deepinspect}
H.~Chen, C.~Fu, J.~Zhao, and F.~Koushanfar, ``Deepinspect: A black-box trojan
  detection and mitigation framework for deep neural networks.'' in
  \emph{IJCAI}, vol.~2, no.~5, 2019, p.~8.

\bibitem{fredrikson2015model}
M.~Fredrikson, S.~Jha, and T.~Ristenpart, ``Model inversion attacks that
  exploit confidence information and basic countermeasures,'' in \emph{SIGSAC},
  2015, pp. 1322--1333.

\bibitem{tran2018spectral}
B.~Tran, J.~Li, and A.~Madry, ``Spectral signatures in backdoor attacks,'' in
  \emph{NeuIPS}, 2018, pp. 8000--8010.

\bibitem{li2021anti}
Y.~Li, X.~Lyu, N.~Koren, L.~Lyu, B.~Li, and X.~Ma, ``Anti-backdoor learning:
  Training clean models on poisoned data,'' in \emph{NeurIPS}, vol.~34, 2021.

\bibitem{koh2017understanding}
P.~W. Koh and P.~Liang, ``Understanding black-box predictions via influence
  functions,'' in \emph{ICML}, 2017, pp. 1885--1894.

\bibitem{shao2021right}
X.~Shao, A.~Skryagin, W.~Stammer, P.~Schramowski, and K.~Kersting, ``Right for
  better reasons: Training differentiable models by constraining their
  influence functions,'' in \emph{AAAI}, vol.~35, no.~11, 2021, pp. 9533--9540.

\bibitem{kong2021resolving}
S.~Kong, Y.~Shen, and L.~Huang, ``Resolving training biases via influence-based
  data relabeling,'' in \emph{ICLR}, 2021.

\bibitem{goldblum2022dataset}
M.~Goldblum, D.~Tsipras, C.~Xie, X.~Chen, A.~Schwarzschild, D.~Song, A.~Madry,
  B.~Li, and T.~Goldstein, ``Dataset security for machine learning: Data
  poisoning, backdoor attacks, and defenses,'' \emph{IEEE TPAMI}, 2022.

\bibitem{carlini2021poisoning}
N.~Carlini and A.~Terzis, ``Poisoning and backdooring contrastive learning,''
  in \emph{ICLR}, 2021.

\bibitem{pan2023asset}
M.~Pan, Y.~Zeng, L.~Lyu, X.~Lin, and R.~Jia, ``Asset: Robust backdoor data
  detection across a multiplicity of deep learning paradigms,''
  \emph{arXiv:2302.11408}, 2023.

\bibitem{steinhardt2018robust}
J.~Steinhardt, \emph{Robust learning: Information theory and algorithms}, 2018.

\bibitem{agarap2018deep}
A.~F. Agarap, ``Deep learning using rectified linear units (relu),''
  \emph{arXiv:1803.08375}, 2018.

\bibitem{he2015delving}
K.~He, X.~Zhang, S.~Ren, and J.~Sun, ``Delving deep into rectifiers: Surpassing
  human-level performance on imagenet classification,'' in \emph{ICCV}, 2015,
  pp. 1026--1034.

\bibitem{jang2016categorical}
E.~Jang, S.~Gu, and B.~Poole, ``Categorical reparameterization with
  gumbel-softmax,'' \emph{arXiv:1611.01144}, 2016.

\bibitem{paszke2019pytorch}
A.~Paszke, S.~Gross, F.~Massa, A.~Lerer, J.~Bradbury, G.~Chanan, T.~Killeen,
  Z.~Lin, N.~Gimelshein, L.~Antiga \emph{et~al.}, ``Pytorch: An imperative
  style, high-performance deep learning library,'' \emph{NeurIPS}, vol.~32, pp.
  8026--8037, 2019.

\bibitem{liu2021investigating}
R.~Liu, J.~Gao, J.~Zhang, D.~Meng, and Z.~Lin, ``Investigating bi-level
  optimization for learning and vision from a unified perspective: A survey and
  beyond,'' \emph{arXiv:2101.11517}, 2021.

\bibitem{grazzi2020iteration}
R.~Grazzi, L.~Franceschi, M.~Pontil, and S.~Salzo, ``On the iteration
  complexity of hypergradient computation,'' in \emph{ICML}, 2020, pp.
  3748--3758.

\bibitem{li2020rethinking}
Y.~Li, T.~Zhai, B.~Wu, Y.~Jiang, Z.~Li, and S.~Xia, ``Rethinking the trigger of
  backdoor attack,'' \emph{arXiv:2004.04692}, 2020.

\bibitem{qiu2021deepsweep}
H.~Qiu, Y.~Zeng, S.~Guo, T.~Zhang, M.~Qiu, and B.~Thuraisingham, ``Deepsweep:
  An evaluation framework for mitigating dnn backdoor attacks using data
  augmentation,'' in \emph{ASIACCS}, 2021, pp. 363--377.

\bibitem{krizhevsky2009learning}
A.~Krizhevsky, G.~Hinton \emph{et~al.}, ``Learning multiple layers of features
  from tiny images,'' 2009.

\bibitem{stallkamp2012man}
J.~Stallkamp, M.~Schlipsing, J.~Salmen, and C.~Igel, ``Man vs. computer:
  Benchmarking machine learning algorithms for traffic sign recognition,''
  \emph{Neural networks}, vol.~32, pp. 323--332, 2012.

\bibitem{kumar2009attribute}
N.~Kumar, A.~C. Berg, P.~N. Belhumeur, and S.~K. Nayar, ``Attribute and simile
  classifiers for face verification,'' in \emph{ICCV}, 2009, pp. 365--372.

\bibitem{russakovsky2015imagenet}
O.~Russakovsky, J.~Deng, H.~Su, J.~Krause, S.~Satheesh, S.~Ma, Z.~Huang,
  A.~Karpathy, A.~Khosla, M.~Bernstein \emph{et~al.}, ``Imagenet large scale
  visual recognition challenge,'' \emph{IJCV}, vol. 115, no.~3, pp. 211--252,
  2015.

\bibitem{ruder2016overview}
S.~Ruder, ``An overview of gradient descent optimization algorithms,''
  \emph{arXiv:1609.04747}, 2016.

\bibitem{kingma2014adam}
D.~P. Kingma and J.~Ba, ``Adam: {A} method for stochastic optimization,'' in
  \emph{ICLR}, Y.~Bengio and Y.~LeCun, Eds., 2015.

\bibitem{liu2019variance}
L.~Liu, H.~Jiang, P.~He, W.~Chen, X.~Liu, J.~Gao, and J.~Han, ``On the variance
  of the adaptive learning rate and beyond,'' in \emph{ICLR}, 2019.

\bibitem{nguyen2020input}
T.~A. Nguyen and A.~Tran, ``Input-aware dynamic backdoor attack,''
  \emph{NeurIPS}, vol.~33, pp. 3454--3464, 2020.

\bibitem{liu2019abs}
Y.~Liu, W.-C. Lee, G.~Tao, S.~Ma, Y.~Aafer, and X.~Zhang, ``Abs: Scanning
  neural networks for back-doors by artificial brain stimulation,'' in
  \emph{SIGSAC}, 2019, pp. 1265--1282.

\bibitem{liu2017trojaning}
Y.~Liu, S.~Ma, Y.~Aafer, W.-C. Lee, J.~Zhai, W.~Wang, and X.~Zhang, ``Trojaning
  attack on neural networks,'' in \emph{NDSS}, 2018.

\bibitem{li2020invisible}
S.~Li, M.~Xue, B.~Zhao, H.~Zhu, and X.~Zhang, ``Invisible backdoor attacks on
  deep neural networks via steganography and regularization,'' \emph{IEEE
  TDSC}, 2020.

\bibitem{saha2020hidden}
A.~Saha, A.~Subramanya, and H.~Pirsiavash, ``Hidden trigger backdoor attacks,''
  in \emph{AAAI}, vol.~34, 2020, pp. 11\,957--11\,965.

\bibitem{liu2020reflection}
Y.~Liu, X.~Ma, J.~Bailey, and F.~Lu, ``Reflection backdoor: A natural backdoor
  attack on deep neural networks,'' in \emph{ECCV, 2020}, 2020, pp. 182--199.

\bibitem{he2016deep}
K.~He, X.~Zhang, S.~Ren, and J.~Sun, ``Deep residual learning for image
  recognition,'' in \emph{CVPR}, 2016, pp. 770--778.

\end{thebibliography}

% \newpage
% \begin{equation}
% \text{placeholder}
% \end{equation}

\newpage

\vspace{-1em}
\section{Appendix}

\vspace{-1em}
\subsection{Detailed Attack Settings in Section \ref{sec:exist?}}
\vspace{-0.5em}
\label{sec:detailed_atk_0}
We detail the 4 attacks in evaluating existing methods:
\begin{packeditemize}
\vspace{-0.5em}
    \item For Label-only attacks, we use Targeted Label-Flipping \cite{tolpegin2020data}. We randomly selected 1000 ``cat'' samples and mislabeled them into ``dog'' samples to launch the attack. Compared to the standard Random Label-Flipping, which is considered in most label-noise defenses, our study finds that Targeted Label-Flipping is much harder to be mitigated. 
    % For this study, we randomly selected 1000 cat samples and mislabeled them into dog samples to launch the attack. 
    
    \item For Feature-only attacks, we consider two representative works, namely, Narcissus\footnote{\url{https://github.com/ruoxi-jia-group/Narcissus}} \cite{zeng2022narcissus}, a clean-label backdoor attack, and Poison Frogs\footnote{\url{https://github.com/LostOxygen/poison_froggo}} \cite{shafahi2018poison}, a targeted poisoning attack. For both attacks, we poison 10\% of the dog-class samples. 
    
    \item For the Label-Feature attacks, we consider BadNets\footnote{\url{https://github.com/verazuo/badnets-pytorch}} \cite{gu2017badnets}. Following the setting in the original paper, we manipulate 5\% of the CIFAR-10 by patching the trigger and changing the label to the target class.
    \vspace{-0.5em}
\end{packeditemize}

\vspace{-1.5em}
\subsection{Detailed Attack Settings in Section \ref{sec:human?}}
\vspace{-0.5em}
\label{sec:detailed_atk}
We detail the 16 attacks in human studies as follows:
\begin{packeditemize}
\vspace{-0.5em}
    % \item For Label-only attacks, as random Label-Flipping and targeted Label-Flipping are of no difference when presenting to human labelers, we randomly shuffle the poisoned sample's label to a different label than their original ones (also known as label noise \cite{xu2021faster}).
    \item For Label-only attacks, Random and Targeted Label-Flipping leads to mislabeled samples. However, since humans can only classify at a sample-wise level, these attacks present the same level of difficulty to humans. Hence, we only evaluate Random Label-Flipping, where we randomly shuffle a poisoned sample's label to a different class.
    % human labelers are unconcerned about the difference between random vs targeted Label-Flipping. Hence we randomly shuffle the labels for select samples (according to the poison ratio) to a label different from the original one (label-noise \cite{xu2021faster}).
    \item For Feature-only attacks, we consider three representative works: Label-Consistent attack \footnote{\url{https://github.com/MadryLab/label-consistent-backdoor-code}}\cite{turner2019label}, Hidden Trigger backdoor\footnote{\url{https://github.com/UMBCvision/Hidden-Trigger-Backdoor-Attacks}} \cite{saha2020hidden}, and the Narcissus attack \cite{zeng2022narcissus}.
    % Among these three attacks, Narcissus achieves the best attack performance. Label-Consistent and Hidden Trigger Backdoor are popular baselines.
    % The Label-Consistent attack adds strong adversarial noise or GAN manipulations to the poisoned samples to make the original semantic features harder to be learned than the backdoor trigger\footnote{\url{https://github.com/MadryLab/label-consistent-backdoor-code}}. The Hidden Trigger backdoor poisons the samples using a technique similar to Poison Frogs \cite{shafahi2018poison} that collides the poisoned samples and the target class at a feature space level. For the Narcissus attack, the original work has presented different adaptive designs of triggers to achieve better stealthiness\footnote{\url{https://github.com/UMBCvision/Hidden-Trigger-Backdoor-Attacks}}.
    We include three different designs of triggers from the original paper: vanilla Narcissus with $l_{\infty}=16/255$; watermarked Narcissus, which used a mask to constrain the trigger within a pre-defined shape; and Narcissus-Smooth trigger, which is constrained to be low-frequency.
    % optimized within a low-pass filter following the original implementations.
    
    \item  For Label-Feature attacks, we study a comprehensive list of dirty-label backdoor attacks with different trigger designs: 1) Patch-based triggers that adopt an arbitrary trigger pattern in small regions: BadNets, and Masked random noise (Masked-Noise) \cite{xu2021detecting}, Watermark \cite{qiu2021deepsweep}; 2) Blending-based triggers that blend the trigger into the original image for better stealthiness: Blended random noise (Blended-Noise) \cite{chen2017targeted} and Refool\footnote{\url{https://github.com/DreamtaleCore/Refool}} \cite{liu2020reflection}; 3) Attack based on elastic transformation: WaNet\footnote{\url{https://github.com/THUYimingLi/BackdoorBox}} \cite{nguyen2021wanet}; 4) Optimized triggers within norm-balls: $l_{0}$ and $l_{2}$ invisible triggers \cite{li2020invisible}; 5) Optimized low-frequency triggers: the smooth trigger\footnote{\url{https://github.com/YiZeng623/frequency-backdoor}} \cite{zeng2021rethinking}; 6) Input-specific triggers with trigger patterns customized to individual inputs: the input aware backdoor (IAB)\footnote{\url{https://github.com/THUYimingLi/BackdoorBox}} \cite{nguyen2020input}.
\end{packeditemize}

\begin{table*}[t!]
\centering
% \vspace{-2em}
\resizebox{0.9\textwidth}{!}{
\begin{tabular}{l cc|ccc|ccccccc}
\hline
\multicolumn{1}{c|}{}&
  \multicolumn{2}{c|}{\textbf{Label-only}} &
  \multicolumn{3}{c|}{\textbf{Feature-only}} &
  \multicolumn{6}{c}{\textbf{Label-Feature}} &
   \\ \cline{2-13}
\multicolumn{1}{c|}{\multirow{-2}{*}{}} &
  \begin{tabular}[c]{@{}c@{}}Targeted Label-\\ Flipping\cite{tolpegin2020data}\end{tabular} &
  \begin{tabular}[c]{@{}c@{}}Random Label-\\ Flipping\cite{ren2018learning}\end{tabular} &
  \begin{tabular}[c]{@{}c@{}}Clean-\\ Label\cite{turner2019label}\end{tabular} &
  \begin{tabular}[c]{@{}c@{}}Narcissus\\ Backdoor\cite{zeng2022narcissus}\end{tabular} &
  \begin{tabular}[c]{@{}c@{}}Poison\\ Frog\cite{shafahi2018poison}\end{tabular} &
  \begin{tabular}[c]{@{}c@{}}BadNets\\ One-Tar\cite{gu2017badnets}\end{tabular} &
  \begin{tabular}[c]{@{}c@{}}Smooth\\ One-Tar\cite{zeng2021rethinking}\end{tabular} &
  \begin{tabular}[c]{@{}c@{}}IAB\\ One-Tar\cite{nguyen2020input}\end{tabular} &
  \begin{tabular}[c]{@{}c@{}}Blended\\ One-Tar\cite{chen2017targeted}\end{tabular} &
  \begin{tabular}[c]{@{}c@{}}BadNets\\ All-to-all\cite{gu2017badnets}\end{tabular} &
  \begin{tabular}[c]{@{}c@{}}Smooth\\ All-to-all\cite{zeng2021rethinking}\end{tabular} &
  \begin{tabular}[c]{@{}c@{}}Blended\\ All-to-all\cite{chen2017targeted}\end{tabular} \\
  \hline

\multicolumn{1}{c|}{\begin{tabular}[c]{@{}c@{}}Poison ratio\end{tabular}}&
  \begin{tabular}[c]{@{}c@{}}Tar: 2\% \end{tabular} &
  \begin{tabular}[c]{@{}c@{}} All: 5\%\end{tabular} &
  \begin{tabular}[c]{@{}c@{}} Tar: 1\%\end{tabular} &
  \begin{tabular}[c]{@{}c@{}} Tar: 1\%\end{tabular} &
  \begin{tabular}[c]{@{}c@{}} Tar: 1\%\end{tabular} &
  \begin{tabular}[c]{@{}c@{}} Tar: 9.1\%\end{tabular} &
  \begin{tabular}[c]{@{}c@{}} Tar: 9.1\%\end{tabular} &
  \begin{tabular}[c]{@{}c@{}} Tar: 9.1\%\end{tabular} &
  \begin{tabular}[c]{@{}c@{}} Tar: 9.1\%\end{tabular} &
  \begin{tabular}[c]{@{}c@{}} All: 5\%\end{tabular} &
  \begin{tabular}[c]{@{}c@{}} All: 5\%\end{tabular} &
  \begin{tabular}[c]{@{}c@{}} All: 5\%\end{tabular} \\ 
\hline
\multicolumn{1}{l|}{\AlgName }&
  \cellcolor[HTML]{D9EAD3}\textbf{0$\pm$0} &
  \cellcolor[HTML]{D9EAD3}\textbf{0$\pm$0} &
  \cellcolor[HTML]{D9EAD3}\textbf{0$\pm$0} &
  \cellcolor[HTML]{D9EAD3}\textbf{0$\pm$0} &
  \cellcolor[HTML]{D9EAD3}\textbf{0$\pm$0} &
  \cellcolor[HTML]{D9EAD3}\textbf{0$\pm$0} &
  \cellcolor[HTML]{D9EAD3}\textbf{0$\pm$0} &
  \cellcolor[HTML]{D9EAD3}\textbf{0$\pm$0} &
  \cellcolor[HTML]{D9EAD3}\textbf{0$\pm$0} &
  \cellcolor[HTML]{D9EAD3}\textbf{0$\pm$0} &
  \cellcolor[HTML]{D9EAD3}\textbf{0$\pm$0} &
  \cellcolor[HTML]{D9EAD3}\textbf{0$\pm$0} 
  \\ \hline
  
\multicolumn{1}{c|}{\begin{tabular}[c]{@{}c@{}}Poison ratio\end{tabular}}&
  \begin{tabular}[c]{@{}c@{}}Tar: 9\%\end{tabular} &
  \begin{tabular}[c]{@{}c@{}}All: 10\%\end{tabular} &
  \begin{tabular}[c]{@{}c@{}}Tar: 5\%\end{tabular} &
  \begin{tabular}[c]{@{}c@{}}Tar: 5\%\end{tabular} &
  \begin{tabular}[c]{@{}c@{}}Tar: 5\%\end{tabular} &
  \begin{tabular}[c]{@{}c@{}}Tar: 16.7\%\end{tabular} &
  \begin{tabular}[c]{@{}c@{}}Tar: 16.7\%\end{tabular} &
  \begin{tabular}[c]{@{}c@{}}Tar: 16.7\%\end{tabular} &
  \begin{tabular}[c]{@{}c@{}}Tar: 16.7\%\end{tabular} &
  \begin{tabular}[c]{@{}c@{}}All: 10\%\end{tabular} &
  \begin{tabular}[c]{@{}c@{}}All: 10\%\end{tabular} &
  \begin{tabular}[c]{@{}c@{}}All: 10\%\end{tabular} \\ 
\hline
\multicolumn{1}{l|}{\AlgName }&
  \cellcolor[HTML]{D9EAD3}\textbf{0$\pm$0} &
  \cellcolor[HTML]{D9EAD3}\textbf{0$\pm$0} &
  \cellcolor[HTML]{D9EAD3}\textbf{0$\pm$0} &
  \cellcolor[HTML]{D9EAD3}\textbf{0$\pm$0} &
  \cellcolor[HTML]{D9EAD3}\textbf{0$\pm$0} &
  \cellcolor[HTML]{D9EAD3}\textbf{0$\pm$0} &
  \cellcolor[HTML]{D9EAD3}\textbf{0$\pm$0} &
  \cellcolor[HTML]{D9EAD3}\textbf{0$\pm$0} &
  \cellcolor[HTML]{D9EAD3}\textbf{0$\pm$0} &
  \cellcolor[HTML]{D9EAD3}\textbf{0$\pm$0} &
  \cellcolor[HTML]{D9EAD3}\textbf{0$\pm$0} &
  \cellcolor[HTML]{D9EAD3}\textbf{0$\pm$0} 
  \\ \hline

\end{tabular}
 }
 \vspace{-0.9em}
\caption{NCR results of our method under the CIFAR-10 settings with lower poison ratios.}
\label{tab:lowpoi}
\vspace{-1em}
\end{table*}

\begin{table*}[t!]
\centering
\resizebox{0.9\textwidth}{!}{
\begin{tabular}{l cc|ccc|ccccccc}
\hline
\multicolumn{1}{c|}{}&
  \multicolumn{2}{c|}{\textbf{Label-only}} &
  \multicolumn{3}{c|}{\textbf{Feature-only}} &
  \multicolumn{6}{c}{\textbf{Label-Feature}} &
   \\ \cline{2-13}
\multicolumn{1}{c|}{\multirow{-2}{*}{}} &
  \begin{tabular}[c]{@{}c@{}}Targeted Label-\\ Flipping\cite{tolpegin2020data}\end{tabular} &
  \begin{tabular}[c]{@{}c@{}}Random Label-\\ Flipping\cite{ren2018learning}\end{tabular} &
  \begin{tabular}[c]{@{}c@{}}Clean-\\ Label\cite{turner2019label}\end{tabular} &
  \begin{tabular}[c]{@{}c@{}}Narcissus\\ Backdoor\cite{zeng2022narcissus}\end{tabular} &
  \begin{tabular}[c]{@{}c@{}}Poison\\ Frog\cite{shafahi2018poison}\end{tabular} &
  \begin{tabular}[c]{@{}c@{}}BadNets\\ One-Tar\cite{gu2017badnets}\end{tabular} &
  \begin{tabular}[c]{@{}c@{}}Smooth\\ One-Tar\cite{zeng2021rethinking}\end{tabular} &
  \begin{tabular}[c]{@{}c@{}}IAB\\ One-Tar\cite{nguyen2020input}\end{tabular} &
  \begin{tabular}[c]{@{}c@{}}Blended\\ One-Tar\cite{chen2017targeted}\end{tabular} &
  \begin{tabular}[c]{@{}c@{}}BadNets\\ All-to-all\cite{gu2017badnets}\end{tabular} &
  \begin{tabular}[c]{@{}c@{}}Smooth\\ All-to-all\cite{zeng2021rethinking}\end{tabular} &
  \begin{tabular}[c]{@{}c@{}}Blended\\ All-to-all\cite{chen2017targeted}\end{tabular} \\
  \hline

\multicolumn{1}{c|}{\begin{tabular}[c]{@{}c@{}}Poison ratio\end{tabular}}&
  \begin{tabular}[c]{@{}c@{}}Tar: 40\% \end{tabular} &
  \begin{tabular}[c]{@{}c@{}}All: 40\%\end{tabular} &
  \begin{tabular}[c]{@{}c@{}}Tar: 40\%\end{tabular} &
  \begin{tabular}[c]{@{}c@{}}Tar: 40\% \end{tabular} &
  \begin{tabular}[c]{@{}c@{}}Tar: 40\%\end{tabular} &
  \begin{tabular}[c]{@{}c@{}}Tar: 40.2\%\end{tabular} &
  \begin{tabular}[c]{@{}c@{}}Tar: 40.2\%\end{tabular} &
  \begin{tabular}[c]{@{}c@{}}Tar: 40.2\%\end{tabular} &
  \begin{tabular}[c]{@{}c@{}}Tar: 40.2\%\end{tabular} &
  \begin{tabular}[c]{@{}c@{}}All: 40\%\end{tabular} &
  \begin{tabular}[c]{@{}c@{}}All: 40\%\end{tabular} &
  \begin{tabular}[c]{@{}c@{}}All: 40\%\end{tabular} \\ 
  \hline
\hline
  
\multicolumn{1}{l|}{DCM} &
  \cellcolor[HTML]{cfe2f3}\textbf{63.0} &
  94.3 &
  37.8 &
  \cellcolor[HTML]{FFCCCC}\textbf{102.0} &
  \cellcolor[HTML]{FFCCCC}\textbf{112.2} &
  98.3 &
  \cellcolor[HTML]{cfe2f3}\textbf{27.5} &
  \cellcolor[HTML]{FFCCCC}\textbf{104.6} &
  \cellcolor[HTML]{cfe2f3}\textbf{54.6} &
  78.2 &
  \cellcolor[HTML]{cfe2f3}\textbf{27.6} &
  63.8 \\
  
\multicolumn{1}{l|}{MI-DCM$^*$} &
  71.2$\pm$10.5 &
  \cellcolor[HTML]{FFCCCC}\textbf{100.7$\pm$8.8} &
  93.5$\pm$0 &
  \cellcolor[HTML]{FFCCCC}\textbf{362.0$\pm$88.6} &
  \cellcolor[HTML]{FFCCCC}\textbf{311.0$\pm$76.9} &
  \cellcolor[HTML]{FFCCCC}\textbf{131.5$\pm$15.6} &
  \cellcolor[HTML]{FFCCCC}\textbf{168.6$\pm$20.9} &
  \cellcolor[HTML]{FFCCCC}\textbf{163.8$\pm$30.1} &
  \cellcolor[HTML]{FFCCCC}\textbf{153.6$\pm$25.5} &
  \cellcolor[HTML]{FFCCCC}\textbf{117.0$\pm$19.6} &
  \cellcolor[HTML]{FFCCCC}\textbf{153.8$\pm$16.8} &
  \cellcolor[HTML]{FFCCCC}\textbf{137.3$\pm$26.3} \\
  
\multicolumn{1}{l|}{SF-Least$^*$ }&
  97.4$\pm$1.67 &
  \cellcolor[HTML]{cfe2f3}\textbf{31.2$\pm$4.63} &
  86.3$\pm$3.45 &
  \cellcolor[HTML]{cfe2f3}\textbf{43.6$\pm$0} &
  \cellcolor[HTML]{FFCCCC}\textbf{106.3$\pm$9.88} &
  \cellcolor[HTML]{cfe2f3}\textbf{32.4$\pm$0.0} &
  28.3$\pm$1.67 &
  \cellcolor[HTML]{cfe2f3}\textbf{44.6$\pm$3.53} &
  60.3$\pm$2.71 &
  \cellcolor[HTML]{cfe2f3}\textbf{56.7$\pm$0} &
  58.9$\pm$0 &
  \cellcolor[HTML]{cfe2f3}\textbf{63.2$\pm$6.72}\\

\multicolumn{1}{l|}{Loss-Scan$^*$ }&
  \cellcolor[HTML]{FFCCCC}\textbf{173.9$\pm$16.9} &
  \cellcolor[HTML]{FFCCCC}\textbf{443.1$\pm$57.8} &
  \cellcolor[HTML]{FFCCCC}\textbf{163.6$\pm$23.4} &
  \cellcolor[HTML]{FFCCCC}\textbf{188.0$\pm$41.1} &
  \cellcolor[HTML]{FFCCCC}\textbf{134.0$\pm$23.8} &
  83.6$\pm$16.3 &
  \cellcolor[HTML]{FFCCCC}\textbf{116.7$\pm$22.9} &
  \cellcolor[HTML]{FFCCCC}\textbf{106.9$\pm$31.6} &
  \cellcolor[HTML]{FFCCCC}\textbf{142.6$\pm$13.8} &
  96.7$\pm$19.6 &
  85.8$\pm$9.8 &
  91.6$\pm$16.6 \\
  
\multicolumn{1}{l|}{Self-IF$^*$ }&
  \cellcolor[HTML]{FFCCCC}\textbf{135.9$\pm$23.6} &
  \cellcolor[HTML]{FFCCCC}\textbf{154.6$\pm$33.7} &
  \cellcolor[HTML]{cfe2f3}\textbf{0.0$\pm$0.0} &
  \cellcolor[HTML]{FFCCCC}\textbf{336.7$\pm$85.7} &
  \cellcolor[HTML]{cfe2f3}\textbf{56.8$\pm$14.6} &
  39.8$\pm$6.8 &
  59.6$\pm$10.0 &
  \cellcolor[HTML]{FFCCCC}\textbf{117.6$\pm$22.4} &
  88.4$\pm$26.3 &
  \cellcolor[HTML]{FFCCCC}\textbf{132.4$\pm$33.6} &
  \cellcolor[HTML]{FFCCCC}\textbf{154.9$\pm$8.7} &
  \cellcolor[HTML]{FFCCCC}\textbf{146.5$\pm$25.6}  \\
\hline

\multicolumn{1}{l|}{\AlgName }&
  \cellcolor[HTML]{D9EAD3}\textbf{0$\pm$0} &
  \cellcolor[HTML]{D9EAD3}\textbf{0$\pm$0} &
  \cellcolor[HTML]{D9EAD3}\textbf{0$\pm$0} &
  \cellcolor[HTML]{D9EAD3}\textbf{0$\pm$0} &
  \cellcolor[HTML]{D9EAD3}\textbf{0$\pm$0} &
  \cellcolor[HTML]{D9EAD3}\textbf{0$\pm$0} &
  \cellcolor[HTML]{D9EAD3}\textbf{0$\pm$0} &
  \cellcolor[HTML]{D9EAD3}\textbf{0$\pm$0} &
  \cellcolor[HTML]{D9EAD3}\textbf{0$\pm$0} &
  \cellcolor[HTML]{D9EAD3}\textbf{0$\pm$0} &
  \cellcolor[HTML]{D9EAD3}\textbf{0$\pm$0} &
  \cellcolor[HTML]{D9EAD3}\textbf{0$\pm$0} 
  \\ \hline
  
\end{tabular}
 }
 \vspace{-0.9em}
\caption{NCR results under the CIFAR-10 settings with higher poison ratios.}
\label{tab:highpoi}
\vspace{-1em}
\end{table*}

\begin{table*}[t!]
\centering
% \vspace{-2em}
\resizebox{0.9\textwidth}{!}{
\begin{tabular}{l cc|ccc|ccccccc}
\hline
\multicolumn{1}{c|}{}&
  \multicolumn{2}{c|}{\textbf{Label-only}} &
  \multicolumn{3}{c|}{\textbf{Feature-only}} &
  \multicolumn{6}{c}{\textbf{Label-Feature}} &
   \\ \cline{2-13}
\multicolumn{1}{c|}{\multirow{-2}{*}{}} &
  \begin{tabular}[c]{@{}c@{}}Targeted Label-\\ Flipping\cite{tolpegin2020data}\end{tabular} &
  \begin{tabular}[c]{@{}c@{}}Random Label-\\ Flipping\cite{ren2018learning}\end{tabular} &
  \begin{tabular}[c]{@{}c@{}}Clean-\\ Label\cite{turner2019label}\end{tabular} &
  \begin{tabular}[c]{@{}c@{}}Narcissus\\ Backdoor\cite{zeng2022narcissus}\end{tabular} &
  \begin{tabular}[c]{@{}c@{}}Poison\\ Frog\cite{shafahi2018poison}\end{tabular} &
  \begin{tabular}[c]{@{}c@{}}BadNets\\ One-Tar\cite{gu2017badnets}\end{tabular} &
  \begin{tabular}[c]{@{}c@{}}Smooth\\ One-Tar\cite{zeng2021rethinking}\end{tabular} &
  \begin{tabular}[c]{@{}c@{}}IAB\\ One-Tar\cite{nguyen2020input}\end{tabular} &
  \begin{tabular}[c]{@{}c@{}}Blended\\ One-Tar\cite{chen2017targeted}\end{tabular} &
  \begin{tabular}[c]{@{}c@{}}BadNets\\ All-to-all\cite{gu2017badnets}\end{tabular} &
  \begin{tabular}[c]{@{}c@{}}Smooth\\ All-to-all\cite{zeng2021rethinking}\end{tabular} &
  \begin{tabular}[c]{@{}c@{}}Blended\\ All-to-all\cite{chen2017targeted}\end{tabular} \\
  \hline
  
\multicolumn{1}{c|}{\begin{tabular}[c]{@{}c@{}}Poison ratio\end{tabular}}&
  \begin{tabular}[c]{@{}c@{}}Tar: 50\%\end{tabular} &
  \begin{tabular}[c]{@{}c@{}}All: 50\%\end{tabular} &
  \begin{tabular}[c]{@{}c@{}}Tar: 50\%\end{tabular} &
  \begin{tabular}[c]{@{}c@{}}Tar: 50\%\end{tabular} &
  \begin{tabular}[c]{@{}c@{}}Tar: 50\%\end{tabular} &
  \begin{tabular}[c]{@{}c@{}}Tar: 50\%\end{tabular} &
  \begin{tabular}[c]{@{}c@{}}Tar: 50\% \end{tabular} &
  \begin{tabular}[c]{@{}c@{}}Tar: 50\% \end{tabular} &
  \begin{tabular}[c]{@{}c@{}}Tar: 50\%\end{tabular} &
  \begin{tabular}[c]{@{}c@{}}All: 50\%\end{tabular} &
  \begin{tabular}[c]{@{}c@{}}All: 50\%\end{tabular} &
  \begin{tabular}[c]{@{}c@{}}All: 50\%\end{tabular} \\ 
\hline
\multicolumn{1}{l|}{\AlgName }&
  \textbf{35.2$\pm$21.4} &
  \textbf{23.6$\pm$11.6} &
  \textbf{46.1$\pm$15.4} &
  \textbf{58.9$\pm$20.1} &
  \textbf{38.6$\pm$16.2} &
  \textbf{53.6$\pm$14.3} &
  \textbf{54.2$\pm$15.3} &
  \textbf{63.1$\pm$19.2} &
  \textbf{46.4$\pm$10.8} &
  \textbf{56.3$\pm$13.7} &
  \textbf{53.6$\pm$13.3} &
  \textbf{43.7$\pm$10.3} 
  \\ \hline
  
\end{tabular}
 }
 \vspace{-0.9em}
\caption{NCR results of our method when poisons are of the majority under the CIFAR-10 settings.}
\label{tab:highestpoi}
\vspace{-1.5em}
\end{table*}

\vspace{-1.9em}
\subsection{Pseudocode of the Training Stage}
\label{sec:pseudo}
\vspace{-0.5em}

\vspace{-1.5em}
\begin{algorithm}[h!]
\SetKwInput{KwParam}{Parameters}
\algsetup{linenosize=\tiny}
\small
    \caption{Training Algorithm of One Sifter}
    \label{algo:AlgoS}
    \SetNoFillComment
    \KwIn{
    $\theta$ (Classifier Model); $\psi$ (weight-assigning network);
    \\ \quad \quad \quad \ 
    $\Gamma$ (meta-gradient-samplers);
    \\ \quad \quad \quad \ 
    $D$ (dataset that requires sifting);
    }
    \KwOut{$(\theta^*,\psi^*)$ (pair of parameters for one Sifter);
    }
    \KwParam{
    $\alpha,\beta > 0$ (step sizes)
    }
    \BlankLine
    \tcc{A.Warming-up $\theta$, i.e., normal training with all weights setting to 1}
    \For{each $n$-size mini-batch in $D$}{
         $\theta=\theta-\alpha \times \frac{1}{n}\sum_{i=1}^{n}  
        %  \nabla L_{\theta}\left(\theta,z_i\right)
         \frac{\partial L_i(\theta)}{ \partial \theta} 
         $ 
         \;
      }
    \tcc{B. Updating Process}
    \For{each $n$-size mini-batch in $D$}{
          \tcc{1. Virtual-update}
          $\text{Formulate gradients according to (\ref{eq:layers})}$ \;
          \tcc{2. Gradient Sampling}
          $\text{Gradient sampler updating and sampling (\ref{eq:sample})}$ \;
          \tcc{3. Meta-update}
          $\text{Updating $\psi$ with sampled hypergradients (\ref{eqn:sample_grad})}$ \;
          \tcc{4. Actual-update}
          $\text{Updating $\theta$ according to (\ref{eq:actual})}$ \;
      }
      \Return $(\theta^*,\psi^*)$
\end{algorithm}

\vspace{-1.9em}

% \subsection{Detailed Derivation of $\mathbf{g}_\psi$ in Section \ref{sec:overallalgo}}
% \label{sec:meta-gradient}

\vspace{-1em}
\subsection{Detailed Adaptive Designs in Section \ref{sec:advatk}}
\label{sec:adpatksetting}
\vspace{-0.5em}
\noindent
\underline{\textbf{Poisoning the Majority:}}
This adaptive design simply poisons more than 50\% of the samples following existing attack design.
This attack design is effective as it breaks our basic assumption that the poisoned samples of each class account for less than 50\% (Section \ref{sec:probform}). This attack design is less practical as manipulating the majority of the samples is hard.
% For example, to poison the training set of CLIP \cite{radford2021learning} requires the attacker to at least have access and be able to manipulate 200 million image-text pairs from the Internet.

\noindent
\underline{\textbf{Adversarial Noise using a Clean Model:}}
Following the noise synthesis formulation in Eqn. (\ref{eqn:adp_atk_1}), for each given poisoned sample in the original $D$, $z_{poi}=(x,y)$, we use SGD with the Adam optimizer and a learning rate of 0.01 for 100 rounds. 
We do not introduce additional norm constraints on the perturbation
as we find that the converged perturbation does not introduce large visual artifacts.
As we consider a white-box attack scenario, the model structure is the same as what we adopted for the feature extractor in Sifters.

\noindent
\underline{\textbf{Adversarial Noise using Sifters:}}
Following the noise synthesis formulation in Eqn. (\ref{eqn:adp_atk_2}), for each given poisoned sample, $z_{poi}=(x,y)$, we use stochastic gradient ascent with the Adam optimizer and a learning rate of 0.01 for 100 rounds. Note that we consider the original $D$ to follow the same attack settings (poison ratio, trigger design, label manipulation, if applicable) as listed in Table \ref{tab:cifar}, and the adaptive noise is adopted to perturb the original poisoned samples.
% The updated $\tilde{D}$ in which the poisoned samples are perturbed by the optimized adversarial noise is then supplied to the defender.}
% We consider one representative attack from each attack category for evaluation, namely, Targeted Label Flipping, Narcissus Backdoor, and BadNets One-Tar. 
% Other than the limited round of synthesis, we do not introduce additional constraints (e.g., $l_p$-norm constraints) as visual stealthiness depicted by distance is not our concern in this evaluation. 

% Note that the adaptive designs based on adversarial noise (using a clean model or trained Sifters) are based on existing attack settings. In other words, w
% We consider the original $D$ to follow the same attack settings (poison ratio, trigger design, label manipulation, if applicable) as listed in Table \ref{tab:cifar}, and the adaptive noise is adopted to perturb the original poisoned samples in $D$. We consider one representative attack from each attack category for evaluation, namely, Targeted Label Flipping, Narcissus Backdoor, and BadNets One-Tar. 

% \rev{
% Following the above three adaptive attack designs, the updated $\tilde{D}$ will be supplied to the defender. 
% The defender will train another 5 Sifters from scratch (each of the Sifters intake all the samples in $\tilde{D}$ for training) and then score each sample and aggregate the scores for data selection as detailed in Section \ref{sec:overallalgo}. For evaluation, we use the same base set budget (1000) as the main evaluation and measure the NCR of the base set.
% }

\vspace{-1em}
\subsection{Additional Results (CIFAR-10)}
\label{sec:addres}
\vspace{-0.5em}

\noindent
\textbf{Lower Poison Ratios.}
\label{sec:lowpoi}
% \rev{
% \ruoxi{this part's clarity can be improved. Start by saying the in Table \ref{tab:attacks} in the main text, we follow the original poison ratios.}
% Table \ref{tab:attacks} in the main text adopted the original poison ratios following the original papers of these attacks.
Table \ref{tab:attacks} in the main text uses the poison ratios reported in the original papers for these attacks.
The attacks can achieve similar attack effects but with a smaller poison ratio. Primarily, many defense works have found attacks with lower poison ratios are harder to be identified \cite{xiang2022post,liu2019abs}. Thus, we present additional results of \AlgName on CIFAR-10 with smaller poison ratios in Table \ref{tab:lowpoi}. 
In particular, for each evaluated attack, we re-examine \AlgName with two settings of smaller poison ratios (half of or quarter of the original poison ratio) listed in Table \ref{tab:lowpoi}. We found \AlgName can still reliably provide high-precision sifting results.
% }

\begin{figure}[t!]
  \centering
  % \vspace{-2em}
  \includegraphics[width=0.8\linewidth]{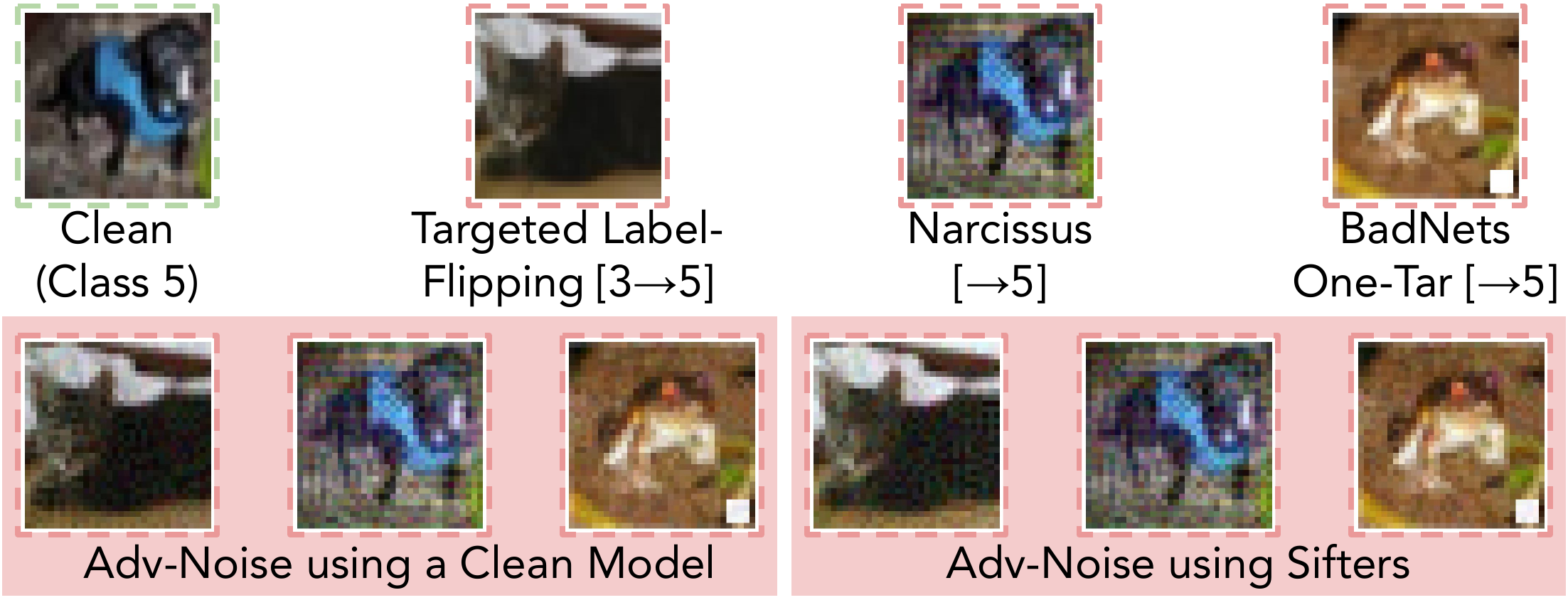}
  \vspace{-1em}
  \caption{
  Visual results of the original poisoned samples and the adversarial noise disguised poisoned samples (CIFAR-10).
  }
  \label{fig:adv_visual}
  \vspace{-1.9em}
\end{figure}

\noindent
\textbf{Higher Poison Ratios.}
\label{sec:highpoi}
% \rev{
In addition to lower poison ratios, we also evaluate whether $\AlgName$'s effectiveness can extend to high poison ratios. 
% higher ones are also recognized as a significant challenge in the field.
We tested the competitors and our proposed solution on corrupted datasets with higher poison ratios (40\%) in Table \ref{tab:highpoi}.
% \ruoxi{be specific about how much higher?} and depict the results in Table \ref{tab:highpoi}. 
% A higher poison ratio resulting some methods being ineffective due to the creation of more diverse poisoning features and a shift in the averaging representative feature space mapping of the dataset. This 
A higher poison ratio results in a significant decrease in the performance of methods that utilize model output features, such as DCM, MI-DCM, and SF-Least. This is because as the poison ratio gets higher, the features of poisoned samples are getting more diverse. At the same time, the inclusion of a larger number of poisoned samples in the optimization process results in a heightened difficulty in reducing the average loss of poison samples. As a result, the efficacy of the Loss-Scan method is also impacted. However, we found \AlgName generates a completely clean base set consistently under the evaluated higher poison ratios.

\noindent
\textbf{Poisoning the Majority.}
\label{sec:highestpoi}
% \rev{
Elaborated in Section \ref{sec:probform}, we assume that the number of poison samples should be less than that of clean samples. This assumption is plausible under normal circumstances. Now we study how our method will perform in an extreme case where the attacker controls more than half of the samples. We evaluate the performance of our method through a series of experiments with each attack using a large poison ratio over 50\% in Table \ref{tab:highestpoi}. Noting this attack setting is a plausible but less practical adaptive attack as discussed in Section \ref{sec:advatk}. The results show that our defense can no longer sift out a purely clean base set since the underlying assumption is not met. Nevertheless, poisoning the majority of the dataset in practice is hard, e.g., to poison the training set of CLIP \cite{radford2021learning} requires the attacker to at least have access and be able to manipulate 200 million image-text pairs.
% \minz{
% In Section \ref{sec:probform}, we have a basic assumption that the number of poison samples will be less than that of clean samples. This assumption is fully established under normal circumstances. However, if in extreme cases where this assumption is broken. That is to say, the attacker attacked more than half of the samples, so how will our method perform? we evaluated the performance of our method through a series of experiments, the results of which are documented in Table \ref{tab:highestpoi}. From the results, we can see that our defense is no more sifiting out a pure base set since the underlying assumptions are broken. Nevertheless, more than half of the samples are poison in the actual scenario is very difficult to meet, therefore the danger of this condition is not as serious as it seems.
% }

\begin{table}[t!]
\centering
% \vspace{-2em}
\resizebox{0.6\columnwidth}{!}{
\begin{tabular}{c|c|c|c}
\hline
\multicolumn{1}{c|}{} &
  \begin{tabular}[c]{@{}c@{}}Targeted Label-\\ Flipping\cite{tolpegin2020data}\end{tabular} &
  \begin{tabular}[c]{@{}c@{}}Narcissus\\ Backdoor\cite{zeng2022narcissus}\end{tabular} &
  \begin{tabular}[c]{@{}c@{}}BadNets\\ One-Tar\cite{gu2017badnets}\end{tabular} \\ \hline
  % \textbf{\begin{tabular}[c]{@{}c@{}}Attack\\ Settings\end{tabular}} &
  % \begin{tabular}[c]{@{}c@{}}$\left [ 3\rightarrow 5 \right ]$;\\ Tar: 16.67\%\end{tabular} &
  % \begin{tabular}[c]{@{}c@{}}$\left [5 \right ]$;\\ Tar: 10\%\end{tabular} &
  % \begin{tabular}[c]{@{}c@{}}$\left [5 \right ]$;\\ Tar: 33\%\end{tabular}  \\ 
  % \hline
  % \textbf{\begin{tabular}[c]{@{}c@{}}Original Attack\\ Results (\%)\end{tabular}} &
  % \begin{tabular}[c]{@{}c@{}}ACC: 91.77\\ ASR: 27.56\\ Tar-ACC: 83.78\end{tabular} &
  % \begin{tabular}[c]{@{}c@{}}ACC: 93.26\\ ASR: 100\end{tabular} &
  % \begin{tabular}[c]{@{}c@{}}ACC: 84.03\\ ASR: 95.78\end{tabular} \\
  % \hline
  \hline
  \textbf{\begin{tabular}[c]{@{}c@{}}Cln-Model-Based\\Adv-Noise (\%)\end{tabular}} &
  \begin{tabular}[c]{@{}c@{}}ACC: 92.01\\ ASR: 6.35\\ Tar-ACC: 88.08\end{tabular} &
  \begin{tabular}[c]{@{}c@{}}ACC: 93.31\\ ASR: 5.78\end{tabular} &
  \begin{tabular}[c]{@{}c@{}}ACC: 85.36\\ ASR: 88.92\end{tabular} \\ 
  \hline
  \textbf{\begin{tabular}[c]{@{}c@{}}\AlgName\end{tabular}} &
  \cellcolor[HTML]{D9EAD3}\textbf{0$\pm$0} &
  \cellcolor[HTML]{D9EAD3}\textbf{0$\pm$0} &
  \cellcolor[HTML]{D9EAD3}\textbf{0$\pm$0}   \\ 
  \hline
  \hline
  \textbf{\begin{tabular}[c]{@{}c@{}}Sifter-Based\\Adv-Noise (\%)\end{tabular}} &
  \begin{tabular}[c]{@{}c@{}}ACC: 91.56\\ ASR: 45.71\\ Tar-ACC: 82.51\end{tabular} &
  \begin{tabular}[c]{@{}c@{}}ACC: 93.19\\ ASR: 9.81\end{tabular} &
  \begin{tabular}[c]{@{}c@{}}ACC: 84.33\\ ASR: 91.64\end{tabular} \\ 
  \hline
  \textbf{\begin{tabular}[c]{@{}c@{}} \AlgName \end{tabular}} &
  \cellcolor[HTML]{D9EAD3}\textbf{0$\pm$0} &
  \cellcolor[HTML]{D9EAD3}\textbf{0$\pm$0} &
  \cellcolor[HTML]{D9EAD3}\textbf{0$\pm$0}   \\ 
  \hline
\end{tabular}}
\vspace{-.9em}
\caption{
Adversarial-noise-based adaptive attacks' effects on the original attack results and the results of \AlgName.}
\label{tab:adv_result}
\vspace{-1.9em}
\end{table}

\begin{table*}[t!]
% \vspace{-2em}
\centering
\resizebox{0.9\textwidth}{!}{
\begin{tabular}{l cc|cc|ccccccc||c}
\hline
\multicolumn{1}{c|}{}&
  \multicolumn{2}{c|}{\textbf{Label-only}} &
  \multicolumn{2}{c|}{\textbf{Feature-only}} &
  \multicolumn{7}{c||}{\textbf{Label-Feature}} &
   \\ \cline{2-12}
\multicolumn{1}{c|}{\multirow{-2}{*}{}} &
  \begin{tabular}[c]{@{}c@{}}Targeted Label-\\ Flipping\cite{tolpegin2020data}\end{tabular} &
  \begin{tabular}[c]{@{}c@{}}Random Label-\\ Flipping\cite{ren2018learning}\end{tabular} &
%   \begin{tabular}[c]{@{}c@{}}Clean-\\ Label\cite{turner2019label}\end{tabular} &
  \begin{tabular}[c]{@{}c@{}}Narcissus\\ Backdoor\cite{zeng2022narcissus}\end{tabular} &
  \begin{tabular}[c]{@{}c@{}}Poison\\ Frog\cite{shafahi2018poison}\end{tabular} &
  \begin{tabular}[c]{@{}c@{}}BadNets\\ One-Tar\cite{gu2017badnets}\end{tabular} &
  \begin{tabular}[c]{@{}c@{}}Smooth\\ One-Tar\cite{zeng2021rethinking}\end{tabular} &
  \begin{tabular}[c]{@{}c@{}}IAB\\ One-Tar\cite{nguyen2020input}\end{tabular} &
  \begin{tabular}[c]{@{}c@{}}Blended\\ One-Tar\cite{chen2017targeted}\end{tabular} &
  \begin{tabular}[c]{@{}c@{}}BadNets\\ All-to-all\cite{gu2017badnets}\end{tabular} &
  \begin{tabular}[c]{@{}c@{}}Smooth\\ All-to-all\cite{zeng2021rethinking}\end{tabular} &
  \begin{tabular}[c]{@{}c@{}}Blended\\ All-to-all\cite{chen2017targeted}\end{tabular} &
  \multirow{-2}{*}{\textbf{Overhead (s)}} \\ \hline

\multicolumn{1}{l|}{DCM} &
  \cellcolor[HTML]{FFCCCC}\textbf{103.1} &
  \cellcolor[HTML]{FFCCCC}\textbf{107.5} &
%   \cellcolor[HTML]{cccccc}\textbf{NA} &
  37.7 &
  75.5 &
  29.6 &
  44.4 &
  87.0 &
  \cellcolor[HTML]{cfe2f3}\textbf{44.4} &
  38.5 &
  \cellcolor[HTML]{FFCCCC}\textbf{121.0} &
  \cellcolor[HTML]{FFCCCC}\textbf{108.0} &
  10 \\
\multicolumn{1}{l|}{MI-DCM$^*$} &
  28.1$\pm$0 &
  \cellcolor[HTML]{FFCCCC}\textbf{112.5$\pm$0} &
%   \cellcolor[HTML]{cccccc}\textbf{NA} &
  \cellcolor[HTML]{FFCCCC}\textbf{1000$\pm$0} &
  \cellcolor[HTML]{FFCCCC}\textbf{528.3$\pm$0} &
  51.9$\pm$0 &
  \cellcolor[HTML]{FFCCCC}\textbf{196.3$\pm$0} &
  \cellcolor[HTML]{FFCCCC}\textbf{105.1} &
  \cellcolor[HTML]{FFCCCC}\textbf{108.9$\pm$0} &
  \cellcolor[HTML]{FFCCCC}\textbf{118.5$\pm$0} &
  30.5$\pm$0 &
  41.2$\pm$0 &
  480+300 \\
\multicolumn{1}{l|}{SF-Least$^*$ }&
  \cellcolor[HTML]{cfe2f3}\textbf{0.0$\pm$0} &
  \cellcolor[HTML]{cfe2f3}\textbf{18.5$\pm$1.73} &
%   \cellcolor[HTML]{cccccc}\textbf{NA} &
  75.47$\pm$0 &
  \cellcolor[HTML]{FFCCCC}\textbf{113.21$\pm$0} &
  74.1$\pm$0 &
  62.96$\pm$0 &
  63.0$\pm$0 &
  54.8$\pm$0 &
  \cellcolor[HTML]{cfe2f3}\textbf{1.0$\pm$0} &
  \cellcolor[HTML]{cfe2f3}\textbf{3.33$\pm$0.58} &
  \cellcolor[HTML]{cfe2f3}\textbf{3.33$\pm$0} &
  480+15 \\

\multicolumn{1}{l|}{Loss-Scan$^*$ }&
  \cellcolor[HTML]{FFCCCC}\textbf{347.9$\pm$56.4} &
  \cellcolor[HTML]{FFCCCC}\textbf{507.0$\pm$0.0} &
%   \cellcolor[HTML]{cccccc}\textbf{NA} &
  \cellcolor[HTML]{cfe2f3}\textbf{32.6$\pm$2.35} &
  \cellcolor[HTML]{cfe2f3}\textbf{10$\pm$0.0} &
  \cellcolor[HTML]{cfe2f3}\textbf{23.7$\pm$6.73} &
  \cellcolor[HTML]{cfe2f3}\textbf{23.7$\pm$5.45} &
  \cellcolor[HTML]{cfe2f3}\textbf{45.0$\pm$6.47} &
  63.7$\pm$2.67 &
  \cellcolor[HTML]{FFCCCC}\textbf{325.0$\pm$56.1} &
  \cellcolor[HTML]{FFCCCC}\textbf{285.5$\pm$104.3} &
  \cellcolor[HTML]{FFCCCC}\textbf{271.6$\pm$57.9} &
  110 \\

\multicolumn{1}{l|}{Self-IF$^*$ }&
  \cellcolor[HTML]{FFCCCC}\textbf{128.1$\pm$5.41} &
  \cellcolor[HTML]{FFCCCC}\textbf{107.5$\pm$0} &
%   \cellcolor[HTML]{cccccc}\textbf{NA} &
  \cellcolor[HTML]{FFCCCC}\textbf{125.8$\pm$10.9} &
  \cellcolor[HTML]{FFCCCC}\textbf{245.3$\pm$0} &
  37.0$\pm$7.41 &
  51.9$\pm$0 &
  84.0$\pm$0 &
  54.8$\pm$1.21 &
  \cellcolor[HTML]{FFCCCC}\textbf{110.5$\pm$0} &
  \cellcolor[HTML]{FFCCCC}\textbf{105.5$\pm$0} &
  \cellcolor[HTML]{FFCCCC}\textbf{102.2$\pm$0} &
  \cellcolor[HTML]{FFE599}\textbf{480+17432} \\
\hline
\multicolumn{1}{l|}{\AlgName }&
  \cellcolor[HTML]{D9EAD3}\textbf{0$\pm$0} &
  \cellcolor[HTML]{D9EAD3}\textbf{0$\pm$0} &
  \cellcolor[HTML]{D9EAD3}\textbf{0$\pm$0} &
%   \cellcolor[HTML]{D9EAD3}\textbf{0$\pm$0} &
  \cellcolor[HTML]{D9EAD3}\textbf{0$\pm$0} &
  \cellcolor[HTML]{D9EAD3}\textbf{0$\pm$0} &
  \cellcolor[HTML]{D9EAD3}\textbf{0$\pm$0} &
  \cellcolor[HTML]{D9EAD3}\textbf{0$\pm$0} &
  \cellcolor[HTML]{D9EAD3}\textbf{0$\pm$0} &
  \cellcolor[HTML]{D9EAD3}\textbf{0$\pm$0} &
  \cellcolor[HTML]{D9EAD3}\textbf{0$\pm$0} &
  \cellcolor[HTML]{D9EAD3}\textbf{0$\pm$0} &
  5$\times$80 \\ \hline
\end{tabular}
}
\vspace{-.9em}
\caption{NCR results under the GTSRB settings.}
\label{tab:gtsrb}
\vspace{-1em}
\end{table*}

\begin{table*}[t!]
\centering
\resizebox{0.7\textwidth}{!}{
\begin{tabular}{l cc|cc|ccccc||c}
\hline
\multicolumn{1}{c|}{}&
  \multicolumn{2}{c|}{\textbf{Label-only}} &
  \multicolumn{2}{c|}{\textbf{Feature-only}} &
  \multicolumn{5}{c||}{\textbf{Label-Feature}} &
   \\ \cline{2-10}
\multicolumn{1}{c|}{\multirow{-2}{*}{}} &
  \begin{tabular}[c]{@{}c@{}}Targeted Label-\\ Flipping\cite{tolpegin2020data}\end{tabular} &
  \begin{tabular}[c]{@{}c@{}}Random Label-\\ Flipping\cite{ren2018learning}\end{tabular} &
%   \begin{tabular}[c]{@{}c@{}}Clean-\\ Label\cite{turner2019label}\end{tabular} &
  \begin{tabular}[c]{@{}c@{}}Narcissus\\ Backdoor\cite{zeng2022narcissus}\end{tabular} &
  \begin{tabular}[c]{@{}c@{}}Poison\\ Frog\cite{shafahi2018poison}\end{tabular} &
  \begin{tabular}[c]{@{}c@{}}BadNets\\ One-Tar\cite{gu2017badnets}\end{tabular} &
  \begin{tabular}[c]{@{}c@{}}IAB\\ One-Tar\cite{nguyen2020input}\end{tabular} &
  \begin{tabular}[c]{@{}c@{}}Blended\\ One-Tar\cite{chen2017targeted}\end{tabular} &
  \begin{tabular}[c]{@{}c@{}}BadNets\\ All-to-all\cite{gu2017badnets}\end{tabular} &
  \begin{tabular}[c]{@{}c@{}}Blended\\ All-to-all\cite{chen2017targeted}\end{tabular} &
  \multirow{-2}{*}{\textbf{Overhead (s)}} \\ \hline
  
\multicolumn{1}{l|}{DCM} &

  \cellcolor[HTML]{FFCCCC}\textbf{173.3} &
  \cellcolor[HTML]{cfe2f3}\textbf{97.1} &
%   \cellcolor[HTML]{cccccc}\textbf{NA} &
  \cellcolor[HTML]{FFCCCC}\textbf{135.0} &
  90.0 &
   74.1 &
   \cellcolor[HTML]{cfe2f3}\textbf{24.0} &
  88.9 &
  \cellcolor[HTML]{FFCCCC}\textbf{148.1} &
  \cellcolor[HTML]{FFCCCC}\textbf{135.1} &
  10 \\
\multicolumn{1}{l|}{MI-DCM$^*$} &
  \cellcolor[HTML]{cfe2f3}\textbf{38.4$\pm$0} &
  \cellcolor[HTML]{FFCCCC}\textbf{100.3$\pm$5.83} &
%   \cellcolor[HTML]{cccccc}\textbf{NA} &
 \cellcolor[HTML]{cfe2f3}\textbf{74.1$\pm$0} &
  86.4$\pm$21.4 &
  \cellcolor[HTML]{FFCCCC}\textbf{167.5$\pm$4.33} &
  81.1$\pm$0 &
  \cellcolor[HTML]{FFCCCC}\textbf{142.5$\pm$26.0} &
 55.5$\pm$9.6 &
  \cellcolor[HTML]{cfe2f3}\textbf{41.3$\pm$2.25} &
  2100+300 \\
\multicolumn{1}{l|}{SF-Least$^*$} &
  \cellcolor[HTML]{FFCCCC}\textbf{160.3$\pm$11.1} &
%   \cellcolor[HTML]{cccccc}\textbf{NA} &
  \cellcolor[HTML]{FFCCCC}\textbf{187.3$\pm$8.69} &
  \cellcolor[HTML]{cfe2f3}\textbf{74.1$\pm$0} &
  \cellcolor[HTML]{FFCCCC}\textbf{111.1$\pm$0} &
  87.5$\pm$4.33 &
  78.0$\pm$4.33 &
  \cellcolor[HTML]{FFCCCC}\textbf{122.5$\pm$4.33} &
  \cellcolor[HTML]{cfe2f3}\textbf{40.3$\pm$1.26} &
  92.8$\pm$4.37 &
  2100+15 \\
\multicolumn{1}{l|}{Loss-Scan$^*$} &
  \cellcolor[HTML]{FFCCCC}\textbf{167.9$\pm$56.3} &
%   \cellcolor[HTML]{cccccc}\textbf{NA} &
  \cellcolor[HTML]{FFCCCC}\textbf{438.5$\pm$118.3} &
  90.0$\pm$0 &
  \cellcolor[HTML]{cfe2f3}\textbf{60.0$\pm$0} &
  \cellcolor[HTML]{cfe2f3}\textbf{63.0$\pm$9.87} &
  54.1$\pm$6.78 &
  \cellcolor[HTML]{cfe2f3}\textbf{56.7$\pm$6.57} &
  \cellcolor[HTML]{FFCCCC}\textbf{193.5$\pm$56.1} &
  \cellcolor[HTML]{FFCCCC}\textbf{192.8$\pm$16.9} &
  140 \\
\multicolumn{1}{l|}{Self-IF$^*$ }&
  \cellcolor[HTML]{FFCCCC}\textbf{179.5$\pm$61.8} &
  \cellcolor[HTML]{FFCCCC}\textbf{108.7$\pm$6.83} &
%   \cellcolor[HTML]{cccccc}\textbf{NA} &
  \cellcolor[HTML]{cfe2f3}\textbf{74.1$\pm$0} &
  86.4$\pm$21.4 &
  85.0$\pm$11.6 &
  63.0$\pm$11.6 &
  57.5$\pm$11.5 &
  \cellcolor[HTML]{FFCCCC}\textbf{111.2$\pm$12.8} &
  \cellcolor[HTML]{FFCCCC}\textbf{121.2$\pm$8.39} &
  \cellcolor[HTML]{FFE599}\textbf{2100+21045} \\

\hline
\multicolumn{1}{l|}{\AlgName }&
  \cellcolor[HTML]{D9EAD3}\textbf{0$\pm$0} &
  \cellcolor[HTML]{D9EAD3}\textbf{0$\pm$0} &
%   \cellcolor[HTML]{D9EAD3}\textbf{0$\pm$0} &
  \cellcolor[HTML]{D9EAD3}\textbf{0$\pm$0} &
  \cellcolor[HTML]{D9EAD3}\textbf{0$\pm$0} &
  \cellcolor[HTML]{D9EAD3}\textbf{0$\pm$0} &
  \cellcolor[HTML]{D9EAD3}\textbf{0$\pm$0} &
  \cellcolor[HTML]{D9EAD3}\textbf{0$\pm$0} &
  \cellcolor[HTML]{D9EAD3}\textbf{0$\pm$0} &
  \cellcolor[HTML]{D9EAD3}\textbf{0$\pm$0} &
  5$\times$80 \\ \hline
\end{tabular}
}
\vspace{-.8em}
\caption{NCR results under the PubFig settings.}
\label{tab:pubfig}
\vspace{-1.5em}
\end{table*}

\noindent
\textbf{Adaptive Attacks via Adversarial Noise.}
% We show the detailed attack results using the two adaptive designs based on adversarial noise in Section \ref{sec:advatk}. 
For the clean-model-based design, we first train $\theta_{\text{cln}}$ using the clean portion of $D$, achieving an average training loss of $0.001$. Then, we use this model to synthesize the noise following Eqn. (\ref{eqn:adp_atk_1}), for each $z_{poi}$. The average loss over these poisoned samples dropped from $5.135$ to $0.0015$ (with $\tilde{z}_{poi}$). For the Sifter-based design, we plug 5 trained Sifters (which can sift out a clean base set of size 1000 from $D$) into Eqn. (\ref{eqn:adp_atk_2}) for noise synthesis. Then, with the updated dataset $\tilde{D}$, we find the old five Sifters trained using $D$ will select only poisons for the target class to form the base set. Both results indicate that the synthesized noise aligns with our optimization goals, as we expected.
The poisoned samples perturbed by the synthesized noises are visualized in Figure \ref{fig:adv_visual}. 
The attack performance on models trained over $\tilde{D}$ and the sifting results with \AlgName are shown in Table \ref{tab:adv_result}. Despite the noises being effective in achieving low loss on the clean model or being selected by the old Sifters, \AlgName remains a 0\% NCR when applied to $\tilde{D}$ that contains adaptively designed poisons. 
% The results remark that it is hard to adaptively attack \AlgName.

\vspace{-1.5em}
\subsection{Additional Results (GTSRB, PubFig)}
\label{sec:addresult}
\vspace{-0.5em}
We further study the sifting effectiveness of different methods over the popular traffic signs dataset GTSRB in Table \ref{tab:gtsrb} and the real-world face dataset PubFig in Table \ref{tab:pubfig}. 
These datasets are more representative of ML tasks encountered in real-world applications. 
Both GTSRB and PubFig are imbalanced, unlike CIFAR-10. 
Therefore,  we maintain the same class distribution when sifting the samples from each class.
Again, similar to the observation in the main text, none of the baselines can constantly produce a fully clean base set for these two datasets.
Even with datasets more challenging than CIFAR-10, \AlgName continues to deliver a completely clean base set with zero variance in each evaluated attack setting and also is faster than the other baselines.

\vspace{-1.5em}
\subsection{Ablation Study}
\label{sec:ablation}
\vspace{-0.5em}
% In this ablation study, we examine the effects caused by changes in various hyperparameters employed by \AlgName. 
Throughout this study, we compare the empirical upper bound on the size of the clean subset sifted under varying parameter settings. Note that this upper bound value implies that for given parameters, \AlgName can sift out a clean base set (of size equal to the upper bound) with 100\% precision.

\begin{table}[h!]
\centering
\vspace{-.8em}
\resizebox{0.75\linewidth}{!}{
\begin{tabular}{c|cccc} 
\hline
\textbf{\# Sifter}    & 1   & 3   & 5   & 10   \\ 
\hline
\textbf{Clean subset size} & 155($\times$10) & 274($\times$10) & 517($\times$10) & 533($\times$10)  \\
\hline
\end{tabular}
}
\vspace{-.9em}
\caption{Sifting performance vs. number of Sifters.}
% CIFAR-10, BadNets (5\%).}
\label{tab:numsifters}
\vspace{-1.5em}
\end{table}

First, we study the effect of adding additional Sifters to select the clean base set (Table \ref{tab:numsifters}). We consider the CIFAR-10 \cite{krizhevsky2009learning} for all ablation studies. The hyperparameter settings remain as declared in Table \ref{tab:datasets}, except we use the ResNet-18 \cite{he2016deep} architecture. We poison the dataset with the BadNets \cite{gu2017badnets} in this study with a 5\% poison ratio into target class 5. 
We observe that as the number of Sifters increases, the upper bound on the base set size also increases. The subset size grows sharply from one to five Sifters, i.e., from 155$(\times 10)$ to 517$(\times 10)$. Although the subset size keeps growing after five Sifters, the pace is comparatively gradual.
% (where ten Sifters can form a balanced, clean base set of size 533$(\times 10)$). 
% It is important to note that we use five Sifters for our experiments in the evaluation section of the paper. 

\begin{figure}[h!]
  \centering
  \vspace{-1.em}
  \includegraphics[width=0.86\linewidth]{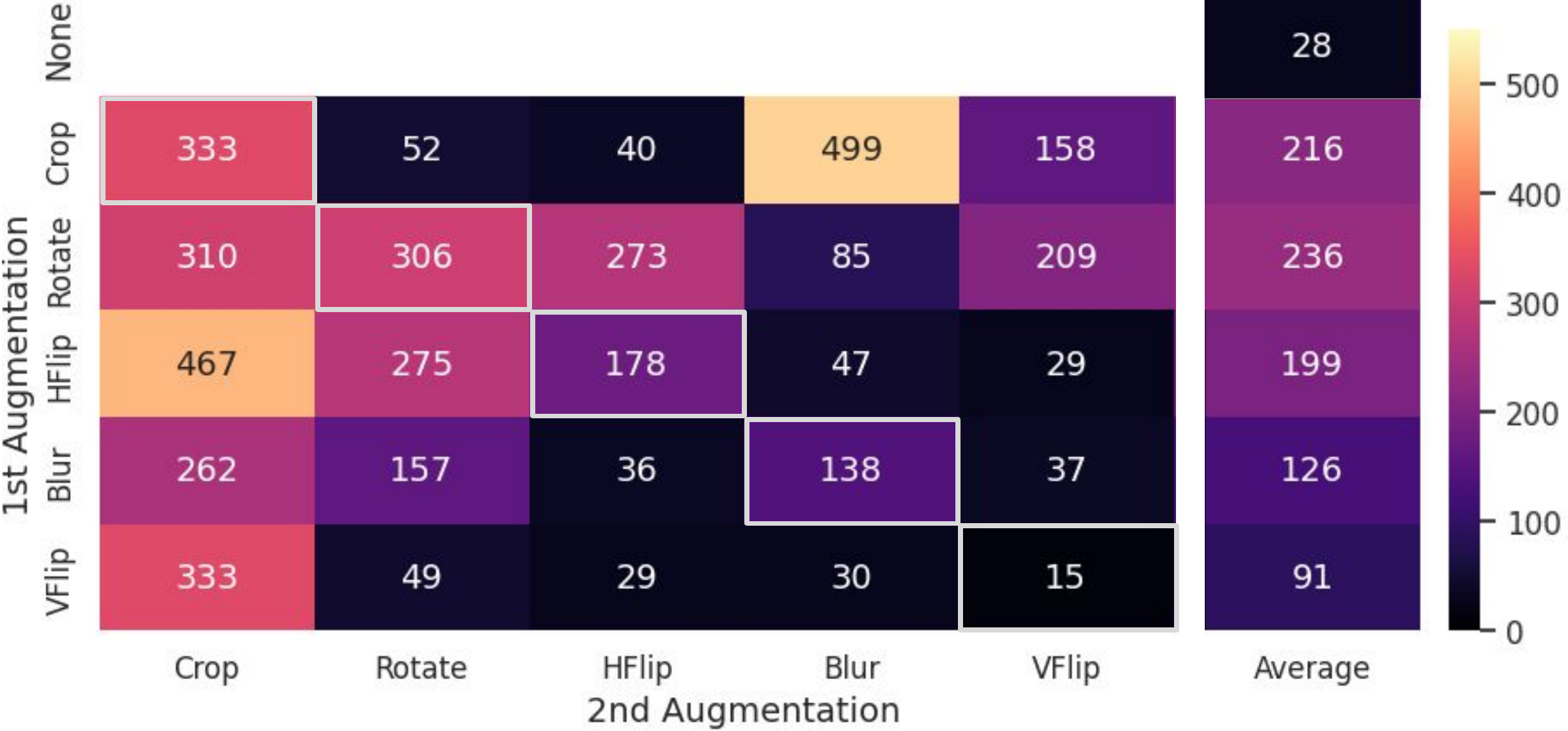}
  \vspace{-1.1em}
  \caption{Clean base set size with different sample-dilution settings. 
  % For all columns but the last, diagonal entries correspond to using only one single transformation, and off-diagonals correspond to the composition of two transformations (applied sequentially). 
  The last column reflects the average over the row. 
  % CIFAR-10, BadNets (5\%).
  }
  \label{fig:heatmap}
  \vspace{-1.3em}
\end{figure}

Secondly, we discuss the impact of different random augmentations used in sample-dilution towards the size of achievable clean base set (Figure \ref{fig:heatmap}). 
% \rev{
We adopt the widely used augmentation settings for all experiments throughout this paper. Specifically, for Random Crop, the image size is 32px, and the padding size is 4px; for Gaussian Blur, the kernel size is 3px; for Random Horizontal \& Vertical Flip, the probability is 0.5. 
From Figure \ref{fig:heatmap}, we observe that different augmentations impact the size of the clean base set greatly. 
For example, when only applying Random Horizontal Flip, we can acquire a clean set of 178$(\times 10)$ samples. However, the sole application of Random Vertical Flip results in a size of only 15$(\times 10)$ samples. We also found combinations of the augmentations lead to better performance, e.g., combining Random Crop with Gaussian blur increases the size from 333$(\times 10)$ to 499$(\times 10)$. In our evaluation, we use no more than four random augmentations in the sample-dilution since too many augmentations may lead to large information loss. Note that, without any augmentation being adopted, our method can still sift out a clean base set of 28$(\times 10)$. In comparison, other baselines' best results with or without sample dilution is only 15$(\times 10)$, Table \ref{tab:otheraug}. It also goes without saying that these baseline approaches have inconsistent performance against different types of attacks, and they might even have worse performance against other settings of attacks (Table \ref{table:fail},\ref{tab:cifar},\ref{tab:imagenet},\ref{tab:gtsrb},\ref{tab:pubfig}).
% } 

% Additionally, we ablate the effect of the sample-dilution to the other baselines in Table \ref{tab:otheraug}. We find that the sample-dilution is not an decisive factor that can help the evaluated base lines to achieve high-precision results.

% \begin{table}[h]
% \centering
% \resizebox{0.75\linewidth}{!}{
% \begin{tabular}{c|cc|cc|cc}
% \hline
% \multicolumn{1}{c|}{}&
%  \multicolumn{2}{c|}{\textbf{Without dilution}} &
%  \multicolumn{2}{c|}{\textbf{With dilution}} &
%   \multicolumn{2}{c}{\textbf{Change}} \\
% \multicolumn{1}{c|}{}&
%  Start to bad &
%  NCR &
%  Start to bad &
%  NCR &
%  Start to bad &
%  NCR \\
%   \hline
%     DCM &     7 & 111.0 & 3 & 102.0 & \cellcolor[HTML]{FFCCCC}\textbf{-4}& \cellcolor[HTML]{cfe2f3}\textbf{-9.0} \\
%     MI-DCM$^*$ &    12 & 114.0 & 9 & 99.1 & \cellcolor[HTML]{FFCCCC}\textbf{-3}& \cellcolor[HTML]{cfe2f3}\textbf{-14.9}\\
%     SF-Least$^*$ & 11 & 72.0 & 15 & 55.0 & \cellcolor[HTML]{cfe2f3}\textbf{+4}& \cellcolor[HTML]{cfe2f3}\textbf{-17.0}\\
%     Loss-Scan$^*$& 6 & 99.0 & 2 & 285.0 & \cellcolor[HTML]{FFCCCC}\textbf{-4}& \cellcolor[HTML]{FFCCCC}\textbf{+186.0}\\
%     Self-IF$^*$ & 8  & 89.0 & 14  & 166.0 & \cellcolor[HTML]{cfe2f3}\textbf{+6}&\cellcolor[HTML]{FFCCCC}\textbf{+77.0}\\ \hline
% \end{tabular}
% }
% \caption{Baseline results with sample-dilution on CIFAR-10 Badnets, 5\% poison rates.}
% \label{tab:otheraug}
% %\vspace{-1.5em}
% \end{table}

\begin{table}[t!]
\centering
\vspace{-.3em}
\resizebox{0.6\linewidth}{!}{
\begin{tabular}{c|cc|cc|cc}
\hline
\multicolumn{1}{c|}{}&
 Without dilution &
 With dilution &
  Change \\
  \hline
    DCM &     7$(\times 10)$  & 3$(\times 10)$  & \cellcolor[HTML]{FFCCCC}\textbf{-4$(\times 10)$}  \\
    MI-DCM$^*$ &    12$(\times 10)$  & 9$(\times 10)$  & \cellcolor[HTML]{FFCCCC}\textbf{-3$(\times 10)$} \\
    SF-Least$^*$ & 11$(\times 10)$  & 15$(\times 10)$  & \cellcolor[HTML]{cfe2f3}\textbf{+4$(\times 10)$} \\
    Loss-Scan$^*$& 6$(\times 10)$  & 2$(\times 10)$  & \cellcolor[HTML]{FFCCCC}\textbf{-4$(\times 10)$} \\
    Self-IF$^*$ & 8$(\times 10)$   & 14$(\times 10)$   & \cellcolor[HTML]{cfe2f3}\textbf{+6$(\times 10)$}\\ \hline
\end{tabular}
}
\vspace{-.9em}
\caption{Baseline results with sample-dilution.}
% CIFAR-10, BadNets (5\%).}
\label{tab:otheraug}
\vspace{-1.9em}
\end{table}

\vspace{-1em}
\subsection{Visual Examples of the Selected Points.}
\vspace{-.5em}

\begin{figure}[h]
  \centering
  \vspace{-1em}
\includegraphics[width=0.9\linewidth]{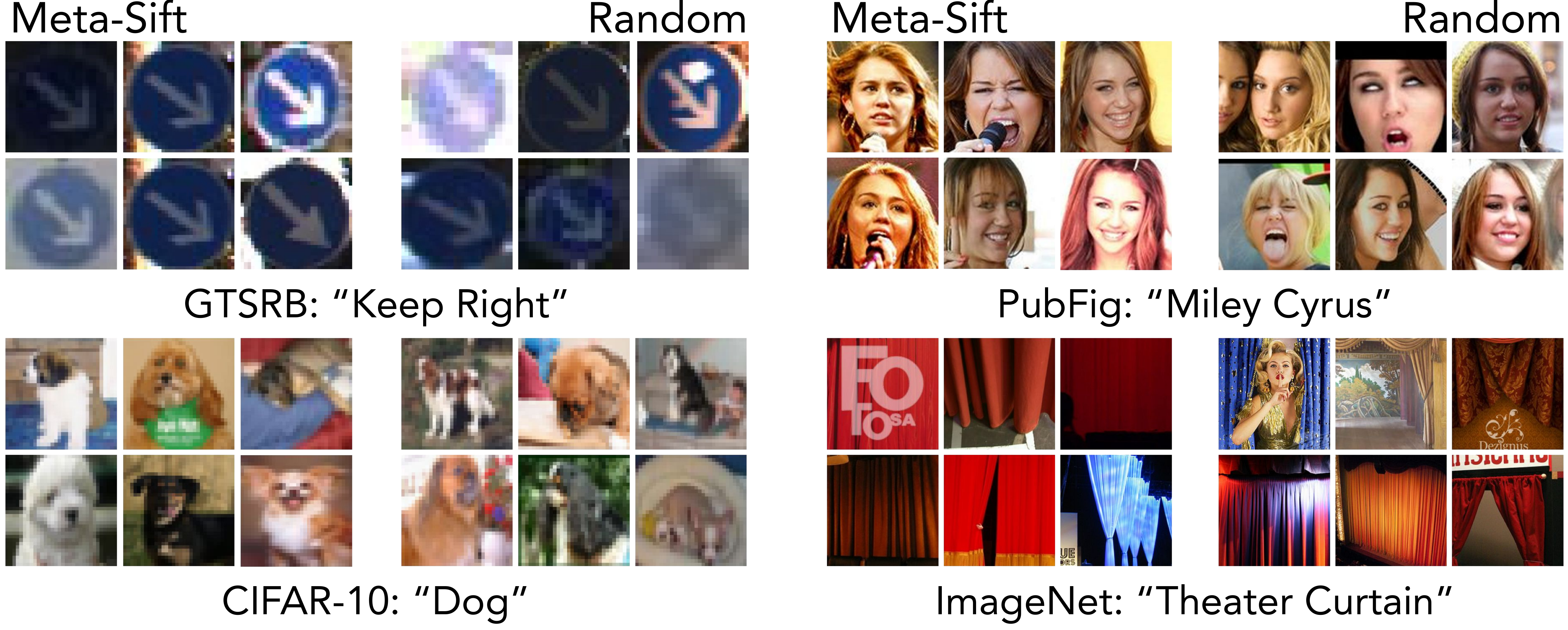}
  \vspace{-1em}
  \caption{Selected sample comparison. 
  % GTSRB: class 38, ``Keep Right''; CIFAR-10: class 5, ``Dog''; PubFig: class 60, ``Miley Cyrus''; ImageNet: class 854, ``Theater Curtain.''
  }
  \label{fig:example_cifar}
  % \vspace{-1em}
\end{figure}

% Figure \ref{fig:example_cifar} depicts the visual examples of the selected data using the \AlgName or randomly sampled clean samples. 
% It seems the samples selected by \AlgName contain slightly better semantic information towards the label than random selected clean samples. However, it is still hard to conclude the sample consistency directly from visual observations.
Figure \ref{fig:example_cifar} show examples of data selected by the \AlgName or random-sampled. The \AlgName-selected samples exhibit slightly better semantic information for the label In comparison. However, it is hard to draw conclusions about sample consistency based solely on visual observations

% \begin{table}
% \centering
% \begin{tabular}{c|cccc} 
% \hline
% \textbf{Model}             & Resnet 18 & Resnet 34 & Resnet 50 & Resnet 101  \\ 
% \hline
% \textbf{Clean subset size} & 467       & 499       & 534       & 310         \\
% \hline
% \end{tabular}
% \caption{Models}
% \end{table}

\end{document}